      [1999/12/01 v1.4c Il Nuovo Cimento]
\documentclass{cimento}

\usepackage{graphicx}
\usepackage{epsfig}
\usepackage{amsmath}
\usepackage{latexsym}
\usepackage{amssymb}
\usepackage{dsfont}
\usepackage{color}
\usepackage{multirow}
\usepackage{enumitem}
\newcommand{\di}{ {\rm d} }
\newcommand{\ud}{\mathrm{d}}

\newcommand{\uM}{\mathcal{M}}

\newcommand{\uTr}{\mathrm{Tr}}

\newcommand{\uslash}{/\!\!\!}

\newcommand{\usigma}{\boldsymbol{\sigma}}

\newcommand{\upi}{\boldsymbol{\pi}}

\newcommand{\uS}{\boldsymbol{S}}
\newcommand{\us}{\boldsymbol{s}}
\newcommand{\uk}{\boldsymbol{k}}

\newcommand{\uzero}{\boldsymbol{0}}

\newcommand{\sgn}{\text{sgn}}

\usepackage{graphicx}  
\title{Models for TMDs and numerical methods}
\author{B.~Pasquini~\from{pavia}\thanks{Lecturer in the school.}
\atque
C.~Lorc\'e~\from{orsay}}
\instlist{\inst{pavia} Dipartimento di Fisica, Universit\`a degli Studi di Pavia, and INFN, Sezione di Pavia, I-27100 Pavia, Italy
  \inst{orsay} IPNO, Universit\'e Paris-Sud 11, CNRS/IN2P3, 
 91406 Orsay, France\\
and LPT, Universit\'e Paris-Sud 11, CNRS, 91406 Orsay, France
}
\PACSes{\PACSit{13.88+e}{}
\PACSit{13.85.Ni}{}
\PACSit{13.60.-r}{}
\PACSit{13.85.Qk}{}
}
\begin{document}

\maketitle

\begin{abstract}
We present a systematic study  of the transverse-momentum dependent 
parton distributions (TMDs) in the framework of quark models.
In particular, we review
the general formalism for modeling the quark TMDs using the representation 
in terms of overlap of  nucleon light-cone wave functions.
Such a formalism can be applied to a large class of quark models.
We will discuss the  building blocks of these different models, and we will use
as explicit example for the calculation of the TMDs a light-cone 
constituent quark model.
Within this model,
 we also propose a phenomenological study of different observables,
trying to learn how  model parameters related to particular assumptions on the nucleon 
structure can be tuned to  describe available experimental data.
\end{abstract}

\section{Introduction}

\begin{figure}[t!]
\begin{center}
\epsfig{file=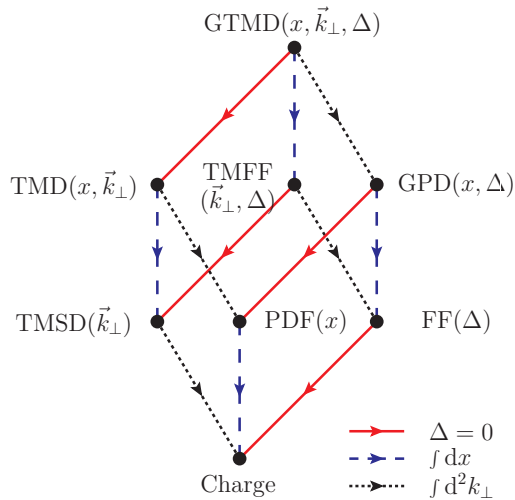,  width=0.5\columnwidth}
\end{center}
\caption{\footnotesize{Representation of the projections of the GTMDs into 
parton distributions and form factors.
The arrows correspond to different reductions in the hadron and quark momentum
space: the solid (red) arrows  give the forward limit in the hadron momentum,
the dotted (black) arrows correspond to integrating over the quark 
transverse momentum and the dashed (blue) arrows project out the longitudinal momentum of quarks.}}
\label{fig1}
\end{figure}

The investigation of how the composite structure of a hadron, consisting of 
nearly massless constituents,  results from the underlying quark-gluon dynamics  is a 
challenging problem, as it is of non-perturbative nature, and displays many facets. 
What are the longitudinal momentum distributions of partons in a fast moving unpolarized 
or polarized hadron? What amount of transverse momentum do these partons carry, and 
how large is the resulting amount of orbital angular momentum? What is the spatial distribution 
of quarks inside a hadron as seen by a vector probe (coupling to the charge of the system), 
by an axial vector probe (coupling to the axial charge), or even seen by a more complicated probe?
\newline
\noindent
Two new types of distributions such as the transverse-momentum dependent parton distributions (TMDs) and the generalized parton distributions (GPDs)  
allow to address and quantify such questions.
At leading-twist, the TMDs are defined as the diagonal matrix elements of the
parton density matrix in the longitudinal and transverse-momentum space~\cite{Collins:1981uk}-\cite{Boer:1997mf}.
Taking into account all the independent  spin-polarization states of the 
nucleon and partons, one obtains 8 independent TMDs describing 
the correlations in the three-dimensional momentum space between the nucleon  and  quark spin, and between the quark orbital angular momentum and both the nucleon and quark spin.
On the other hand, GPDs are probability amplitudes related to the off-diagonal matrix elements 
of the parton density matrix in the longitudinal momentum space~\cite{Dittes88}-\cite{Boffi:2007yc}.
After a Fourier transform to the impact-parameter space, they also provide a three-dimensional  
picture of the hadron in a mixed momentum-coordinate space~\cite{Soper:1976jc}-\cite{Burkardt:2002hr}.
\newline
\noindent
The information encoded in the TMDs and the GPDs can be collected in
 a unified framework through the generalized transverse-momentum 
dependent parton distributions (GTMDs); for a classification see Refs.~\cite{Meissner:2008ay,Meissner:2009ww}.
The GTMDs contain the most general one-body information about partons, corresponding to the full one-quark density matrix in momentum space~\cite{Lorce:2011dv}.
GTMDs depend on the three-momentum of the quarks as well as on the four-momentum $\Delta$ which is transferred by the probe to the nucleon.
The Fourier transform of the GTMDs to the impact-parameter space allows to define
the Wigner distributions of the parton-hadron system, which represent the quantum-mechanical analogues of the classical phase-space distributions~\cite{Ji:2003ak,Belitsky:2003nz,Lorce:2011kd,BR05}.
Wigner distributions provide 5-dimensional (two position and three momentum coordinates) images
of the nucleon  as seen in the infinite-momentum frame.
As such they contain the full correlations between the quark transverse position and three-momentum. In particular limits, they
reduce to different 
parton distributions and form factors as shown in Fig.~\ref{fig1}.
The different arrows in this figure represent particular projections in hadron and quark momentum space, and give  the links between the matrix elements of different reduced matrices.
In the forward limit $\Delta=0,$ they reduce to the TMDs, while
 the integration over the quark transverse momentum
$\uk_\perp$  leads to the GPDs. 
The common limit of TMDs and GPDs is given by the standard parton distribution functions (PDFs).
The integration over the longitudinal-momentum fraction $x$  brings to the lower plane of the box in Fig.~\ref{fig1},  and at each vertex we find the restricted version of the operator 
defining the distributions in the upper plane. For example, 
in correspondence of the TMDs we encounter the 
transverse-momentum dependent spin densities (TMSD), while from the GPDs we 
obtain the form factors (FFs). Both FFs and TMSDs have the charges as common limit.
Each of these observable is sensitive to specific properties of the nucleon 
structure, and contributes from different perspective 
to form a global and comprehensive view~\cite{Boer:2011fh}.
In these lectures we will concentrate on the model calculation of TMDs, 
restricting ourselves to the quark contribution.
 We will exploit the relations with different observables 
and parton distributions in the box of Fig.~\ref{fig1}
to test different model assumptions and their implications for
the partonic structure of the nucleon.
A variety low-energy QCD-inspired models have been employed to model the TMDs~\cite{Lorce:2011dv},~\cite{Jakob:1997wg}-\cite{DelDotto:2011ue}.
Although they all have in common that they strongly
oversimplify the complexity of the QCD dynamics in hadrons,
studies in different models based on often complementary 
assumptions, help to unravel non-perturbative aspects of TMDs. 
Insights into non-perturbative properties are of particular
interest when confirmed in various models.
The practical value of model results is that they can be used
to predict new observables, or to guide educated Ans\"atze 
for fits of TMD parametrizations.
Especially in the context of TMDs one should not underestimate
the conceptual importance of model calculations. 
Model calculations demonstrated the existence of effects 
\cite{Brodsky:2002cx}, paved the way towards an understanding 
of universality in the fragmentation process \cite{Metz:2002iz},
established new TMDs \cite{Afanasev:2003ze,Metz:2004je},
see \cite{Metz:2004ya} for a review.

The manuscript is organized as follows.
In sect.~\ref{section-2} we review the definitions of the leading-twist TMDs
and  introduce a convenient representation of the quark-quark correlator in terms of the net-polarization states of the quark and the hadron. 
In sect.~\ref{section-3} we discuss model relations among TMDs which hold in a large class of quark models.
In particular, we review the derivation given in Ref.~\cite{Lorce:2011zt}
to explain the physical origin of such relations.
There are two linear relations and a quadratic relation which are flavour independent and involve polarized TMDs, while a further linear relation is flavour dependent and involves both polarized and unpolarized TMDs.
The relations among polarized TMDs connect the distributions of quarks inside the nucleon for different configurations of the polarization states of the hadron and the parton. As a consequence, it is natural to expect that they can originate from rotational invariance of the polarization states of the system. 
Rotations are more easily discussed in the basis of canonical spin. Therefore, instead of working in the standard basis of light-cone helicity, we introduce in sects.~\ref{section-41} and \ref{section-42} the tensor correlator defining the TMDs in the canonical-spin basis. Such a representation is used in sect.~\ref{section-43} to discuss the consequences of rotational symmetries of the system. In such a way we will be able to identify the key ingredients for the existence of relations among polarized TMDs in quark models.

In order to complete the discussion, including the flavour-dependent relation among polarized and unpolarized TMDs, we need to introduce specific assumptions about the spin-isospin structure of the nucleon state. Therefore, in sect.~\ref{section-5} we discuss the consequences of rotational invariance using the explicit representation of the TMDs in terms of three-quark (3Q) wave functions. In particular, in sect.~\ref{section-51} we derive the overlap representation of the TMDs in terms of light-cone wave functions (LCWFs), while the corresponding representation in terms of wave function in the canonical-spin basis is given in sect.~\ref{section-52}. In sect.~\ref{section-53} we discuss the constraints of rotational symmetry on the nucleon wave function and, as a result, we give an alternative derivation of the flavour-independent relations among TMDs. Finally, in sect.~\ref{section-54} we discuss the constraint due to $SU(6)$ symmetry of the spin-flavour dependent part of the nucleon wave function. This additional ingredient allows us to explain the origin of the flavour-dependent relation.
The formal derivation of the relations among TMDs is made explicit within different quark models in sect.~\ref{models}. There we review different quark models which have been used in the literature for the calculation of TMDs. In particular, we discuss the key ingredients of the models, showing how and to which extent the conditions which lead to relations among TMDs are realized. 
In sect.~\ref{section:su6br}
 we also study the phenomenological implications 
of the breaking of  $SU(6)$ symmetry.
We will work in the framework of a light-cone constituent quark model, giving the essential step for the explicit calculation of the TMDs in the model in sect.~\ref{sec:calculation}.
In sect.~\ref{results-tmds} we  discuss the model results for T-even TMDs and related spin densities in the transverse-momentum space.
In  sect.~\ref{sec:observables} we propose a phenomenological study of different observables, fitting the model parameter related to $SU(6)$-symmetry breaking 
to the double spin asymmetry in inclusive deep inelastic scattering (DIS).
The results from this fit are then exploited to obtain predictions for other observables which are particularly sensitive to $SU(6)$-breaking effects, like the double spin asymmetry for semi-inclusive deep inelastic scattering (SIDIS) and the neutron electric form factor. A summary of our findings is given in the final section. Technical details and further explanations about the representation in terms of nucleon wave functions are collected in three appendices.

\section{Transverse-momentum dependent parton distributions}
\label{section-2}

\subsection{Definitions}

In this section, we review the formalism for the definition of TMDs, following the conventions of refs.~\cite{Mulders:1995dh,Boer:1997nt,Goeke:2005hb,Bacchetta:2006tn}. Introducing two lightlike four-vectors $n_\pm$ satisfying $n_+\cdot n_-=1$, we write the light-cone components of a generic four-vector $a$ as $\left[a^+,a^-,\boldsymbol a_\perp\right]$ with $a^\pm=a\cdot n_\mp$. The density of quarks can be defined from the following quark-quark correlator~\footnote{In order to simplify the notation, we will omit the colour 
index, keeping in mind that the quark-quark correlator is diagonal in the colour space.}
\begin{equation}
\Phi_{ab}(x,\uk_\perp,S)=\int\frac{\ud\xi^-\,\ud^2\xi_\perp}{(2\pi)^3}\,e^{i\left(k^+\xi^--\uk_\perp\cdot\boldsymbol\xi_\perp\right)}\langle P,S|\overline\psi_b(0)\mathcal U^{n_-}_{(0,+\infty)}\mathcal U^{n_-}_{(+\infty,\xi)}\psi_a(\xi)|P,S\rangle\big|_{\xi^+=0},
\label{correlator}
\end{equation}
where $k^+=xP^+$, $\psi$ is the quark field operator with $a,b$ indices in the Dirac space, and $\mathcal U$ is the Wilson line which ensures colour gauge invariance \cite{Bomhof:2004aw}. The target state is characterized by its four-momentum $P$ and covariant spin four-vector $S$ satisfying $P^2=M^2$, $S^2=-1$, and $P\cdot S=0$. We choose a frame where the hadron momentum has no transverse components, $P=\left[P^+,\tfrac{M^2}{2P^+},\uzero_\perp\right]$, and so $S=\left[S_z\,\tfrac{P^+}{M},-S_z\,\tfrac{M}{2P^+},\uS_\perp\right]$ with $\uS^2=1$. From now on, we replace the dependence on the covariant spin four-vector $S$ by the dependence on the unit three-vector $\uS=\left(\uS_\perp,S_z\right)$.

TMDs enter the general Lorentz-covariant decomposition of the correlator $\Phi_{ab}(x,\uk_\perp,\uS)$ which, at twist-two level and for a spin-$1/2$ target, reads
\begin{multline}
\Phi(x,\uk_\perp,\uS)=\frac{1}{2}\Big\{f_1\,\uslash n_+-\tfrac{\epsilon_T^{ij}\,k_\perp^iS_\perp^j}{M}\,f_{1T}^\perp\,\uslash n_++S_z\,g_{1L}\,\gamma_5\uslash n_++\tfrac{\uk_\perp\cdot\uS_\perp}{M}\,g_{1T}\,\gamma_5\uslash n_+\\
+h_{1T}\,\tfrac{[\uslash S_\perp,\uslash n_+]}{2}\,\gamma_5+S_z\,h_{1L}^\perp\,\tfrac{[\uslash k_\perp,\uslash n_+]}{2M}\,\gamma_5+\tfrac{\uk_\perp\cdot\uS_\perp}{M}\,h_{1T}^\perp\,\tfrac{[\uslash k_\perp,\uslash n_+]}{2M}\,\gamma_5+ih_1^\perp\,\tfrac{[\uslash k_\perp,\uslash n_+]}{2M}\Big\},
\end{multline}
where $\epsilon_T^{12}=-\epsilon_T^{21}=1$, and the transverse four-vectors are defined as $a_\perp=\left[0,0,\boldsymbol a_\perp\right]$. The nomenclature of the distribution functions follows closely that of Ref.~\cite{Mulders:1995dh}, sometimes referred to as ``Amsterdam notation'': $f$ refers to unpolarized quarks; $g$ and $h$ to longitudinally and transversely polarized quark, respectively; a subscript $1$ is given to the twist-two functions; subscripts $L$ or $T$ refer to the connection with the hadron spin being longitudinal or transverse; and a symbol $\perp$ signals the explicit presence of transverse momenta with an uncontracted index. Among these eight distributions, the so-called Boer-Mulders function $h_1^\perp$~\cite{Boer:1997nt} and Sivers function $f_{1T}^\perp$~\cite{Sivers:1989cc} are T-odd, \emph{i.e.} they change sign under ``naive time-reversal'', which is defined as usual time-reversal but without interchange of initial and final states. All the TMDs depend on $x$ and $\uk^2_\perp$. These functions can be individually isolated by performing traces of the correlator with suitable Dirac matrices. Using the abbreviation $\Phi^{[\Gamma]}\equiv\uTr[\Phi\Gamma]/2$, we have
\begin{align}
\Phi^{[\gamma^+]}(x,\uk_\perp,\uS)&=f_1-\tfrac{\epsilon_T^{ij}\,k_\perp^iS_\perp^j}{M}\,f_{1T}^\perp,
\label{vector}\\
\Phi^{[\gamma^+\gamma_5]}(x,\uk_\perp,\uS)&=S_z\,g_{1L}+\tfrac{\uk_\perp\cdot\uS_\perp}{M}\,g_{1T},\\
\Phi^{[i\sigma^{j+}\gamma_5]}(x,\uk_\perp,\uS)&=S_\perp^j\,h_1+S_z\,\tfrac{k_\perp^j}{M}\,h^\perp_{1L}+S^i_\perp\,\tfrac{2k^i_\perp k^j_\perp-\uk^2_\perp\delta^{ij}}{2M^2}\,h^\perp_{1T}+\tfrac{\epsilon_T^{ji}\,k_\perp^i}{M}\,h_1^\perp,
\label{tensor}
\end{align}
where $j=1,2$ is a transverse index, and
\begin{equation}
h_1=h_{1T}+\tfrac{\uk^2_\perp}{2M^2}\,h^\perp_{1T}.
\end{equation}

The correlation function $\Phi^{[\gamma^+]}(x,\uk_\perp,\uS)$ is just the unpolarized quark distribution, which integrated over $\uk_\perp$ gives the familiar light-cone momentum distribution $f_1(x)$. All the other TMDs characterize the strength of different spin-spin and spin-orbit correlations. The precise form of this correlation is given by the prefactors of the TMDs in Eqs.~\eqref{vector}-\eqref{tensor}. In particular, the TMDs $g_{1L}$ and $h_1$ describe the strength of a correlation between a longitudinal/transverse target polarization and a longitudinal/transverse parton polarization. After integration over $\uk_\perp$, they reduce to the helicity and transversity distributions, respectively. By definition, the spin-orbit correlations described by $f_{1T}^\perp$, $g_{1T}$, $h_1^\perp$, $h_{1L}^\perp$ and $h_{1T}^\perp$ involve the transverse parton momentum and the polarization of both the parton and the target, and vanish upon integration over $\uk_\perp$.

If one calculates these distributions in the light-cone gauge $A^+=0$ using advanced boundary condition for the transverse component of the gauge field, the gauge links in the quark-quark correlator can be ignored \cite{Ji:2002aa}-\cite{Boer:2003cm}. However, in this case the wave function amplitudes are not real and, apart from the structural information on the hadron, they also have an imaginary phase mimicking the final state interactions~\cite{Brodsky:2009dv,Brodsky:2010vs}.

\subsection{Helicity and four-component bases}
\label{sect:II-B}

The physical meaning of the correlations encoded in TMDs becomes especially transparent when using for the quark field operator the Fourier expansion in momentum space in terms of light-cone Fock operators. To this aim, we first introduce the Fourier expansion in momentum space of
the quark field operator $\psi^q$ of flavour $q$~(see emph{e.g.}~\cite{Brodsky:1997de})
\begin{eqnarray}\label{Free-quark-field}
\psi^q(z^-, \mathbf{z}_\perp) & = &
\int\frac{dk^{+}d^2\mathbf{k}_\perp}{16\pi^3k^{+}}\Theta(k^+)
\sum_\lambda\{q_\lambda(\tilde k)u_{LC}(\tilde k,\lambda)
\exp(-ik^+z^- + i\mathbf{k}_\perp \cdot
\mathbf{z}_\perp)
\nonumber\\
& &  + \bar{q}^{\dagger }_\lambda(\tilde k)v_{LC}(\tilde k)\exp(+ik^+z^- - i\mathbf{k}_\perp \cdot
\mathbf{z}_\perp)\},
\end{eqnarray}
 where $\tilde k=(k^+,\mathbf{k}_\perp)$, $\lambda$,
represents the parton helicity, and $q$ and $\bar{q}^\dagger$ are the annihilation operator of the quark fields and the creation 
operators of the antiquark fields, respectively, which fulfill
 the anticommutations relations
\begin{eqnarray}\label{Anticommutations-rules}
\{q_{\lambda'}(\tilde k'), q^{\dagger}_\lambda(\tilde k)\} & = &\{\bar{q}_{\lambda'}(\tilde k'), \bar{q}^{\dagger}_\lambda(\tilde k)\}
=16\pi^{3}k^+\delta(k'^+ - k^+)\delta^{(2)}(\mathbf{k'}_\perp -
\mathbf{k}_\perp)\delta_{q'q}\delta_{\lambda'\lambda}.
\end{eqnarray}
In Eq.~(\ref{Free-quark-field}),
the quark LC spinors  
$u_{LC}(\tilde k,\lambda)$ and
$v_{LC}(\tilde k,\lambda)$
are the free Dirac light-cone spinor of the quark and antiquark, respectively
(see Appendix~\ref{app:1}).

For simplicity, in this study we will consider  only  the quark 
contribution to TMDs, ignoring the contribution from gauge fields and therefore reducing the gauge links in Eq.~(\ref{correlator}) to the identity. As a consequence, we will not be able to discuss the T-odd TMDs, which vanish identically 
in absence of gauge-field degrees of freedom.
Furthermore,  we decompose the correlator~\eqref{correlator}
into the different quark flavour contributions
\begin{equation}
\Phi=\sum_q\Phi_q.
\end{equation} 
Following the lines of refs.~\cite{Diehl:2000xz}-\cite{Pasquini:2008ax}, 
we obtain at twist-two level 
\begin{equation}\label{corrTMD}
\Phi^{[\Gamma]}_q(x,\uk_\perp,\uS)=\frac{1}{\mathcal N}\,\langle P,\uS|\sum_{\lambda'\lambda}q^\dag_{\lambda'}(\tilde k)\,q_\lambda(\tilde k)
\,M^{[\Gamma]\lambda'\lambda}|P,\uS\rangle,
\end{equation}
where $\mathcal N=\left[2x(2\pi)^3\right]^2\delta^{(3)}(\uzero)$ and 
$M^{[\Gamma]\lambda'\lambda}=\overline u_{LC}(k,\lambda')\Gamma u_{LC}(k,\lambda)/2k^+$.
As a result, 
using~\eqref{vector}-\eqref{tensor},
the quark contribution to the T-even TMDs is given by
\begin{gather}
f^q_1(x,\uk^2_\perp)=\langle P,\Lambda|V^q(\tilde k)|P,\Lambda\rangle,\label{start}\\
\Lambda\,g^q_{1L}(x,\uk^2_\perp)=\langle P,\Lambda|A^q(\tilde k)|P,\Lambda\rangle,\\
\frac{\uk_\perp\cdot\uS_\perp}{M}\,g^q_{1T}(x,\uk^2_\perp)=\langle P,S_\perp|A^q(\tilde k)|P,S_\perp\rangle,\\
\Lambda\,\frac{\uk_\perp}{M}\,h^{\perp q}_{1L}(x,\uk^2_\perp)=
\frac{1}{2}\sum_\Lambda{\rm sign}\,(\Lambda)\langle P,\Lambda|\boldsymbol T^q_\perp(\tilde k)|P,\Lambda\rangle,\\
\uS_\perp\,h^q_{1T}(x,\uk^2_\perp)+\frac{\uk_\perp\cdot\uS_\perp}{M}\frac{\uk_\perp}{M}\,h^{\perp q}_{1T}(x,\uk^2_\perp)=\frac{1}{2}\sum_{S_\perp}{\rm sign}\,(S_\perp)\langle P,S_\perp|\boldsymbol T^q_\perp(\tilde k)|P,S_\perp\rangle.\label{disentangle}
\end{gather}
where the quark operators $V^q$, $A^q$, and $\boldsymbol T_\perp^q$ are defined
 as
\begin{gather}
V^q(\tilde k)=\sum_\lambda q^\dag_\lambda(\tilde k)q_\lambda(\tilde k),\\
A^q(\tilde k)=\sum_\lambda\sgn(\lambda)q^\dag_\lambda(\tilde k)q_\lambda(\tilde k),\\
T^q_R(\tilde k)=\left[T^q_L(\tilde k)\right]^\dag=2 q^\dag_+(\tilde k)q_-(\tilde k).
\end{gather} 
The quark operators $V^q$ and $A^q$ have a probabilistic interpretation since they are written just in terms of number operators $N^q_\lambda=q^\dag_\lambda(\tilde k)q_\lambda(\tilde k)$. The operator $\boldsymbol T_\perp^q$ has also a probabilistic interpretation but only when written in terms of transversely polarized operators
\begin{equation}
\us_\perp\cdot\boldsymbol T_\perp^q=\sum_{s_\perp}\sgn(s_\perp)
q^\dag_{s_\perp}(\tilde k)q_{s_\perp}(\tilde k),
\end{equation}
where for a generic direction $\us=(\sin\theta_s\cos\phi_s,\sin\theta_s\sin\phi_s,\cos\theta_s)$, the creation operators for a polarized quark can be expressed in terms of creation operators for a longitudinally polarized quark as
\begin{equation}
\begin{pmatrix}q^\dag_{+\us},&q^\dag_{-\us}\end{pmatrix}=\begin{pmatrix}q^\dag_+,&q^\dag_-\end{pmatrix}u(\theta_s,\phi_s),
\end{equation}
where the $SU(2)$ rotation matrix $u(\theta,\phi)$ is given by
\begin{equation}\label{su2rot}
u(\theta,\phi)=\begin{pmatrix}\cos\tfrac{\theta}{2}\,e^{-i\phi/2}&-\sin\tfrac{\theta}{2}\,e^{-i\phi/2}\\\sin\tfrac{\theta}{2}\,e^{i\phi/2}&\cos\tfrac{\theta}{2}\,e^{i\phi/2}\end{pmatrix}.
\end{equation}
The operators $V^q(\tilde k)$, $A^q(\tilde k)$, and $\us_\perp\cdot\boldsymbol T_\perp^q(\tilde k)$ give the number, the effective longitudinal polarization, and the effective transverse polarization of quarks with flavour $q$ and light-cone momentum $\tilde k$, respectively.

We can further disentangle \eqref{disentangle} by choosing appropriate target and quark polarizations
\begin{gather}
h^q_1(x,\uk^2_\perp)=\frac{1}{4}\left[\sum_{S_x}{\rm sign}\,(S_x)\langle P,S_x|T^q_x(\tilde k)|P,S_x\rangle+\sum_{S_y}{\rm sign}\,(S_y)\langle P,S_y|T^q_y(\tilde k)|P,S_y\rangle\right],\\
\frac{k_x^2-k_y^2}{M^2}\,h^{\perp q}_{1T}(x,\uk^2_\perp)=\frac{1}{2}\left[\sum_{S_x}{\rm sign}\,(S_x)\langle P,S_x|T^q_x(\tilde k)|P,S_x\rangle-\sum_{S_y}{\rm sign}\,(S_y)\langle P,S_y|T^q_y(\tilde k)|P,S_y\rangle\right]\\
\text{or}\nonumber\\
\frac{k_xk_y}{M^2}\,h^{\perp q}_{1T}(x,\uk^2_\perp)=\frac{1}{2}\sum_{S_x}{\rm sign}\,(S_x)\langle P,S_x|T^q_y(\tilde k)|P,S_x\rangle=\frac{1}{2}\sum_{S_y}{\rm sign}\,(S_y)\langle P,S_y|T^q_x(\tilde k)|P,S_y\rangle,
\end{gather}
where the nucleon polarization state $|P,\uS\rangle$ 
in a generic direction $\uS=(\sin\theta_S\cos\phi_S,\sin\theta_S\sin\phi_S,\cos\theta_S)$  can be written in terms of longitudinally polarized states $|P,\Lambda\rangle$ as
\begin{equation}
\begin{pmatrix}|P,+\uS\rangle,&|P,-\uS\rangle\end{pmatrix}=\begin{pmatrix}|P,+\rangle,&|P,-\rangle\end{pmatrix}u(\theta_S,\phi_S).
\end{equation}

In the literature, one often represents correlators in terms of helicity amplitudes which treat in a symmetric way both quark and target polarizations
\begin{equation}\label{helicityamplitude}
\Phi^q_{\Lambda'\lambda',\Lambda\lambda}(x,\uk_\perp)=\frac{1}{\mathcal N}\,\langle P,\Lambda'|q^\dag_{\lambda'}(\tilde k)\,q_\lambda(\tilde k)|P,\Lambda\rangle.
\end{equation}

Thanks to light-cone parity invariance, the helicity amplitudes 
are not all independent and are constrained by the following  relation
\begin{equation}\label{helicityamplitude2}
\Phi^q_{\Lambda'\lambda',\Lambda\lambda}(\tilde k)=
\Phi^q_{-\Lambda'-\lambda',-\Lambda-\lambda}(\tilde k_P),
\end{equation}
where $\tilde k_P=(x,-k_x,k_y)$.
From the definitions in Eqs.~\eqref{start}-\eqref{disentangle} it follows that
it is possible to express each TMD as linear combination of helicity amplitudes
as described in Table~\ref{table-1}.
In particular we note that each TMD involves combinations of 
helicity amplitudes with a definite value of orbital angular momentum transfer $\Delta l_z$ between the initial and final state. This is a consequence of the
 conservation of the total angular momentum which gives the constraint
$\Delta l_z=(\Delta\Lambda-\Delta\lambda)/2$, with $\Delta\Lambda=\Lambda'-\Lambda$ and analogously $\Delta\lambda=\lambda'-\lambda$.
\begin{table}[h!]
\begin{center}
\caption{Relations between the 8 leading-twist TMDs and helicity amplitudes.
The table can be used to obtain  the TMDs as functions of the helicity amplitudes as well as the inverse relation.
The entries in the table give  the coefficients in the linear 
combination which relates the two sets in the first column and in the first row.
The helicity amplitudes are also classified in terms of the transfer of orbital angular momentum $\Delta l_z$ between the initial and final state.}
\label{table-1}
\begin{tabular}{c|ccc|cccc|c}
\hline
&\multicolumn{3}{c|}{$\Delta l_z=0$}&\multicolumn{4}{c|}{$\Delta l_z=+1$}
&$\Delta l_z=+2$\\
&$\Phi^q_{++,++}$&$\Phi^q_{+-,+-}$&$\Phi^q_{++,--}$&$\Phi^q_{++,-+}$
&$\Phi^q_{+-,--}$&$\Phi^q_{--,-+}$&$\Phi^q_{+-,++}$&$\Phi^q_{+-,-+}$\\
\hline
$f^q_1$&$1$&$1$&$\cdot$&$\cdot$&$\cdot$&$\cdot$&$\cdot$&$\cdot$\\
$g^q_{1L}$&$1$&$-1$&$\cdot$&$\cdot$&$\cdot$&$\cdot$&$\cdot$&$\cdot$\\
$h^q_1$&$\cdot$&$\cdot$&$1$&$\cdot$&$\cdot$&$\cdot$&$\cdot$&$\cdot$\\
$f^{\perp q}_{1T}$&$\cdot$&$\cdot$&$\cdot$&$-i\frac{Mk_R}{\uk^2_\perp}$&$-i\frac{Mk_R}{\uk^2_\perp}$&$\cdot$&$\cdot$&$\cdot$\\
$g^q_{1T}$&$\cdot$&$\cdot$&$\cdot$&$\frac{Mk_R}{\uk^2_\perp}$&$-\frac{Mk_R}{\uk^2_\perp}$&$\cdot$&$\cdot$&$\cdot$\\
$h^{\perp q}_1$&$\cdot$&$\cdot$&$\cdot$&$\cdot$&$\cdot$&$-i\frac{Mk_R}{\uk^2_\perp}$&$-i\frac{Mk_R}{\uk^2_\perp}$&$\cdot$\\
$h^{\perp q}_{1L}$&$\cdot$&$\cdot$&$\cdot$&$\cdot$&$\cdot$&$-\frac{Mk_R}{\uk^2_\perp}$&$\frac{Mk_R}{\uk^2_\perp}$&$\cdot$\\
$h^{\perp q}_{1T}$&$\cdot$&$\cdot$&$\cdot$&$\cdot$&$\cdot$&$\cdot$&$\cdot$&$2\left(\frac{Mk_R}{\uk^2_\perp}\right)^2$\\
\hline
\end{tabular}
\end{center}
\end{table}

We can collect the information encoded in Table~\ref{table-1} by introducing the following matrix form
\begin{eqnarray}\label{amplTMDs}\\
&&
\Phi^q_{\Lambda'\lambda',\Lambda\lambda}\nonumber\\
&&=
\begin{pmatrix}
\tfrac{1}{2}\left(f^q_1+g^q_{1L}\right)&-\tfrac{k_R}{2M}\left(ih_1^{\perp q}-h_{1L}^{\perp q}\right)&\tfrac{k_L}{2M}\left(if_{1T}^{\perp q}+g^q_{1T}\right)&h^q_1\\
\tfrac{k_L}{2M}\left(ih_1^{\perp q}+h_{1L}^{\perp q}\right)&\tfrac{1}{2}\left(f^q_1-g^q_{1L}\right)&\tfrac{k_L^2}{2M^2}\,h_{1T}^{\perp q}&\tfrac{k_L}{2M}\left(if_{1T}^{\perp q}-g^q_{1T}\right)\\
-\tfrac{k_R}{2M}\left(if_{1T}^{\perp q}-g^q_{1T}\right)&\tfrac{k_R^2}{2M^2}\,h_{1T}^{\perp q}&\tfrac{1}{2}\left(f^q_1-g^q_{1L}\right)&-\tfrac{k_R}{2M}\left(ih_1^{\perp q}+h_{1L}^{\perp q}\right)\\
h^q_1&-\tfrac{k_R}{2M}\left(if_{1T}^{\perp q}+g^q_{1T}\right)&\tfrac{k_L}{2M}\left(ih_1^{\perp q}-h_{1L}^{\perp q}\right)&\tfrac{1}{2}\left(f^q_1+g^q_{1L}\right)
\end{pmatrix},\nonumber
\end{eqnarray}
where the  entries in the rows correspond to $(\Lambda'\lambda')=(++),(+-),(-+),(--)$ 
and the entries in the column are likewise $(\Lambda\lambda)=(++),(+-),(-+),(--)$.

Alternatively, we can represent the quark correlator \eqref{corrTMD} in the four-component basis. 
There are  only four twist-two Dirac structures $\Gamma_\text{twist-2}=\{\gamma^+,i\sigma^{1+}\gamma_5,i\sigma^{2+}\gamma_5,\gamma^+\gamma_5\}$, corresponding to the four kinds of transition the light-cone helicity of the active quark can undergo (see \emph{e.g.}~\cite{Boffi:2002yy}-\cite{Pasquini:2005dk})
\begin{equation}\label{correspondence}
M^{[\gamma^+]\lambda'\lambda}=\delta^{\lambda'\lambda},\quad M^{[i\sigma^{j+}\gamma_5]\lambda'\lambda}=(\sigma_j)^{\lambda'\lambda},\quad M^{[\gamma^+\gamma_5]\lambda'\lambda}=(\sigma_3)^{\lambda'\lambda}
\end{equation}
with $\sigma_i$ the three Pauli matrices. For further convenience, we associate a four-component vector~\footnote{Note this is \emph{not} a Lorentz four-vector but Einstein's summation convention still applies.} to every quantity with superscript $\Gamma$
\begin{equation}
a^{[\Gamma]}\mapsto a^\nu=\left(a^0,a^1,a^2,a^3\right)\equiv\left(a^{[\gamma^+]},a^{[i\sigma^{1+}\gamma_5]},a^{[i\sigma^{2+}\gamma_5]},a^{[\gamma^+\gamma_5]}\right).
\end{equation}
With this notation, the correspondence \eqref{correspondence} takes the simple form
\begin{equation}
M^{\nu\lambda'\lambda}=(\bar\sigma^\nu)^{\lambda'\lambda},
\end{equation}
where $\bar\sigma^\nu=(\mathds{1},\vec \sigma)$.  The symbols $\bar\sigma^\mu$ and $\sigma_\mu$ satisfy the relations
\begin{equation}\label{basisidentities}
\frac{1}{2}\,(\bar\sigma^\mu)^{\lambda'\lambda}(\sigma_\mu)_{\tau\tau'}=\delta^{\lambda'}_{\tau'}\delta^\lambda_\tau,\qquad\frac{1}{2}\uTr\left[\bar\sigma^\mu\sigma_\nu\right]=\frac{1}{2}\sum_{\lambda'\lambda}(\bar\sigma^\mu)^{\lambda'\lambda}(\sigma_\nu)_{\lambda\lambda'}=\delta^\mu_\nu.
\end{equation}
One can think of $\bar\sigma^{\nu\lambda'\lambda}\equiv(\bar\sigma^\nu)^{\lambda'\lambda}$ as the matrix of a mere change of basis, the one being labeled by the couple $\lambda\lambda'$ and the other by $\nu$
\begin{equation}
a^\nu=\sum_{\lambda'\lambda}\bar\sigma^{\nu\lambda'\lambda}\,a_{\lambda\lambda'}.
\end{equation}
In this basis, we can introduce
 the tensor correlator $\Phi^{\mu\nu}_q(x,\uk_\perp)$ which is related to helicity amplitudes as follows
\begin{equation}\label{basischange}
\Phi^{\mu\nu}_q=\frac{1}{2}\sum_{\Lambda'\Lambda\lambda'\lambda}(\bar\sigma^\mu)^{\Lambda\Lambda'}(\bar\sigma^\nu)^{\lambda'\lambda}\,\Phi^q_{\Lambda'\lambda',\Lambda\lambda},\qquad\Phi^q_{\Lambda'\lambda',\Lambda\lambda}=\frac{1}{2}\,\Phi^{\mu\nu}_q(\sigma_\mu)_{\Lambda'\Lambda}(\sigma_\nu)_{\lambda\lambda'}.
\end{equation}
The tensor correlator is then given by the following combinations of TMDs
\begin{align}
\Phi^{\mu\nu}_q&=\begin{pmatrix}
f^q_1&\frac{k_y}{M}\,h^{\perp q}_1&-\frac{k_x}{M}\,h^{\perp q}_1&0\\
\frac{k_y}{M}\,f^{\perp q}_{1T}&h^q_1+\frac{k^2_x-k^2_y}{2M^2}\,h^{\perp q}_{1T}&\frac{k_xk_y}{M^2}\,h^{\perp q}_{1T}&\frac{k_x}{M}\,g^q_{1T}\\
-\frac{k_x}{M}\,f^{\perp q}_{1T}&\frac{k_xk_y}{M^2}\,h^{\perp q}_{1T}&h^q_1-\frac{k^2_x-k^2_y}{2M^2}\,h^{\perp q}_{1T}&\frac{k_y}{M}\,g^q_{1T}\\
0&\frac{k_x}{M}\,h^{\perp q}_{1L}&\frac{k_y}{M}\,h^{\perp q}_{1L}&g^q_{1L}
\end{pmatrix}\nonumber\\
&=\begin{pmatrix}
f^q_1&\tfrac{k_y}{M}\,h_1^{\perp q}&-\tfrac{k_x}{M}\,h_1^{\perp q}&0\\
\tfrac{k_y}{M}\,f_{1T}^{\perp q}&h_{1T}^{+q}\,\hat k_x^2+h_{1T}^{-q}\,\hat k_y^2&\left(h_{1T}^{+q}-h_{1T}^{-q}\right)\hat k_x\hat k_y&\tfrac{k_x}{M}\,g_{1T}^q\\
-\tfrac{k_x}{M}\,f_{1T}^{\perp q}&\left(h_{1T}^{+q}-h_{1T}^{-q}\right)\hat k_x\hat k_y&h_{1T}^{-q}\,\hat k_x^2+h_{1T}^{+q}\,\hat k_y^2&\tfrac{k_y}{M}\,g_{1T}^q\\
0&\tfrac{k_x}{M}\,h_{1L}^{\perp q}&\tfrac{k_y}{M}\,h_{1L}^{\perp q}&g_{1L}^q
\end{pmatrix},\label{tensor2}
\end{align}
where we introduced the notations $h_{1T}^{\pm q}=h^q_1\pm\tfrac{\uk_\perp^2}{2M^2}\,h_{1T}^{\perp q}$ and $\hat k_i=k_i/k_\perp$ with $k_\perp=|\uk_\perp|$. The component $\Phi^{00}_q$ gives the density of quarks in the target irrespective of any polarization, \emph{i.e.} the density of unpolarized quarks in the unpolarized target. The components $\Phi^{0j}_q$ give the net density of quarks with light-cone polarization in the direction $\boldsymbol e_j$ in the unpolarized target, while the components $\Phi^{i0}_q$ give the net density of unpolarized quarks in the target with light-cone polarization in the direction $\boldsymbol e_i$. Finally, the components $\Phi^{ij}_q$ give the net density of quarks with light-cone polarization in the direction $\boldsymbol e_j$ in the target with light-cone polarization in the direction $\boldsymbol e_i$. The density of quarks with definite light-cone polarization in the direction $\us$ inside the target with definite light-cone polarization in the direction $\uS$ is then obviously given by $\Phi_q(x,\uk_\perp,\uS,\us)=\frac{1}{2}\,\bar S_\mu\Phi_q^{\mu\nu}\bar s_\nu$, where we have introduced the four-component vectors $\bar S_\mu=(1,\uS)$ and $\bar s_\nu=(1,\us)$.

\section{Model relations}
\label{section-3}

In QCD, the eight TMDs are all independent. It appeared however in a large panel of low-energy models that relations among some TMDs exist. At twist-two level, there are three flavour-independent relations~\footnote{Other expressions can be found in the literature, but are just combinations of the relations \eqref{rel1}-\eqref{rel3}.}, two are linear and one is quadratic in the TMDs
\begin{gather}
g^q_{1L}-\left[h^q_1+\tfrac{\uk_\perp^2}{2M^2}\,h_{1T}^{\perp q}\right]=0,\label{rel1}\\
g^q_{1T}+h_{1L}^{\perp q}=0,\label{rel2}\\
\left(g^q_{1T}\right)^2+2h^q_1\,h_{1T}^{\perp q}=0.\label{rel3}
\end{gather}
A further flavour-dependent relation involves both polarized and unpolarized TMDs
\begin{equation}\label{rel4}
\mathcal D^qf_1^q+g_{1L}^q=2h_1^q,
\end{equation}
where, for a proton target, the flavour factors with $q=u,d$ are given by $\mathcal D^u=\tfrac{2}{3}$ and $\mathcal D^d=-\tfrac{1}{3}$. As discussed in Ref.~\cite{Avakian:2010br}, at variance with the relations \eqref{rel1}-\eqref{rel3}, the flavour dependence in the relation \eqref{rel4} requires specific assumptions for the spin-isospin structure of the nucleon state, like $SU(6)$ spin-flavour symmetry.

A discussion on how general these relations are can be found in Ref.~\cite{Avakian:2010br}. Let us just mention that they were observed in the bag model \cite{Avakian:2010br}-\cite{Avakian:2009jt}, light-cone constituent quark models \cite{Pasquini:2008ax}, some quark-diquark models \cite{Jakob:1997wg}-\cite{Zhu:2011zza}, the covariant parton model \cite{Efremov:2009ze} and more recently in the light-cone version of the chiral quark-soliton model \cite{Lorce:2011dv}. Note however that there also exist models where the relations are not satisfied, like in some versions of the spectator model~\cite{Bacchetta:2008af} and the quark-target model \cite{Meissner:2007rx}. 

As already emphasized, the model relations \eqref{rel1}-\eqref{rel4} are not expected to hold identically in QCD, since the TMDs in these relations follow different evolution patterns. This implies that even if the relations are satisfied at some (low) scale, they would not hold anymore for other (higher) scales. The interest in these relations is therefore purely phenomenological. Experiments provide more and more data on observables related to TMDs, but need inputs from educated models and parametrizations for the extraction of these distributions. It is therefore particularly interesting to see to what extent the relations \eqref{rel1}-\eqref{rel4} can be useful as \emph{approximate} relations, which provide simplified and intuitive notions for the interpretation of the data. Note that some preliminary calculations in lattice QCD give indications that the relation \eqref{rel2} may indeed be approximately satisfied \cite{Hagler:2009mb}-\cite{Musch:2009ku}.

Using two different approaches, we show in the next sections that the flavour-independent relations \eqref{rel1}-\eqref{rel3} can easily be derived, once the following assumptions are made:
\begin{enumerate}
\item the probed quark behaves as if it does not interact directly with the other partons (\emph{i.e.} one works within the standard impulse approximation) and there are no explicit gluons;
\item the quark light-cone and canonical polarizations are related by a rotation with axis orthogonal to both light-cone and $\hat k_\perp$ directions;
\item the target has spherical symmetry in the canonical-spin basis.
\end{enumerate}
From these assumptions, one realizes that the flavour-independent relations have essentially a geometrical origin, as was already suspected in the context of the bag model almost a decade ago \cite{Efremov:2002qh}. We note however that the spherical symmetry is a sufficient but not necessary condition for the validity of the individual flavour-independent relations. As discussed in the following section, a subset of relations can be derived using less restrictive conditions, like axial symmetry about a specific direction.

For the flavour-dependent relation \eqref{rel4}, we need a further condition for the spin-flavour dependent part of the nucleon wave function. Specifically, we require 
\begin{enumerate}[resume]
\item $SU(6)$ spin-flavour symmetry of the wave function.
\end{enumerate}
As shown in sect.~\ref{models}, it is found that all the models satisfying the relations also satisfy the above conditions. We are not aware of any model calculation which satisfies some or all the three flavour-independent relations and at the same time breaks at least one of the conditions 1-3. However, this is not a priori excluded.

\section{Amplitude approach}
\label{section-4}

The first derivation of the TMD relations stays at the level of the amplitudes. As we have seen, the TMDs can be expressed in simple terms using light-cone polarization. On the other hand, rotational symmetry is easier to handle in terms of canonical polarization, which is the natural one in the instant form. We therefore write the TMDs in the canonical-spin basis, and then impose spherical symmetry. But before that, we need to know how to connect light-cone helicity to canonical spin.

\subsection{Connection between light-cone helicity and canonical spin}
\label{section-41}

Relating in general light-cone helicity with canonical spin is usually quite complicated, as the dynamics is involved. Fortunately, the common approach in quark models is to assume that the target can be described by quarks without mutual interactions. In this case the connection simply reduces to a rotation in polarization space with axis orthogonal to both $\hat k_\perp$ and $\boldsymbol e_z$ directions. The quark creation operator with canonical spin $\sigma$ can then be written in terms of quark creation operators with light-cone helicity $\lambda$ as follows
\begin{equation}\label{genMelosh}
q^\dag_\sigma=\sum_\lambda D^{(1/2)*}_{\sigma\lambda}\,q^\dag_\lambda\qquad\text{with}\qquad D^{(1/2)*}_{\sigma\lambda}=\begin{pmatrix}\cos\tfrac{\theta}{2}&-\hat k_R\,\sin\tfrac{\theta}{2}\\\hat k_L\,\sin\tfrac{\theta}{2}&\cos\tfrac{\theta}{2}\end{pmatrix}.
\end{equation}
Note that the rotation does not depend on the quark flavour. The angle $\theta$ between light-cone and canonical polarizations is usually a complicated function of the quark momentum $k$ and is specific to each model. It contains part of the model dynamics. The only general property is that $\theta\to 0$ as $k_\perp\to 0$. Due to our choice of reference frame where the target has no transverse momentum, the light-cone helicity and canonical spin of the target can be identified, at variance with the quark polarizations.

\subsection{TMDs in canonical-spin basis}
\label{section-42}

The four-component notation introduced in sect.~\ref{sect:II-B} is very convenient for discussing the rotation between canonical spin and light-cone helicity at the amplitude level. Since the light-cone helicity and canonical spin of the target can be identified in our choice of reference frame, we expect the canonical tensor correlator $\Phi^{\mu\nu}_{Cq}$ to be related to the light-cone one in Eq.~\eqref{tensor2} as follows
\begin{equation}\label{expectation}
\Phi^{\mu\nu}_{Cq}=\Phi_q^{\mu\rho}\,O_\rho^{\phantom{\rho}\nu},
\end{equation}
with $O$ some orthogonal matrix $O^TO=\mathds 1$ representing the rotation at the amplitude level. From Eqs.~\eqref{helicityamplitude}, \eqref{basischange}, \eqref{basisidentities} and \eqref{genMelosh} we find that the orthogonal matrix is given by
\begin{align}
O_\rho^{\phantom{\rho}\nu}&=\frac{1}{2}\uTr\left[D^{(1/2)}\sigma_\rho D^{(1/2)\dag}\bar\sigma^\nu\right]\nonumber\\
&=\frac{1}{2}\sum_{\sigma'\sigma\lambda'\lambda}D^{(1/2)}_{\sigma\lambda}\left(\sigma_\rho\right)_{\lambda\lambda'}\,D^{(1/2)*}_{\sigma'\lambda'}\left(\bar\sigma^\nu\right)^{\sigma'\sigma}\nonumber\\
&=\begin{pmatrix}
1&0&0&0\\
0&\hat k_y^2+\hat k_x^2\,\cos\theta&-\hat k_x\hat k_y\left(1-\cos\theta\right)&-\hat k_x\,\sin\theta\\
0&-\hat k_x\hat k_y\left(1-\cos\theta\right)&\hat k_x^2+\hat k_y^2\,\cos\theta&-\hat k_y\,\sin\theta\\
0&\hat k_x\,\sin\theta&\hat k_y\,\sin\theta&\cos\theta
\end{pmatrix}.\label{ortho}
\end{align}
The canonical tensor correlator then takes the form
\begin{equation}\label{Ctensor}
\Phi^{\mu\nu}_{Cq}=\begin{pmatrix}
f^q_1&\tfrac{k_y}{M}\,h_1^{\perp q}&-\tfrac{k_x}{M}\,h_1^{\perp q}&0\\
\tfrac{k_y}{M}\,f_{1T}^{\perp q}&\mathfrak h_{1T}^{+q}\,\hat k_x^2+h_{1T}^{-q}\,\hat k_y^2&\left(\mathfrak h_{1T}^{+q}-h_{1T}^{-q}\right)\hat k_x\hat k_y&\tfrac{k_x}{M}\,\mathfrak g_{1T}^q\\
-\tfrac{k_x}{M}\,f_{1T}^{\perp q}&\left(\mathfrak h_{1T}^{+q}-h_{1T}^{-q}\right)\hat k_x\hat k_y&h_{1T}^{-q}\,\hat k_x^2+\mathfrak h_{1T}^{+q}\,\hat k_y^2&\tfrac{k_y}{M}\,\mathfrak g_{1T}^q\\
0&\tfrac{k_x}{M}\,\mathfrak h_{1L}^{\perp q}&\tfrac{k_y}{M}\,\mathfrak h_{1L}^{\perp q}&\mathfrak g_{1L}^q
\end{pmatrix},
\end{equation}
where we introduced the notations
\begin{align}
\begin{pmatrix}
\mathfrak g^q_{1L}\\ \tfrac{k_\perp}{M}\,\mathfrak h_{1L}^{\perp q}
\end{pmatrix}
&=\begin{pmatrix}\cos\theta&-\sin\theta\\\sin\theta&\cos\theta\end{pmatrix}\begin{pmatrix}
g^q_{1L}\\ \tfrac{k_\perp}{M}\,h_{1L}^{\perp q}
\end{pmatrix},\label{rot1}\\
\begin{pmatrix}
\tfrac{k_\perp}{M}\,\mathfrak g^q_{1T}\\ \mathfrak h_{1T}^{+q}
\end{pmatrix}
&=\begin{pmatrix}\cos\theta&-\sin\theta\\\sin\theta&\cos\theta\end{pmatrix}\begin{pmatrix}
\tfrac{k_\perp}{M}\,g^q_{1T}\\ h_{1T}^{+q}
\end{pmatrix}.\label{rot2}
\end{align}
Comparing Eq.~\eqref{Ctensor} with Eq.~\eqref{tensor2}, we observe that the multipole structure is conserved under the rotation~\eqref{expectation}. The rotation from light-cone to canonical polarizations affects only some of the multipole magnitudes, see Eqs.~\eqref{rot1} and \eqref{rot2}.

Note that the orientation of the axes in the transverse plane has been fixed arbitrarily. There is however a privileged direction given by the active quark transverse momentum $\uk_\perp$. Choosing the orientation of transverse axes so that either $\uk_\perp=k_\perp\,\boldsymbol e_x$ or $\uk_\perp=k_\perp\,\boldsymbol e_y$ simplifies the transformation, as it eliminates the cumbersome factors $\hat k_x$ and $\hat k_y$ in Eqs.~(\ref{ortho}) and (\ref{Ctensor}). Choosing \emph{e.g} the second option, the orthogonal matrix of Eq.~\eqref{ortho} reduces to
\begin{equation}
O_\rho^{\phantom{\rho}\nu}=\begin{pmatrix}
\phantom{0}1\phantom{0}&0&0&0\\
0&\phantom{0}1\phantom{0}&0&0\\
0&0&\cos\theta&-\sin\theta\\
0&0&\sin\theta&\cos\theta
\end{pmatrix},
\end{equation}
and the light-cone and canonical tensor correlators take the following simpler forms
\begin{align}
\Phi^{\mu\nu}_q&=\begin{pmatrix}
f^q_1&\frac{k_\perp}{M}\,h^{\perp q}_1&0&0\\
\frac{k_\perp}{M}\,f^{\perp q}_{1T}&h_{1T}^{-q}&0&0\\
0&0&h_{1T}^{+q}&\frac{k_\perp}{M}\,g^q_{1T}\\
0&0&\frac{k_\perp}{M}\,h^{\perp q}_{1L}&g^q_{1L}
\end{pmatrix},\\
\Phi^{\mu\nu}_{Cq}&=\begin{pmatrix}\label{Ctensorred}
f^q_1&\frac{k_\perp}{M}\,h^{\perp q}_1&0&0\\
\frac{k_\perp}{M}\,f^{\perp q}_{1T}&h_{1T}^{-q}&0&0\\
0&0&\mathfrak h_{1T}^{+q}&\frac{k_\perp}{M}\,\mathfrak g_{1T}^q\\
0&0&\frac{k_\perp}{M}\,\mathfrak h^{\perp q}_{1L}&\mathfrak g^q_{1L}
\end{pmatrix}.
\end{align}

Playing a little bit with Eqs.~\eqref{rot1} and \eqref{rot2}, we find
\begin{equation}\label{lineq}
\begin{pmatrix}
\tfrac{k_\perp}{M}\left(g^q_{1T}+h_{1L}^{\perp q}\right)\\ g^q_{1L}-h_{1T}^{+q}
\end{pmatrix}
=\begin{pmatrix}\cos\theta&-\sin\theta\\\sin\theta&\cos\theta\end{pmatrix}\begin{pmatrix}
\tfrac{k_\perp}{M}\left(\mathfrak g^q_{1T}+\mathfrak h_{1L}^{\perp q}\right)\\ \mathfrak g^q_{1L}-\mathfrak h_{1T}^{+q}
\end{pmatrix},
\end{equation}
and three expressions invariant under the rotation \eqref{expectation}
\begin{align}
\left(\tfrac{k_\perp}{M}\,g^q_{1T}\right)^2+\left(h_{1T}^{+q}\right)^2&=\left(\tfrac{k_\perp}{M}\,\mathfrak g_{1T}^q\right)^2+\left(\mathfrak h_{1T}^{+q}\right)^2,\label{quadeq1}\\
\left(g^q_{1L}\right)^2+\left(\tfrac{k_\perp}{M}\,h_{1L}^{\perp q}\right)^2&=\left(\mathfrak g_{1L}^q\right)^2+\left(\tfrac{k_\perp}{M}\,\mathfrak h_{1L}^{\perp q}\right)^2,\label{quadeq2}\\
g^q_{1L}\,g^q_{1T}+h_{1L}^{\perp q}\,h_{1T}^{+q}&=\mathfrak g_{1L}^q\,\mathfrak g_{1T}^q+\mathfrak h_{1L}^{\perp q}\,\mathfrak h_{1T}^{+q}\label{quadeq3}.
\end{align}
These three invariant expressions have a simple geometric interpretation. The three-component vector $\sum_iS^i\Phi_q^{ij}\equiv\pi^{qj}_{\uS}$ represents the net light-cone polarization of a quark with three-momentum $(xP^+,\uk_\perp)$ and flavour $q$ in a target with net polarization in the $\uS$-direction. From Eq.~\eqref{expectation}, we see that the vector $\upi^q_{C\uS}$ representing the net canonical polarization of the quark is obtained by a rotation of $\upi^q_{\uS}$ 
\begin{equation}
\pi^{qj}_{C\uS}=\sum_{k}\pi^{qk}_{\uS}\,O^{kj}.
\end{equation}
It follows automatically that $\upi^q_{\uS}\cdot\upi^q_{\uS'}$ is invariant under the rotation \eqref{expectation}
\begin{equation}\label{invariance}
\upi^q_{\uS}\cdot\upi^q_{\uS'}=\upi^q_{C\uS}\cdot\upi^q_{C\uS'}.
\end{equation}
Expressions \eqref{quadeq1} and \eqref{quadeq2} are obtained from \eqref{invariance} for the cases $\uS=\uS'=\boldsymbol e_\perp$ and $\uS=\uS'=\boldsymbol e_z$, respectively. They just express the fact that the magnitude of quark net polarization is the same in the light-cone helicity and canonical-spin bases. Expression \eqref{quadeq3} is obtained for the case $\uS=\boldsymbol e_\perp$ and $\uS'=\boldsymbol e_z$. All the remaining cases do not lead to new independent expressions.

\subsection{Spherical symmetry}
\label{section-43}

We are now ready to discuss the implications of spherical symmetry in the canonical-spin basis. Spherical symmetry means that the canonical tensor correlator has to be invariant $O_R^T\Phi_{Cq}O_R=\Phi_{Cq}$ under any spatial rotation $O_R=\left(\begin{smallmatrix}1&0\\0&R\end{smallmatrix}\right)$ with $R$ the ordinary $3\times 3$ rotation matrix. It is equivalent to the statement that the tensor correlator has to commute with all the elements of the rotation group $\Phi_{Cq}O_R=O_R\Phi_{Cq}$. As a result of Schur's lemma, the canonical tensor correlator must have the following structure
\begin{equation}
\Phi^{\mu\nu}_{Cq}=\begin{pmatrix}A^q&0&0&0\\0&B^q&0&0\\0&0&B^q&0\\0&0&0&B^q\end{pmatrix}.
\end{equation}
Comparing this with Eqs.~\eqref{Ctensor} or \eqref{Ctensorred}, we conclude that spherical symmetry implies
\begin{gather}
f^q_1=A^q,\label{constraint0}\\
\mathfrak g_{1L}^q=\mathfrak h_{1T}^{+q}=h_{1T}^{-q}=B^q,\label{constraint1}\\
\mathfrak g_{1T}^q=\mathfrak h_{1L}^{\perp q}=f_{1T}^{\perp q}=h_1^{\perp q}=0.\label{constraint2}
\end{gather}
Clearly, only the monopole structures in the canonical-spin basis are allowed to survive. Furthermore, the Sivers and Boer-Mulders functions $f_{1T}^{\perp q}$ and $h_1^{\perp q}$ vanish identically, as expected from the fact that we are neglecting gauge-field degrees of freedom.

Note however that the monopole structures in the canonical-spin basis generate higher multipole structures in the light-cone helicity basis. It follows that spherical symmetry imposes some relations among the multipole structures in the light-cone helicity basis, and therefore among the TMDs.
Inserting the constraints \eqref{constraint1} and \eqref{constraint2} into Eq.~\eqref{lineq}, we automatically obtain the linear relations
\begin{equation}
\begin{pmatrix}
\tfrac{k_\perp}{M}\left(g^q_{1T}+h_{1L}^{\perp q}\right)\\ g^q_{1L}-h_{1T}^{+q}
\end{pmatrix}
=\begin{pmatrix}
0\\ 0
\end{pmatrix}.
\end{equation}
Using now the constraints from spherical symmetry in Eq.~\eqref{quadeq1}, we obtain the quadratic relation~\eqref{rel3}
\begin{equation}\label{rel3new}
0=\left(\tfrac{k_\perp}{M}\,g^q_{1T}\right)^2+\left(h_{1T}^{+q}\right)^2-\left(h_{1T}^{-q}\right)^2=\tfrac{\uk_\perp^2}{M^2}\left[\left(g^q_{1T}\right)^2+2h^q_1\,h_{1T}^{\perp q}\right].
\end{equation}
The linear relations \eqref{rel1} and \eqref{rel2} being satisfied, the Eqs.~\eqref{quadeq2} and \eqref{quadeq3} do not lead to independent quadratic relations.

We have seen that spherical symmetry is a \emph{sufficient} condition~\footnote{From Eqs.~\eqref{lineq} and \eqref{quadeq1}, one can see that the \emph{minimal} conditions are actually
\begin{gather*}
\mathfrak g_{1L}^q-\mathfrak h_{1T}^{+q}=0,\\
\mathfrak g_{1T}^q+\mathfrak h_{1L}^{\perp q}=0,\\
\left(\tfrac{k_\perp}{M}\,\mathfrak g_{1T}^q\right)^2+\left(\mathfrak h_{1T}^{+q}\right)^2-\left(h_{1T}^{-q}\right)^2=0.
\end{gather*}
They are indeed fulfilled by spherical symmetry, see Eqs.~\eqref{constraint1} and \eqref{constraint2}.} to obtain all three flavour-independent relations. Restricting ourselves to axial symmetries, we find that some of the relations can already be obtained. For example, axial symmetry about $\boldsymbol e_z$ alone implies the quadratic relation \eqref{rel3} and 
\begin{equation}\label{newquadratic}
g^q_{1L}\,g^q_{1T}+h_{1L}^{\perp q} \,h_{1T}^{+q}=0.
\end{equation}
Axial symmetry about $\hat k_\perp\times\boldsymbol e_z$ implies the two linear relations \eqref{rel1} and \eqref{rel2}. The relation \eqref{newquadratic} is naturally also satisfied but is not independent.

\section{Wave-function approach}
\label{section-5}

Many quark models are based on a wave-function approach. We therefore translate here the derivation of the previous section in the language of 3Q wave functions. The advantage is that we can then also discuss the additional $SU(6)$ spin-flavour symmetry needed for the flavour-dependent relation \eqref{rel4}.

\subsection{Overlap representation of the TMDs on the light cone}
\label{section-51}

Restricting ourselves to the 3Q Fock sector, the target state with definite four-momentum $P=[P^+,\tfrac{M^2}{2P^+},\uzero_\perp]$ and light-cone helicity $\Lambda$ can be written as follows
\begin{equation}\label{LCWF}
|P,\Lambda\rangle=\sum_{\lambda_1\lambda_2\lambda_3}\sum_{q_1q_2q_3}\int[\ud x]_3\,[\ud^2k_\perp]_3\,\psi^{\Lambda;q_1q_2q_3}_{\lambda_1\lambda_2\lambda_3}(\tilde k_1,\tilde k_2,\tilde k_3)\,|\{\lambda_i,q_i,\tilde k_i\}\rangle,
\end{equation}
where $\psi^{\Lambda;q_1q_2q_3}_{\lambda_1\lambda_2\lambda_3}(\tilde k_1,\tilde k_2,\tilde k_3)$ is the three-quark light-cone wave function (3Q LCWF) with $\lambda_i$, $q_i$ and $\tilde k_i$ referring to the light-cone helicity, flavour and light-cone momentum of quark $i$, respectively. The total orbital angular momentum of a given component $\psi^\Lambda_{\lambda_1\lambda_2\lambda_3}$ is given by the expression $\ell_z=\Lambda-\lambda_1-\lambda_2-\lambda_3$ with $\Lambda,\lambda_i=\pm\tfrac{1}{2}$. The integration measures in Eq.~\eqref{LCWF} are defined as
\begin{equation}
\begin{split}
[\ud x]_3&\equiv\left[\prod_{i=1}^3\ud x_i\right]\delta\!\!\left(1-\sum_{i=1}^3x_i\right),\\
[\ud^2k_\perp]_3&\equiv\left[\prod_{i=1}^3\frac{\ud^2k_{i\perp}}{2(2\pi)^3}\right]2(2\pi)^3\,\delta^{(2)}\!\!\left(\sum_{i=1}^3\uk_{i\perp}\right).
\end{split}
\end{equation}
Choosing to label the active quark with $i=1$, the TMDs can be obtained by the following overlaps~\footnote{In the 3Q approach, the spectator system consists of two quarks. It is straightforward to generalize the expression for helicity amplitudes to any kind of spectator system, as the latter is integrated out.} of 3Q LCWFs
\begin{subequations}
\begin{align}
f^q_1&=\int\ud[23]\sum_{\lambda_2\lambda_3}\sum_{q_2q_3}\left[|\psi^{+;qq_2q_3}_{+\lambda_2\lambda_3}|^2+|\psi^{+;qq_2q_3}_{-\lambda_2\lambda_3}|^2\right],\\
g^q_{1L}&=\int\ud[23]\sum_{\lambda_2\lambda_3}\sum_{q_2q_3}\left[|\psi^{+;qq_2q_3}_{+\lambda_2\lambda_3}|^2-|\psi^{+;qq_2q_3}_{-\lambda_2\lambda_3}|^2\right],\\
h^q_1&=\int\ud[23]\sum_{\lambda_2\lambda_3}\sum_{q_2q_3}\left(\psi^{+;qq_2q_3}_{+\lambda_2\lambda_3}\right)^*\psi^{-;qq_2q_3}_{-\lambda_2\lambda_3},\\
\tfrac{k_\perp}{M}\,f^{\perp q}_{1T}&=\int\ud[23]\sum_{\lambda_2\lambda_3}\sum_{q_2q_3}2\,\Im m\left[\hat k_R\left(\psi^{+;qq_2q_3}_{+\lambda_2\lambda_3}\right)^*\psi^{-;qq_2q_3}_{+\lambda_2\lambda_3}\right],\label{dipi}\\
\tfrac{k_\perp}{M}\,g^q_{1T}&=\int\ud[23]\sum_{\lambda_2\lambda_3}\sum_{q_2q_3}2\,\Re e\left[\hat k_R\left(\psi^{+;qq_2q_3}_{+\lambda_2\lambda_3}\right)^*\psi^{-;qq_2q_3}_{+\lambda_2\lambda_3}\right],\\
\tfrac{k_\perp}{M}\,h^{\perp q}_1&=\int\ud[23]\sum_{\lambda_2\lambda_3}\sum_{q_2q_3}2\,\Im m\left[\hat k_R\left(\psi^{+;qq_2q_3}_{-\lambda_2\lambda_3}\right)^*\psi^{+;qq_2q_3}_{+\lambda_2\lambda_3}\right],\\
\tfrac{k_\perp}{M}\,h^{\perp q}_{1L}&=\int\ud[23]\sum_{\lambda_2\lambda_3}\sum_{q_2q_3}2\,\Re e\left[\hat k_R\left(\psi^{+;qq_2q_3}_{-\lambda_2\lambda_3}\right)^*\psi^{+;qq_2q_3}_{+\lambda_2\lambda_3}\right],\label{dipf}\\
\tfrac{\uk_\perp^2}{2M^2}\,h^{\perp q}_{1T}&=\int\ud[23]\sum_{\lambda_2\lambda_3}\sum_{q_2q_3}\hat k^2_R\left(\psi^{+;qq_2q_3}_{-\lambda_2\lambda_3}\right)^*\psi^{-;qq_2q_3}_{+\lambda_2\lambda_3},
\end{align}
\end{subequations}
where we used the notation
\begin{equation}
\ud[23]=[\ud x]_3\,[\ud^2k_\perp]_3\,3\,\delta(x-x_1)\,\delta^{(2)}(\uk_\perp-\uk_{1\perp}).
\end{equation}
Clearly, the TMDs associated to the monopole structures ($f^q_1,g^q_{1L},h^q_1$) are represented by overlaps with no global change of orbital angular momentum $\Delta\ell_z=0$, the ones associated to the dipole structures ($f_{1T}^{\perp q},g^q_{1T},h_1^{\perp q},h_{1L}^{\perp q}$) involve a change by one unit of orbital angular momentum $|\Delta\ell_z|=1$ and the one associated to the quadrupole structure ($h_{1T}^{\perp q}$) involves a change by two units of orbital angular momentum $|\Delta\ell_z|=2$.
If one neglects gauge-field degrees of freedom, the brackets in Eqs.~(\ref{dipi})-(\ref{dipf}) are real, and one obtains vanishing T-odd TMDs.

\subsection{Overlap representation of the TMDs in the canonical-spin basis}
\label{section-52}

Most of the quark models being originally formulated in the instant form, it is more natural to work in the canonical-spin basis instead of the light-cone helicity basis. Since we considered a frame where the target has no transverse momentum, there is no difference between target light-cone and canonical polarizations. Assuming that the quark light-cone helicity and canonical spin are connected by the rotation in Eq.~\eqref{genMelosh}, the components of the LCWF in the canonical-spin basis $\psi^\Lambda_{\sigma_1\sigma_2\sigma_3}$ (with $\sigma_i=\uparrow,\downarrow$) and in the light-cone helicity basis $\psi^\Lambda_{\lambda_1\lambda_2\lambda_3}$ (with $\lambda_i=\pm$) are related as follows~\footnote{Note that $\psi^\Lambda_{\sigma_1\sigma_2\sigma_3}$ cannot be identified in general with the usual rest-frame wave function $ \Psi^\Lambda_{\sigma_1\sigma_2\sigma_3}$. They have the same spin structure, but not the same momentum dependence.}
\begin{equation}\label{connection}
\psi^\Lambda_{\lambda_1\lambda_2\lambda_3}=\sum_{\sigma_1\sigma_2\sigma_3}\psi^\Lambda_{\sigma_1\sigma_2\sigma_3}\,D^{(1/2)*}_{\sigma_1\lambda_1}\,D^{(1/2)*}_{\sigma_2\lambda_2}\,D^{(1/2)*}_{\sigma_3\lambda_3}.
\end{equation}
The correspondence between the components in the two polarization bases is given in a more explicit form in app.~\ref{table}. Since $D^{(1/2)\dag} D^{(1/2)}=\mathds{1}$ for the spectator quarks, we find the explicit overlap representations in canonical-spin basis
\begin{subequations}\label{canonicalrep}
\begin{align}
f^q_1&=\int\ud[23]\sum_{\sigma_2\sigma_3}\sum_{q_2q_3}\left[|\psi^{\uparrow;qq_2q_3}_{\uparrow\sigma_2\sigma_3}|^2+|\psi^{\uparrow;qq_2q_3}_{\downarrow\sigma_2\sigma_3}|^2\right],\\
\mathfrak g^q_{1L}&=\int\ud[23]\sum_{\sigma_2\sigma_3}\sum_{q_2q_3}\left[|\psi^{\uparrow;qq_2q_3}_{\uparrow\sigma_2\sigma_3}|^2-|\psi^{\uparrow;qq_2q_3}_{\downarrow\sigma_2\sigma_3}|^2\right],\\
\mathfrak h_1^q&=\int\ud[23]\sum_{\sigma_2\sigma_3}\sum_{q_2q_3}\left(\psi^{\uparrow;qq_2q_3}_{\uparrow\sigma_2\sigma_3}\right)^*\psi^{\downarrow;qq_2q_3}_{\downarrow\sigma_2\sigma_3},\\
\tfrac{k_\perp}{M}\,f^{\perp q}_{1T}&=\int\ud[23]\sum_{\sigma_2\sigma_3}\sum_{q_2q_3}2\,\Im m\left[\hat k_R\left(\psi^{\uparrow;qq_2q_3}_{\uparrow\sigma_2\sigma_3}\right)^*\psi^{\downarrow;qq_2q_3}_{\uparrow\sigma_2\sigma_3}\right],\\
\tfrac{k_\perp}{M}\,\mathfrak g^q_{1T}&=\int\ud[23]\sum_{\sigma_2\sigma_3}\sum_{q_2q_3}2\,\Re e\left[\hat k_R\left(\psi^{\uparrow;qq_2q_3}_{\uparrow\sigma_2\sigma_3}\right)^*\psi^{\downarrow;qq_2q_3}_{\uparrow\sigma_2\sigma_3}\right],\\
\tfrac{k_\perp}{M}\,h^\perp_1&=\int\ud[23]\sum_{\sigma_2\sigma_3}\sum_{q_2q_3}2\,\Im m\left[\hat k_R\left(\psi^{\uparrow;qq_2q_3}_{\downarrow\sigma_2\sigma_3}\right)^*\psi^{\uparrow;qq_2q_3}_{\uparrow\sigma_2\sigma_3}\right],\\
\tfrac{k_\perp}{M}\,\mathfrak h^{\perp q}_{1L}&=\int\ud[23]\sum_{\sigma_2\sigma_3}\sum_{q_2q_3}2\,\Re e\left[\hat k_R\left(\psi^{\uparrow;qq_2q_3}_{\downarrow\sigma_2\sigma_3}\right)^*\psi^{\uparrow;qq_2q_3}_{\uparrow\sigma_2\sigma_3}\right],\\
\tfrac{\uk_\perp^2}{2M^2}\,\mathfrak h^{\perp q}_{1T}&=\int\ud[23]\sum_{\sigma_2\sigma_3}\sum_{q_2q_3}\hat k^2_R\left(\psi^{\uparrow;qq_2q_3}_{\downarrow\sigma_2\sigma_3}\right)^*\psi^{\downarrow;qq_2q_3}_{\uparrow\sigma_2\sigma_3},
\end{align}
\end{subequations}
with $\mathfrak h_{1T}^{+q}=\mathfrak h^q_1+\tfrac{\uk_\perp^2}{2M^2}\,\mathfrak h_{1T}^{\perp q}$ and $h_{1T}^{-q}=\mathfrak h^q_1-\tfrac{\uk_\perp^2}{2M^2}\,\mathfrak h_{1T}^{\perp q}$. The functions $\mathfrak g^q_{1L},\mathfrak g^q_{1T},\mathfrak h^{\perp q}_{1L},\mathfrak h^{+ q}_{1T}$ are again related to the TMDs $g^q_{1L},g^q_{1T},h^{\perp q}_{1L},h^{+ q}_{1T}$ according to Eqs.~\eqref{rot1} and \eqref{rot2}.

\subsection{Spherical symmetry}
\label{section-53}

We now discuss how spherical symmetry restricts the form of the wave function in the canonical-spin basis. Spherical symmetry requires the wave function to be invariant under any rotation, \emph{i.e.}
\begin{equation}
\sum_{\Lambda'\sigma'_1\sigma'_2\sigma'_3}\left[u(\theta,\phi)\right]_{\sigma_1\sigma'_1}\left[u(\theta,\phi)\right]_{\sigma_2\sigma'_2}\left[u(\theta,\phi)\right]_{\sigma_3\sigma'_3}\left[u(\theta,\phi)\right]^*_{\Lambda\Lambda'}\psi^{\Lambda'}_{\sigma'_1\sigma'_2\sigma'_3}=\psi^\Lambda_{\sigma_1\sigma_2\sigma_3},
\end{equation}
with the $SU(2)$ rotation matrix $u(\theta,\phi)$ given by Eq.~\eqref{su2rot}. In particular, invariance under a $(\pi,0)$-rotation leads to
\begin{equation}\label{rotinv1}
\psi^{-\Lambda}_{-\sigma_1-\sigma_2-\sigma_3}=(-1)^{\Lambda+\sigma_1+\sigma_2+\sigma_3}\,\psi^\Lambda_{\sigma_1\sigma_2\sigma_3},
\end{equation}
while invariance under $(0,\phi)$-rotations implies that all components with $\ell_z\neq 0$ have to vanish
\begin{equation}\label{rotinv2}
\psi^\uparrow_{\uparrow\uparrow\uparrow}=\psi^\uparrow_{\downarrow\downarrow\uparrow}=\psi^\uparrow_{\downarrow\uparrow\downarrow}=\psi^\uparrow_{\uparrow\downarrow\downarrow}=\psi^\uparrow_{\downarrow\downarrow\downarrow}=0.
\end{equation}
Taking into account the constraints \eqref{rotinv1} and \eqref{rotinv2} in an arbitrary $(\theta,\phi)$-rotation, one finally gets~\footnote{Note that spherical symmetry neither restricts the number of non-zero components of the wave function in the light-cone helicity basis nor relates them in a simple way, see Table~\ref{3QLCWF} in app.~\ref{table}.}
\begin{equation}\label{rotinv3}
\psi^\uparrow_{\uparrow\uparrow\downarrow}+\psi^\uparrow_{\uparrow\downarrow\uparrow}+\psi^\uparrow_{\downarrow\uparrow\uparrow}=0.
\end{equation}

Again, spherical symmetry implies that the TMDs are either identically zero or proportional to the unpolarized and polarized amplitudes $A^q$ and $B^q$,
\begin{subequations}\label{sphericalTMDs}
\begin{align}
f^q_1&=A^q,\\
g^q_{1L}&=\cos\theta\,B^q,\\
h^q_1&=\frac{\cos\theta+1}{2}\,B^q,\\
\tfrac{k_\perp}{M}\,f^{\perp q}_{1T}&=0,\\
\tfrac{k_\perp}{M}\,g^q_{1T}&=\sin\theta\,B^q,\\
\tfrac{k_\perp}{M}\,h^{\perp q}_1&=0,\\
\tfrac{k_\perp}{M}\,h^{\perp q}_{1L}&=-\sin\theta\,B^q,\\
\tfrac{\uk_\perp^2}{2M^2}\,h^{\perp q}_{1T}&=\frac{\cos\theta-1}{2}\,B^q,
\end{align}
\end{subequations}
with $A^q$ and $B^q$ given by the following overlaps 
\begin{align}
\label{eq:a}
A^q&=\int\ud[23]\sum_{q_2q_3}\left[|\psi^{\uparrow;qq_2q_3}_{\uparrow\uparrow\downarrow}|^2+|\psi^{\uparrow;qq_2q_3}_{\uparrow\downarrow\uparrow}|^2+|\psi^{\uparrow;qq_2q_3}_{\downarrow\uparrow\uparrow}|^2\right],\\
\label{eq:b}
B^q&=\int\ud[23]\sum_{q_2q_3}\left[|\psi^{\uparrow;qq_2q_3}_{\uparrow\uparrow\downarrow}|^2+|\psi^{\uparrow;qq_2q_3}_{\uparrow\downarrow\uparrow}|^2-|\psi^{\uparrow;qq_2q_3}_{\downarrow\uparrow\uparrow}|^2\right]\\
&=\int\ud[23]\sum_{q_2q_3}\left[\left(\psi^{\uparrow;qq_2q_3}_{\uparrow\uparrow\downarrow}\right)^*\psi^{\downarrow;qq_2q_3}_{\downarrow\uparrow\downarrow}+\left(\psi^{\uparrow;qq_2q_3}_{\uparrow\downarrow\uparrow}\right)^*\psi^{\downarrow;qq_2q_3}_{\downarrow\downarrow\uparrow}\right].\nonumber
\end{align}
The TMD relations \eqref{rel1}-\eqref{rel3} then follow trivially. 

\subsection{$SU(6)$ spin-flavour symmetry}
\label{section-54}

Many quark models, in addition of being spherically symmetric, assume also the $SU(6)$ spin-flavour symmetry. 
As a result, the wave function in the canonical spin basis is given by
the product of a symmetric momentum wave function $\phi$ and a spin-isospin component
$\Phi^{\Lambda;q_1q_2q_3}_{\sigma_1\sigma_2\sigma_3}$
\begin{equation}
\psi^{\Lambda;q_1q_2q_3}_{\sigma_1\sigma_2\sigma_3}=\phi(\tilde k_1,\tilde k_2,\tilde k_3)\Phi^{\Lambda;q_1q_2q_3}_{\sigma_1\sigma_2\sigma_3}.
\label{eq:su6}
\end{equation}
The wave function in the spin-flavour space can be written as
\begin{equation}
\Phi^{\Lambda;q_1q_2q_3}_{\sigma_1\sigma_2\sigma_3}=
\frac{1}{\sqrt{2}}\left[\chi^\alpha
\xi^\alpha
+\chi^\beta
\xi^\beta\right]
\end{equation}
with
\begin{align}
\label{eq:spin-isospin}
\chi^\alpha=\frac{1}{\sqrt{2}}\left[\uparrow \uparrow\downarrow-\uparrow\downarrow\uparrow\right],
&\quad
\xi^\alpha=\frac{1}{\sqrt{2}}\left[u u d-udu\right],
\\
\chi^\beta=\frac{1}{\sqrt{6}}\left[2\downarrow\uparrow\uparrow-\uparrow\downarrow\uparrow-\uparrow\uparrow\downarrow\right],&\quad
\xi^\beta=\frac{1}{\sqrt{6}}\left[2duu-udu-uud\right].
\end{align}
As a result, writing explicitly the spin and isospin components of the wave function 
\eqref{eq:su6},  we find the following results in the canonical spin basis
\begin{equation}
\begin{array}{c|ccc}
\psi^{\uparrow;q_1q_2q_3}_{\sigma_1\sigma_2\sigma_3}&\,\,uud\,\,&\,\,udu\,\,&\,\,duu\,\,\\ &&&\\
\hline
\uparrow\uparrow\downarrow&2\phi&-\phi&-\phi\\
\uparrow\downarrow\uparrow&-\phi&2\phi&-\phi\\
\downarrow\uparrow\uparrow&-\phi&-\phi&2\phi
\end{array}
\end{equation}
with $\phi=\phi(\{\tilde k_i\})$  normalized as $\int[\ud x]_3\,[\ud^2k_\perp]_3|\phi|^2=1/6$. This implies that the unpolarized and polarized amplitudes $A^q$ and $B^q$ are simply proportional
\begin{equation}
\label{eq:a-b}
A^u=2A^d=\tfrac{3}{2}\,B^u=-6\,B^d=12\int\ud[23]|\phi|^2,
\end{equation}
and so the flavour-dependent relation \eqref{rel4} follows trivially with $\mathcal D^q=B^q/A^q$.

\section{Quark models}
\label{models}

In this section we review different quark models which have been used for the calculation of TMDs. In particular, we summarize the main ingredients of the models and discuss whether they satisfy the conditions of sect.~\ref{section-3}. In order to facilitate the discussion, we sort the quark models in classes defined as follows: 
\begin{itemize}
\item The light-cone constituent quark model (LCCQM) of Ref.~\cite{Pasquini:2008ax} and the light-cone quark-diquark model (LCQDM) of Refs.~\cite{She:2009jq,Zhu:2011zza,Ma:1991xq} constitute the class of light-cone models;
\item The covariant parton model of Ref.~\cite{Efremov:2009ze} constitutes its own class;
\item The bag model of Refs.~\cite{Avakian:2010br,Avakian:2008dz} and the light-cone version of the chiral quark-soliton model (LC$\chi$QSM) of Refs.~\cite{Lorce:2011dv,Lorce:2006nq,Lorce:2007as} constitute the class of mean-field models; 
\item The quark-diquark models of Refs.~\cite{Jakob:1997wg,Bacchetta:2008af,Goldstein:2002vv,Gamberg:2003ey,Gamberg:2007wm} constitute the class of spectator models.
\end{itemize}
We will not discuss the quark-target model of Ref.~\cite{Meissner:2007rx} as it deals with gluons and therefore does already not satisfy the first condition of sect.~\ref{section-3}.

\subsection{Light-cone models}
\label{sect:LCM}

The class of light-cone models is characterized by the fact that the target state is expanded in the basis of free parton (Fock) states. One usually truncates the expansion and considers only the state with the lowest number of partons. In the LCCQM, this lowest state consists of three valence quarks, while in the LCQDM it consists of a valence quark and a spectator diquark.

It is well known that light-cone helicity and canonical spin of free partons are simply related by the so-called Melosh rotation \cite{Melosh:1974cu}. Its $j=1/2$ and $j=1$ representations \cite{Ahluwalia:1993xa} are given by (see app.~\ref{app:1} for the definition of the spinors and polarization four-vectors) 
\begin{align}
D^{(1/2)*}_{\sigma\lambda}(\tilde k)&=\frac{\overline u_{LC}(k,\lambda)u(k,\sigma)}{2m}\nonumber\\
&=\frac{1}{\sqrt{N}}\begin{pmatrix}\sqrt{2}\,k^++m&-k_R\\k_L&\sqrt{2}\,k^++m\end{pmatrix},\label{meloshspinor}\\
D^{(1)*}_{\sigma\lambda}(\tilde k)&=-\varepsilon^*_{LC}(k,\lambda)\cdot\varepsilon(k,\sigma)\nonumber\\
&=\frac{1}{N}\begin{pmatrix}\left(\sqrt{2}\,k^++m\right)^2&-\sqrt{2}\left(\sqrt{2}\,k^++m\right)k_R&k_R^2\\\sqrt{2}\left(\sqrt{2}\,k^++m\right)k_L&\left(\sqrt{2}\,k^++m\right)^2-\uk_\perp^2&-\sqrt{2}\left(\sqrt{2}\,k^++m\right)k_R\\k_L^2&\sqrt{2}\left(\sqrt{2}\,k^++m\right)k_L&\left(\sqrt{2}\,k^++m\right)^2\end{pmatrix},\label{meloshvector}
\end{align}
where $m$ is the parton mass and $N=(\sqrt{2}\,k^++m)^2+\uk_\perp^2$. The LCWF in the canonical-spin basis being identified in these models with the instant-form wave function, it follows that $\sqrt{2}\,k^+=x\mathcal M_0$ with $\mathcal M_0=\sum_i\omega_i$ the mass of the Fock state and $\omega_i$ the free energy of parton $i$. Comparing now Eqs.~\eqref{meloshspinor} and \eqref{meloshvector} with Eqs.~\eqref{genMelosh} and \eqref{genMelosh2}, we obtain
\begin{equation}
\cos\tfrac{\theta}{2}=\frac{m+x\mathcal M_0}{\sqrt{N}}\qquad\text{and}\qquad\sin\tfrac{\theta}{2}=\frac{k_\perp}{\sqrt{N}}.
\end{equation}
Finally, both the LCCQM and LCQDM consider wave functions with spherical symmetry and $SU(6)$ spin-flavour symmetry. In other words, all the conditions of sect.~\ref{section-3} are satisfied in these models, and so are the TMD relations~\eqref{rel1}-\eqref{rel4}~\footnote{In the derivations of sects.~\ref{section-4} and \ref{section-5}, we tacitly assumed that the rotation connecting light-cone and canonical polarizations depends only on the momentum of the parton under consideration. The Melosh rotation involves the free invariant mass $\mathcal M_0$ and therefore the momenta of all partons in the state. We are in fact not allowed to pull the factor due to the rotation of the active quark out of the integral like in Eq.~\eqref{sphericalTMDs}. Nevertheless, this technical detail does not affect the conclusion about the validity of the relations \eqref{rel1}-\eqref{rel4}. Our tacit assumption, which does not apply to this specific case,  has been introduced to keep the presentation as simple as possible.} .

\subsection{Covariant parton model}

The standard quark-parton model (QPM) refers to the infinite momentum frame (IMF), where the parton mass can be neglected. The covariant parton model is an alternative to the QPM that is not confined to a preferred reference frame. Following the standard assumptions of the QPM, the covariant parton model describes the target system as a gas of quasi-free partons, \emph{i.e.} the partons bound inside the target behave at the interaction with the external probe (at sufficiently high $Q^2$) as free particles having four-momenta on the mass shell. However, since the covariant parton model does not refer specifically to the IMF, the parton mass~\footnote{Note that the parton mass appearing in the model has to be regarded as an effective mass, in the sense that it corresponds to the mass of the free parton behaving at the interaction like the actual bound parton.} $m$ is not neglected. One also assumes explicitly that the parton distributions are spherically symmetric.

The covariant parton model does not refer explicitly to quark canonical spin or light-cone helicity. Instead, it deals with the covariant quark polarization vector. Identifying in the Bjorken limit the Lorentz structures of the hadronic tensor with those of the TMD correlator, the authors of Ref.~\cite{Efremov:2009ze} found that the TMDs are given in the covariant parton model by
\begin{subequations}
\begin{align}
f^q_1&=\tfrac{1}{2}\left[\left(m+xM\right)^2+\uk^2_\perp\right]\int\{\ud\tilde k^1\},\\
g^q_{1L}&=\tfrac{1}{2}\left[\left(m+xM\right)^2-\uk^2_\perp\right]\int\{\ud k^1\},\\
\tfrac{k_\perp}{M}\,f^{\perp q}_{1T}&=0,\\
\tfrac{k_\perp}{M}\,g^q_{1T}&=\left(m+xM\right)k_\perp\int\{\ud k^1\},\\
\tfrac{k_\perp}{M}\,h^{\perp q}_1&=0,\\
\tfrac{k_\perp}{M}\,h^{\perp q}_{1L}&=-\left(m+xM\right)k_\perp\int\{\ud k^1\},\\
\tfrac{\uk_\perp^2}{2M^2}\,h^{\perp q}_{1T}&=-\tfrac{1}{2}\,\uk_\perp^2\int\{\ud k^1\},
\end{align}
\end{subequations}
where $M$ is the target mass, $\{\ud\tilde k^1\}$ and $\{\ud k^1\}$ are the measures associated to the distributions of unpolarized and polarized quarks, respectively. Comparing with Eq.~\eqref{sphericalTMDs}, we find that
\begin{equation}
\cos\tfrac{\theta}{2}=\frac{m+xM}{\sqrt{\left(m+xM\right)^2+\uk_\perp^2}}\qquad\text{and}\qquad\sin\tfrac{\theta}{2}=\frac{k_\perp}{\sqrt{\left(m+xM\right)^2+\uk_\perp^2}},
\end{equation}
which is nothing else than the Melosh rotation. This is consistent with the fact that the active quark is quasi-free in this model. The difference with light-cone models is that the physical mass of the target $M$ is used in the Melosh rotation instead of the free invariant mass $\mathcal M_0$~\footnote{In the covariant parton model, only the active parton is considered on-shell. In light-cone models, all the partons are on-shell so that the Fock state itself is off-shell $\mathcal M_0\neq M$.}. The conditions 1-3 of sect.~\ref{section-3} are therefore satisfied in the covariant parton model, and so are the TMD relations \eqref{rel1}-\eqref{rel3}. Since this model does not use the language of wave functions, the implementation of $SU(6)$ spin-flavour symmetry is more delicate and one has to assume that the unpolarized and polarized distributions become simply proportional in order to recover the TMD relation \eqref{rel4}.

\subsection{Mean-field models}

In mean-field models, the target is considered as made of quarks bound by a classical mean field representing the non-perturbative (long-range) contribution of the gluon field. Accordingly, the positive-frequency part of the quark field appearing in the definition of the correlator \eqref{correlator} is expanded in the basis of the bound-state solutions $e^{-iE_nt}\,\varphi_n(k,\sigma)$ instead of the free Dirac light-cone spinors $e^{-ik\cdot x}\,u_{LC}(k,\lambda)$. Moreover, one truncates the expansion to the lowest mode $\varphi\equiv\varphi_1$ with energy $E_\text{lev}\equiv E_1$.

In these models, the bound-state solution $\varphi(k,\sigma)$ is called the quark wave function. This object is clearly different from the LCWF introduced in sect.~\ref{section-5}. In particular, the former is a spinor while the latter is an ordinary scalar function. It is however possible to relate them. Since we consider twist-$2$ Dirac operators $\Gamma$, only the good components of the spinors are involved in the quark bilinear $\overline\varphi(k,\sigma')\Gamma\varphi(k,\sigma)$. Using $u_G(\lambda)=P_+u_{LC}(k,\lambda)/\sqrt{2^{1/2}k^+}$ (see app.~\ref{app:1}), we find that
\begin{align}
\overline\varphi(k,\sigma')\Gamma\varphi(k,\sigma)&=\varphi^\dag(k,\sigma')P_+\gamma^0\Gamma P_+\varphi(k,\sigma)\nonumber\\
&=\sum_{\lambda\lambda'}F^*_{\lambda'\sigma'}(k')\,F_{\lambda\sigma}(k)\,\frac{\overline u_{LC}(k,\lambda')\Gamma u_{LC}(k,\lambda)}{\sqrt{2}k^+},
\end{align}
where we have defined $F_{\lambda\sigma}(k)=u^\dag_G(\lambda)\varphi(k,\sigma)$. In agreement with \cite{Petrov:2002jr,Diakonov:2005ib} where one boosts explicitly the system in the mean-field approximation to the IMF, $F_{\lambda\sigma}(k)$ can be interpreted as the quark LCWF with $k_z=xM-E_\text{lev}$. The mass $M$ is identified with the nucleon mass $M_N$ in the bag model and with the soliton mass $\uM_N$ in the LC$\chi$QSM.

The mean field is assumed spherically symmetric in the target rest frame. It follows that the lowest quark-state solution in momentum space takes the form 
\begin{equation}\label{sphericalspinor}
\varphi(k,\sigma)=\begin{pmatrix}f(|\uk|)\\\tfrac{\uk\cdot\usigma}{|\uk|}\,g(|\uk|)\end{pmatrix}\chi_\sigma,
\end{equation}
with $\chi_\sigma$ the Pauli spinor. The functions $f$ and $g$ in Eq.~\eqref{sphericalspinor} represent the $s$ ($\ell=0$) and $p$ ($\ell=1$) waves of the bound-state solution. On the other hand, the general 3Q LCWF for a spin-$1/2$ target involves usually $s$-, $p$- and $d$-waves. There is no contradiction between these two statements since $f$ and $g$ describe a single quark in the target and therefore do not represent partial waves of \emph{total} angular momentum. Note also that in the language of 3Q LCWF, the $s$-, $p$- and $d$-waves refer to components with $\ell_z=0$, $\pm 1$ and $\pm 2$, respectively. This is an abuse of language as partial waves should refer to $\ell$ and not $\ell_z$.

The quark LCWF corresponding to \eqref{sphericalspinor} is then given by~\footnote{Replacing $\varphi(k,\sigma)$ by the free Dirac spinor $u(k,\sigma)$, one recovers the Melosh rotation given by Eq.~\eqref{meloshspinor} $u^\dag_G(\lambda)u(k,\sigma)=\sqrt{2^{1/2}k^+}\,D^{(1/2)*}_{\sigma\lambda}(\tilde k)$.}
\begin{equation}
F_{\lambda\sigma}(k)=\frac{1}{\sqrt{2}}\begin{pmatrix}f(|\uk|)+\tfrac{k_z}{|\uk|}\,g(|\uk|)&-\tfrac{k_R}{|\uk|}\,g(|\uk|)\\\tfrac{k_L}{|\uk|}\,g(|\uk|)&f(|\uk|)+\tfrac{k_z}{|\uk|}\,g(|\uk|)\end{pmatrix}_{\sigma\lambda}.
\end{equation}
It describes in particular how canonical spin $\sigma$ and light-cone helicity $\lambda$ are related.
Comparing with Eq.~\eqref{genMelosh}, we find that
\begin{equation}
\cos\tfrac{\theta}{2}=\frac{f(|\uk|)+\tfrac{k_z}{|\uk|}\,g(|\uk|)}{\sqrt{N}}\qquad\text{and}\qquad\sin\tfrac{\theta}{2}=\frac{\tfrac{k_\perp}{|\uk|}\,g(|\uk|)}{\sqrt{N}},
\end{equation}
with $N=f^2(|\uk|)+2\,\tfrac{k_z}{|\uk|}\,f(|\uk|)g(|\uk|)+g^2(|\uk|)$. The 3Q LCWF written as $\prod_{i=1}^{3}F_{\lambda_i\sigma_i}(k_i)$ times the standard $SU(6)$ spin-flavour wave function with target polarization $\Lambda=\sum_i\sigma_i$, is then consistent with spherical symmetry in the canonical-spin basis. All the conditions of sect.~\ref{section-3} being satisfied in mean-field models, the TMD relations \eqref{rel1}-\eqref{rel4} follow automatically.

\subsection{Spectator models}

The basic idea of spectator models is to evaluate the quark-quark correlator $\Phi$ of Eq.~\eqref{correlator} by inserting a complete set of intermediate states and then truncating this set at tree level to a single on-shell spectator diquark state, \emph{i.e.} a state with the quantum numbers of two quarks. The diquark can be either an isospin singlet with spin $0$ (scalar diquark) or an isospin triplet with spin $1$ (axial-vector diquark). The target is then seen as made of an off-shell quark and an on-shell diquark. Spectator models differ by their specific choice of target-quark-diquark vertices, polarization four-vectors associated with the axial-vector diquark, and form factors which take into account in an effective way the composite nature of the target and the spectator diquark. 

As advocated in Ref.~\cite{Brodsky:2000ii}, the parton distributions can conveniently be computed using the language of LCWFs. The scalar quark-diquark LCWF is defined as
\begin{equation}
\psi^\Lambda_\lambda(\tilde k)\propto\overline u_{LC}(k,\lambda)\mathcal Y_s u_{LC}(P,\Lambda)
\end{equation}
with target momentum $P=\left[P^+,\tfrac{M^2}{2P^+},\uzero_\perp\right]$. We do not need to specify all the factors in the definition as we are only interested in the structure of the wave function in the light-cone helicity basis. The scalar vertex is of the Yukawa type $\mathcal Y_s=g_s(k^2)\,\mathds 1$ with $g_s(k^2)$ some form factor. Writing down explicitly the components, one finds
\begin{equation}
\psi^\Lambda_\lambda(\tilde k)\propto\frac{g_s(k^2)}{\sqrt{x}}\begin{pmatrix}m+xM&-k_R\\k_L&m+xM\end{pmatrix}_{\Lambda\lambda}.
\label{melosh-diquark}
\end{equation}
Note the striking resemblance with the Melosh rotation matrix of Eq.~\eqref{meloshspinor}. One can similarly define a rest-frame scalar quark-diquark wave function as
\begin{equation}
\label{wf-restframe}
\Psi^\Lambda_\sigma\propto\overline u(k,\sigma)\mathcal Y_s u(P_\text{rest},\Lambda)=g_s(k^2)\sqrt{2M(E+m)}\,\delta_{\Lambda\sigma}
\end{equation}
with target momentum $P_\text{rest}=\left(M,\uzero\right)$. This wave function is obviously spherically symmetric~\footnote{The rest-frame wave function in Eq.~\eqref{wf-restframe} is expressed in terms of canonical spin and therefore has the same spin structure as the LCWF expressed in the canonical-spin basis. It follows that the constraints due to spherical symmetry discussed in app.~\ref{app-c1} apply also here. Furthermore, the momentum-dependent part of the wave function in the rest frame does not depend on a specific direction.}. Furthermore, Eq.~\eqref{melosh-diquark} suggests that the quark light-cone helicity and canonical spin are simply related by a Melosh rotation, as if the quark was free~\cite{Ellis:2008in}. In other words, the conditions 1-3 of sect.~\ref{section-3} are satisfied in scalar diquark models with Yukawa-like vertex, and so are the TMD relations \eqref{rel1}-\eqref{rel3}.

The axial-vector quark-diquark LCWF is defined as
\begin{equation}
\psi^\Lambda_{\lambda\lambda_D}(\tilde k)\propto\overline u_{LC}(k,\lambda)\varepsilon^*_{LC\mu}(K,\lambda_D)\mathcal Y^\mu_a u_{LC}(P,\Lambda).
\end{equation}
The spectator model of Jakob \emph{et al.} \cite{Jakob:1997wg} assumes the following structure for the axial-vector vertex $\mathcal Y^\mu_a=\frac{g_a(k^2)}{\sqrt{3}}\,\gamma_5\left(\gamma^\mu+\tfrac{P^\mu}{M}\right)$ and the following momentum argument for the polarization four-vector $K=P$. The motivation for such a choice is to ensure that, in the target rest frame, the diquark spin-$1$ states are purely spatial. Indeed, the rest-frame axial-vector quark-diquark wave function reads in this model
\begin{equation}
\Psi^\Lambda_{\sigma\sigma_D}\propto\overline u(k,\sigma)\varepsilon^*_\mu(K,\sigma_D)\mathcal Y^\mu_a u(P_\text{rest},\Lambda)=\frac{g_a(k^2)}{\sqrt{3}}\,\sqrt{2M(E+m)}\left(\boldsymbol\epsilon_{\sigma_D}\cdot\usigma\right)_{\sigma\Lambda}.
\end{equation}
It satisfies the constraints \eqref{sphericalAV} and is therefore spherically symmetric. Writing down explicitly the components of the corresponding LCWF, one finds
\begin{equation}\label{JakobAV}
\begin{gathered}
\psi^+_{+0}\propto\frac{g_a(k^2)}{\sqrt{3x}}\left(m+xM\right),\qquad\psi^+_{-0}\propto-\frac{g_a(k^2)}{\sqrt{3x}}\,k_R,\\
\psi^+_{-+}=-\sqrt{2}\,\psi^+_{+0},\qquad\psi^-_{--}=\sqrt{2}\,\psi^+_{-0},\qquad\psi^+_{+-}=\psi^+_{--}=0,
\end{gathered}
\end{equation}
the other components being given by $\psi^{-\Lambda}_{-\lambda-\lambda_D}=(-1)^{\Lambda+\lambda+\lambda_D}\left(\psi^{\Lambda}_{\lambda\lambda_D}\right)^*$, with $\Lambda,\lambda=\pm\tfrac{1}{2}$ and $\lambda_D=+1, 0, -1$. Again, one recognizes the characteristic factors of the Melosh rotation~\cite{Ellis:2008in}. Comparing the structure of the components of the LCWF in Eq.~\eqref{JakobAV} with the structure of the components of the LCWF given in Table \ref{AVLCWF} after applying the constraints of spherical symmetry in the canonical-spin basis \eqref{sphericalAV}, one concludes that only the quark polarization is rotated. This is in agreement with the fact that the momentum argument of the polarization four-vector $\varepsilon_\mu$ does not have any transverse momentum, and so there is no rotation of the diquark polarization. All the conditions 1-3 of sect.~\ref{section-3} being satisfied in the axial-vector diquark model of Ref.~\cite{Jakob:1997wg}, the TMD relations \eqref{rel1}-\eqref{rel3} follow automatically. The flavour-dependent relation \eqref{rel4} can be obtained by further imposing $SU(6)$ spin-flavour symmetry to the wave function~\footnote{The scalar and axial-vector diquarks represent in principle more than just two quarks. For this reason, they have \emph{a priori} different masses, cutoffs, form factors, \ldots When we impose $SU(6)$ symmetry, we implicitly consider that the quark-diquark picture originates from a 3Q picture. The scalar and axial-vector diquarks then just differ by their spin and flavour structures which are uniquely determined by the $SU(6)$ symmetry.}.

On the contrary, some versions of the spectator model presented by Bacchetta \emph{et al.} in Ref.~\cite{Bacchetta:2008af} do not support any TMD relation. We therefore expect that at least one of the conditions 1-3 of sect.~\ref{section-3} is not satisfied. These versions are based on the axial-vector vertex $\mathcal Y^\mu_a=\frac{g_a(k^2)}{\sqrt{2}}\,\gamma^\mu\gamma_5$ and involve the diquark momentum $K=P-k$ in the polarization four-vector. With these choices, it is found that the condition 3 of sect.~\ref{section-3} is not fulfilled since the corresponding rest-frame wave function does not satisfy the requirements of spherical symmetry
\begin{equation}
\Psi^\Lambda_{\sigma\sigma_D}\not\propto\left(\boldsymbol\epsilon_{\sigma_D}\cdot\usigma\right)_{\sigma\Lambda},
\end{equation}
in accordance with the discussions of Refs.~\cite{Gross:2006fg}-\cite{Gross:2008zza} and the comment in Ref.~\cite{Bacchetta:2008af} that in this approach the partons do not necessarily occupy the lowest-energy available orbital (with quantum numbers $J^P=\tfrac{1}{2}^+$ and $L_z=0$.)

\section{$SU(6)$-symmetry breaking in a light-cone constituent quark model}
\label{section:su6br}
In this section we apply the formalism introduced in the previous sections for the calculation  of T-even TMDs from LCWFs
to a specific model, namely the light-cone constituent quark model.
For recent quark-model calculation of the T-odd TMDs we refer 
to~\cite{Bacchetta:2008af,Gamberg:2007wm,Pasquini:2010af,Courtoy:2009pc,Courtoy:2008vi,Gamberg:2009uk,Cherednikov:2006zn,Lu:2004au}.
First, we specify the LCWF of \eqref{connection} which assumes $SU(6)$ symmetry for the spin-flavour component of the wave function. Then, we add a component of the LCWF which breaks the $SU(6)$ 
symmetry.

In a first step  the percentage of $SU(6)$-breaking terms in the total LCWF is left as free parameter, 
and we discuss in general the effects of these terms on the TMD results.
This calculation can be reproduced numerically using the Mathematica program which can be downloaded from Ref.~\cite{mat1} in both the Windows and Linux 
versions~\footnote{In order to run the program, you also need the auxiliary functions available at the same web address with the name ``Functions-TMD.zip''.}.
In particular, this program allows one to reproduce the model calculation for {\em i)} the T-even TMDs;
{\em ii)} the spin densities in the transverse-momentum space as function of the quark and nucleon polarizations; {\em iii)} the  $ \uk^2_\perp$ dependence of the $x$ moments of the TMDs, with a comparison
of the model results with the Gaussian Ansatz.

In a second step, we study different observables, which are particularly sensitive to $SU(6)$ breaking
and can be used to fix the free parameter of the model 
corresponding to the percentage of $SU(6)$-breaking terms in the LCWF of the nucleon.
Explicit examples of this fitting procedure are given in a second Mathematica file~\cite{mat2}.
In particular, in this program we present different strategies
for fitting the available experimental data
 of three  observables: {\em i)} the inclusive polarized asymmetry $A_1$; 
{\em ii)} the ratio $F_2^n/F_2^p$ of the neutron to proton unpolarized structure function of DIS;
{\em iii)} the nucleon electroweak form factors.
In these notes we will discuss, as an example, the fit of the double spin asymmetry $A_1$ in DIS, leaving
for a future work a systematic study to select the best fitting procedure.

\subsection{Light-cone constituent quark model}
\label{sec:calculation}
The $SU(6)$-symmetric component of the LCWF in the LCCQM has the form in Eq.~\eqref{connection}, with 
the 
$SU(2)$ matrix relating the light-cone helicity and the spin given by the Melosh rotation discussed in 
sect.~\ref{sect:LCM} and the wave function in the canonical spin basis given in Eq.~\eqref{eq:su6}.
For the symmetric momentum wave function one assumes the following  
functional form
\begin{equation}
\phi(\{\tilde k_i\})=
2(2\pi)^3\bigg[\frac{1}{M_0}\frac{\omega_1\omega_2\omega_3}{x_1x_2x_3}\bigg]^{1/2}
\frac{N'}{(M_0^2+\beta^2)^\gamma},
\label{eq:30}
\end{equation}
where $N'$ is a normalization factor, and the scale $\beta$,
the parameter $\gamma$ for the power-law behaviour, and the quark mass $m$ are 
taken from Ref.~\cite{Schlumpf:94a}, \emph{i.e.} $\beta=0.607 $ GeV, $\gamma=3.4$ and $m=0.267$ GeV. According to Schlumpf's analysis~\cite{Schlumpf:94b} these values lead to a very good description of many baryonic properties.

We now introduce an admixture of mixed-symmetry component in  the nucleon wave function given by
\begin{equation}
\label{eq:psi-total}
\Psi^{\Lambda;q_1q_2q_3}_{\sigma_1\sigma_2\sigma_3}=\cos\delta\,\psi^{\Lambda;q_1q_2q_3}_{\sigma_1\sigma_2\sigma_3}+\sin\delta\,
\tilde\psi^{\Lambda;q_1q_2q_3}_{\sigma_1\sigma_2\sigma_3},
\end{equation}
where $\psi^{\Lambda;q_1q_2q_3}_{\sigma_1\sigma_2\sigma_3}$ is the $SU(6)$-symmetric component of Eq.~\eqref{eq:su6} and $\tilde\psi^{\Lambda;q_1q_2q_3}_{\sigma_1\sigma_2\sigma_3} $ is a mixed-symmetry component.
The factor $\cos^2\delta$ and $\sin^2\delta$ give  the percentages 
in the nucleon wave function of  $SU(6)$-symmetric and $SU(6)$-breaking terms, respectively.
For the $SU(6)$-breaking component we take an appropriate combination of mixed-symmetry spin-isospin wave functions with two momentum wave function of mixed symmetry
\begin{equation}
\tilde\psi^{\Lambda;q_1q_2q_3}_{\sigma_1\sigma_2\sigma_3}(\{\tilde k_i\}=
   \frac{1}{2}\left[\phi^\beta\left(\chi^\alpha\xi^\alpha-\chi^\beta\xi^\beta\right)+\phi^\alpha\left(\chi^\alpha\xi^\beta+\chi^\beta\xi^\alpha\right)\right],
\end{equation}
where the spin and isospin wave functions are defined 
as in~\eqref{eq:spin-isospin}, and the momentum-dependent part is given by
\begin{align}
\label{eq:su6-br}
\phi^\alpha(\{\tilde k_i\})=
N^\alpha\frac{\vec \alpha\cdot\vec\beta}{\vec\alpha^2+\vec\beta^2}\phi(\{\tilde k_i\}),
&\quad\quad
\phi^\beta=N^\beta\frac{\vec\beta^2-\vec\alpha^2}{\vec\alpha^2+\vec\beta^2}\phi(\{\tilde k_i\}).
\end{align}
In Eq.~\eqref{eq:su6-br} $N^\alpha$ and $N^\beta$ are normalization factors and
 the Jacobi coordinates  are defined as
\begin{align}
\vec \alpha=\frac{1}{\sqrt{2}}(\vec k_2-\vec k_3),&\quad\quad
\vec \beta=\sqrt{\frac{3}{2}}\,\vec k_1.
\end{align}
As a result, writing explicitly the spin and isospin components of the wave function 
\eqref{eq:su6-br},  we find the following results in the canonical spin basis
\begin{equation}
\begin{array}{c|ccc}
\tilde\psi^{\uparrow;q_1q_2q_3}_{\sigma_1\sigma_2\sigma_3}&\,\,uud\,\,&\,\,udu\,\,&\,\,duu\,\,\\ &&&\\
\hline
\uparrow\uparrow\downarrow&\phi^\beta-\sqrt{3}\phi^\alpha&-2\phi^\beta&\phi^\beta+\sqrt{3}\phi^\alpha\\
\uparrow\downarrow\uparrow&-2\phi^\beta&\phi^\beta+\sqrt{3}\phi^\alpha&\phi^\beta-\sqrt{3}\phi^\alpha\\
\downarrow\uparrow\uparrow&\phi^\beta+\sqrt{3}\phi^\alpha&\phi^\beta-\sqrt{3}\phi^\alpha&-2\phi^\beta
\end{array}
\end{equation}
with the normalizations  $\int[\ud x]_3\,[\ud^2k_\perp]_3|\phi^\alpha|^2=\int[\ud x]_3\,[\ud^2k_\perp]_3|\phi^\beta|^2=
1/6$.

We note that the mixed-symmetry components of the nucleon wave function 
correspond to  a state  with $l_z=0$ and satisfy the relation~\eqref{rotinv3}
derived from rotational invariance.
Therefore, the LCWF overlap representation of the TMDs keeps the form
given in
Eqs.~\eqref{sphericalTMDs}, with the functions $A^q$ and $B^q$ calculated 
from the LCWF in Eq.~\eqref{eq:psi-total}.
In particular, separating the contribution from the $SU(6)$-symmetric component,
the $SU(6)$-breaking term and the interference term we can write
\begin{align}
\label{eq:a-br}
A^q&=\cos^2\delta\, A^q_{{\rm sym}}+2\cos\delta\sin\delta\, A^q_{{\rm int}}
+\sin^2\delta\, A^q_{{\rm br}},\\
\label{eq:b-br}
B^q&=\cos^2\delta\, B^q_{{\rm sym}}+2\cos\delta\sin\delta\, B^q_{{\rm int}}
+\sin^2\delta\, B^q_{{\rm br}}.
\end{align}
$A^q_{{\rm sym}}$ and $B^q_{{\rm sym}}$ are  the contributions 
calculated in Eq.~\eqref{eq:a-b}, while  the contribution from the $SU(6)$-breaking term reads
\begin{align}
A^u_{{\rm br}}&=\frac{A^d_{{\rm br}}}{2}=6[(\phi^\beta)^2+(\phi^\alpha)^2)],\\
B^u_{{\rm br}}&=4(\phi^\beta)^2,\quad B^d_{{\rm br}}=-(\phi^\beta)^2+3\,(\phi^\alpha)^2.
\end{align}
Furthermore, the contribution from the interference between the two components
is given by
\begin{equation}
A^u_{{\rm int}}=-A^d_{{\rm int}}=\frac{3}{5}B^u_{{\rm int}}=3B^d_{{\rm int}}=
6\sqrt{2}\, \phi\,\phi^\beta.
\end{equation}

We immediately notice that the admixture of mixed-symmetry components leads
to the breaking of the flavour-dependent relation~\eqref{rel4}
between the unpolarized and polarized TMDs, while it does not spoil the remaining relations among 
polarized TMDs.
Furthermore, the up and down quark contribution to the TMDs are no longer proportional.
\subsection{Results for TMDs}
\label{results-tmds}
In this section we report a few results for the TMDs in the LCCQM which can be explicitly calculated from the program of Ref.~\cite{mat1}.
In particular we discuss the spin densities in the transverse-momentum plane given by
\begin{eqnarray}
\rho(k_x,k_y,(\lambda,\vec{s}_\perp),(\Lambda,\vec{S}_\perp))
 &=& \frac{1}{2} \Bigg[ f_1^{\phantom{\perp}\!\!}
   + S_\perp^i \epsilon^{ij} k^j \frac{1}{M}\, f_{1T}^\perp
   + \lambda \Lambda\, g_{1L}^{\phantom{\perp\!\!}}
   + \lambda\, S_\perp^i k^i \frac{1}{M}\, g_{1T}^{\phantom{\perp\!\!}} 
\nonumber
\\
 && \hspace{0.7em}
   {}
   + s^i_\perp \epsilon^{ij} k^j \frac{1}{M}\, h_{1}^\perp
   +  \Lambda\, s^i_\perp k^i \frac{1}{M}\, h_{1L}^\perp
\nonumber \\
  \label{piet-distr2}
 && \hspace{0.7em}
   {}+ s^i_\perp S^i_\perp h_1^{\phantom{\perp\!\!}}
 + s^i_\perp (2 k^i k^j - \vec{k}^2_\perp \delta^{ij}) S^j_\perp 
       \frac{1}{2M^2}\, h_{1T}^\perp \Bigg] \, ,
\end{eqnarray}
where, for a generic TMD $j$, we introduced the first-$x$ moment defined as

\begin{equation}
   j(\uk_\perp^2)=\int\di{x}\,
   j(x,\uk_\perp^2).
\label{Eq:integratedTMD}\\
\end{equation}
As discussed in sect.~\ref{section-2}, the unpolarized TMD $f_1$, the helicity TMD $g_{1L}$, and the transversity TMD $h_1$ in Eq.~ (\ref{piet-distr2}) correspond to 
monopole distributions in the momentum space for unpolarized, longitudinally and transversely polarized quarks, respectively.
They can be obtained from the overlap of LCWFs which are diagonal in the 
orbital angular momentum, but probe different transverse momentum and helicity correlations of the quarks inside the nucleon.

\begin{figure}[ht]
	\centering
		\includegraphics[width=6cm]{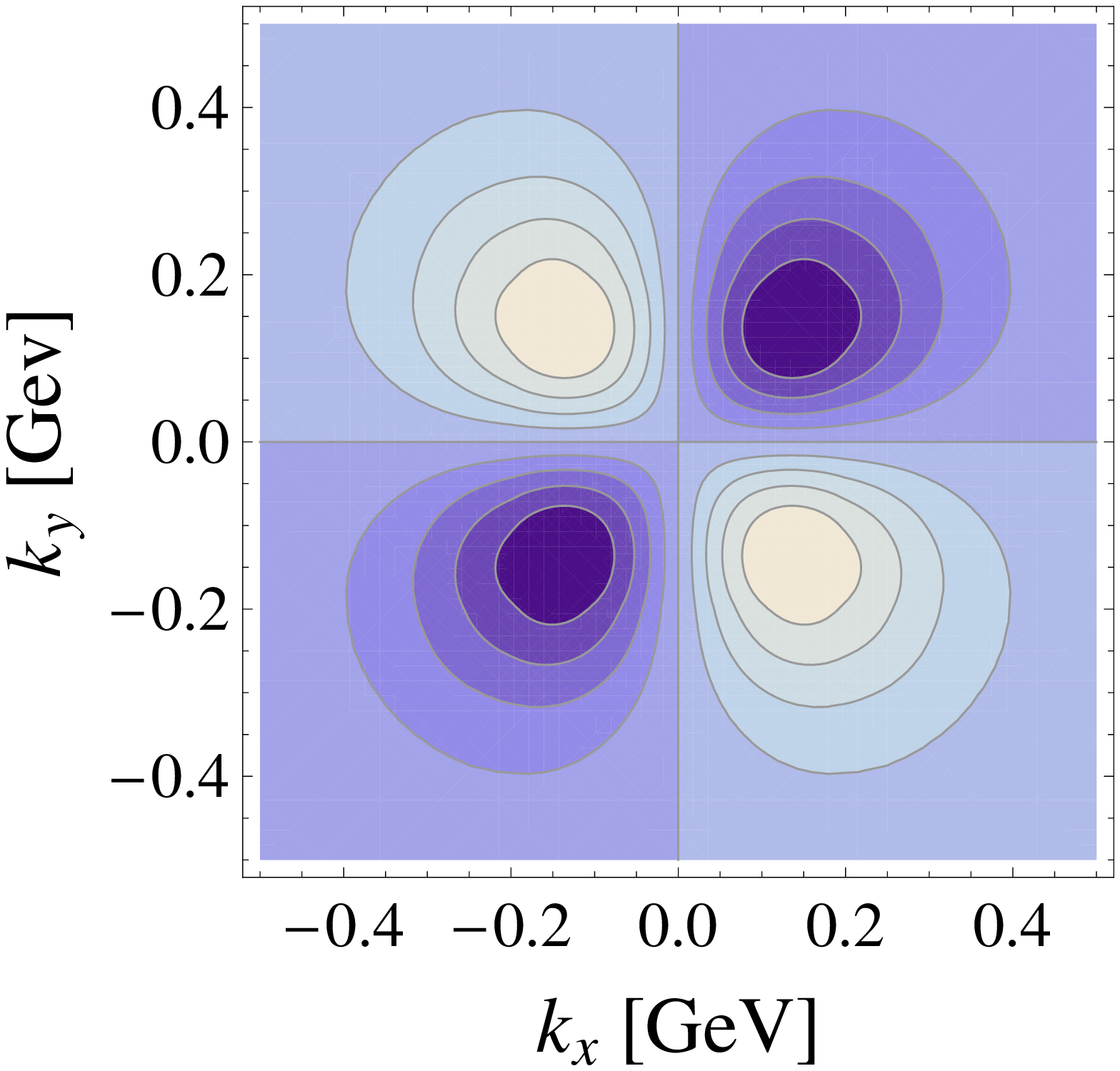}
\includegraphics[width=6cm]{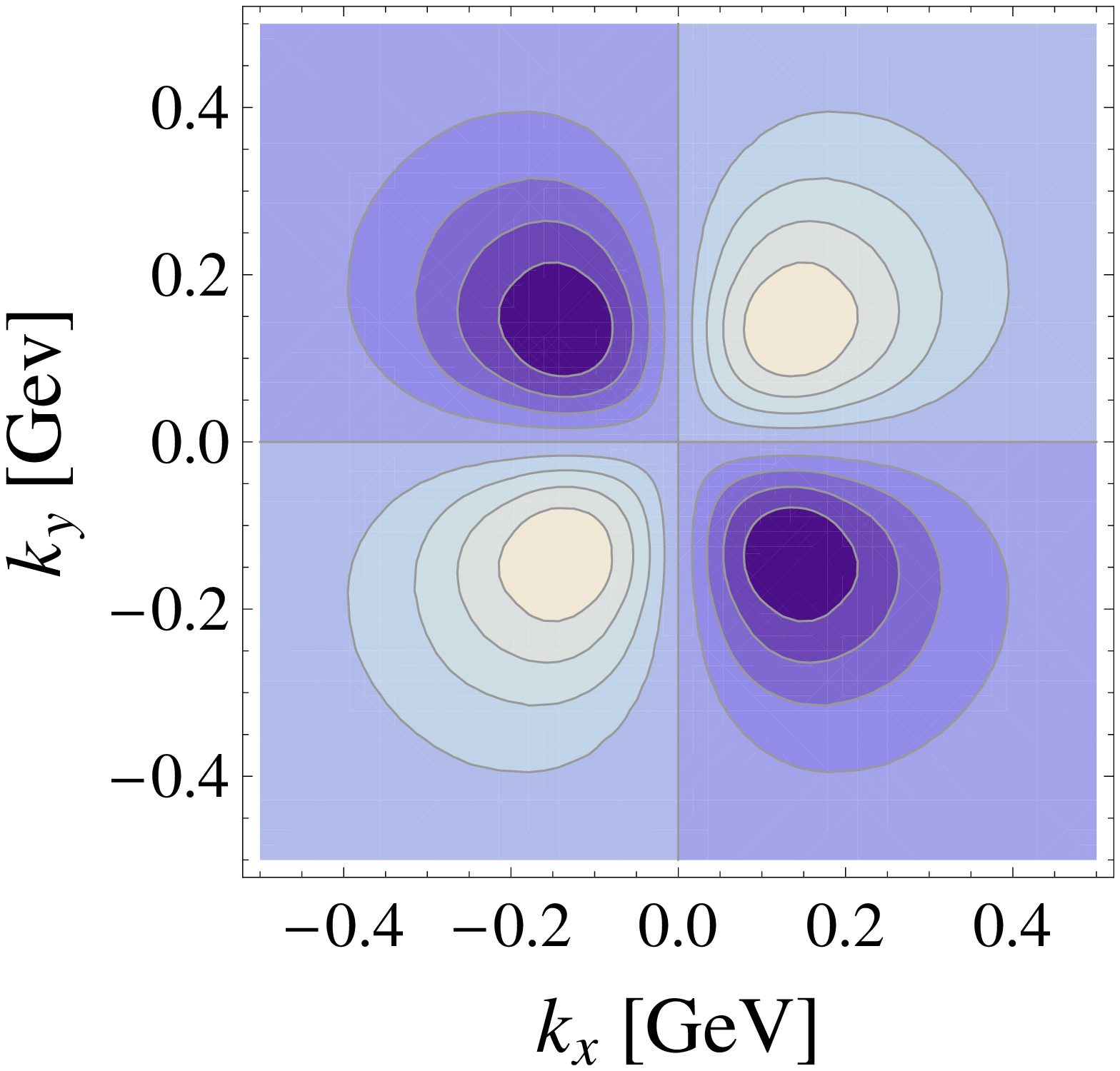}
\caption{Density of quarks in the $\uk_\perp$ plane
for net transverse polarization of quarks and proton in perpendicular 
directions.
The left and right  panel shows the results for up and down quarks, respectively.
\protect\label{fig2}}
\end{figure}
All the other TMDs require a transfer of orbital angular momentum
between the initial and final state.
The  $h_{1T}^\perp$ TMD describes the distortion
due to the 
transverse polarizations  in perpendicular directions
 of the quark and the nucleon~\cite{Miller:2007ae}. In this case, the nucleon helicity flips in the direction 
 opposite to the quark helicity, with a mismatch of two units 
for the orbital angular 
momentum of the initial and final LCWFs.
In Fig.~\ref{fig2} we show the results for both up and down quarks in the case of a $SU(6)$-symmetric LCWF.

The results for  definite quark and nucleon polarizations in transverse directions are obtained by adding the monopole
contributions  from $f_1$ and $h_1$ in Eq.~\eqref{piet-distr2}.
In the case of quark and nucleon transversely polarized in perpendicular directions ($\rho(k_x,k_y,s_x,S_y)=\rho(k_x,k_y,s_y,S_x)$),
one has from Eq.~\eqref{piet-distr2} the sum of the monopole contributions 
from $f_1$ and $h_1$, and of the quadrupole contribution from $h_{1T}^\perp$, ignoring the contributions from the T-odd TMDs which are vanishing in absence of gluon degrees of freedom.
Since the quadrupole distortion induced by $h_{1T}^\perp$ is quite small with respect to $f_1$ and $h_1$, the density for definite polarization differs slightly from a monopole distribution.
\begin{figure}[h]
\centerline{\hspace{-1 cm}\epsfig{file=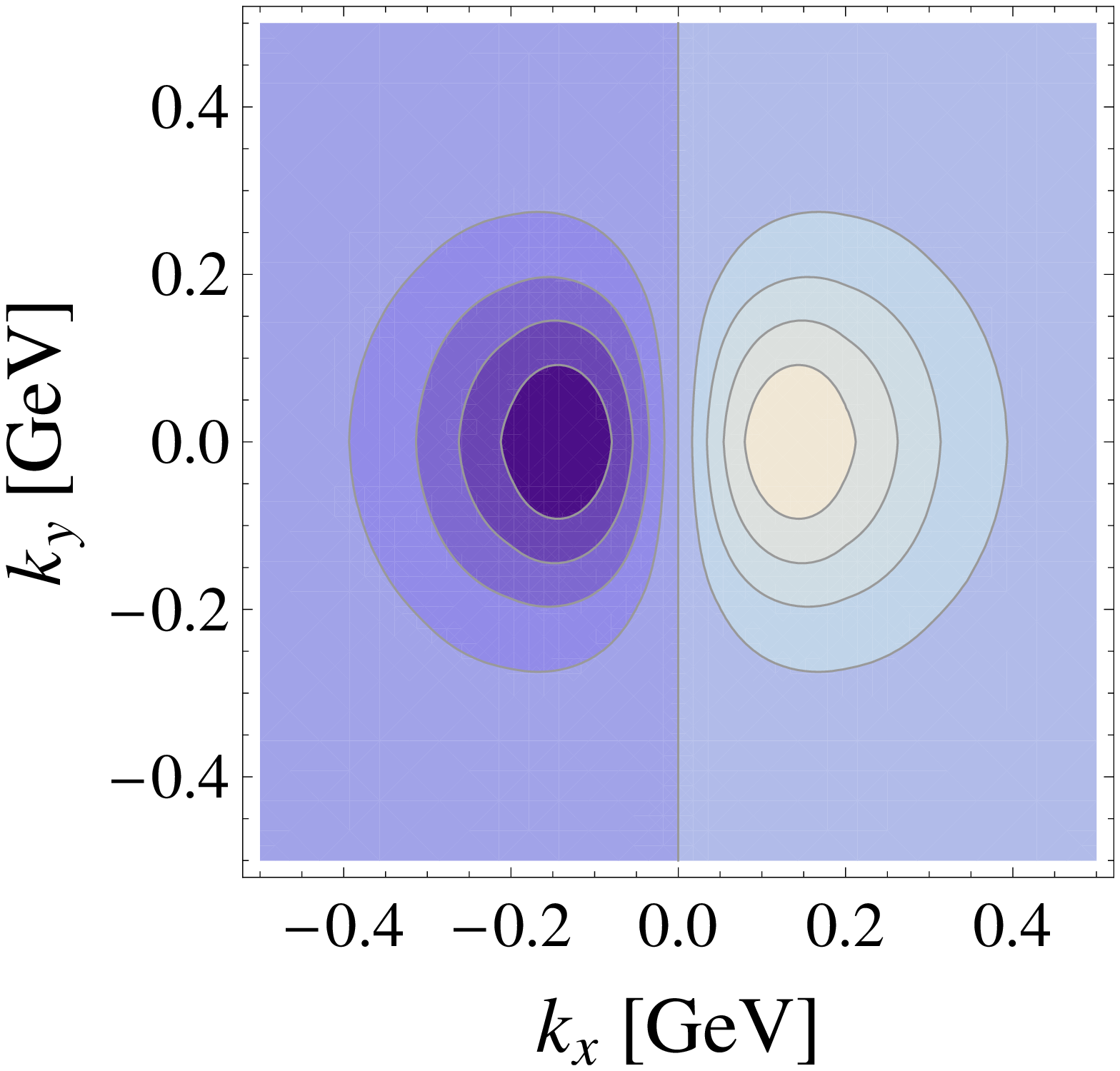, width=5.5 cm}
\epsfig{file=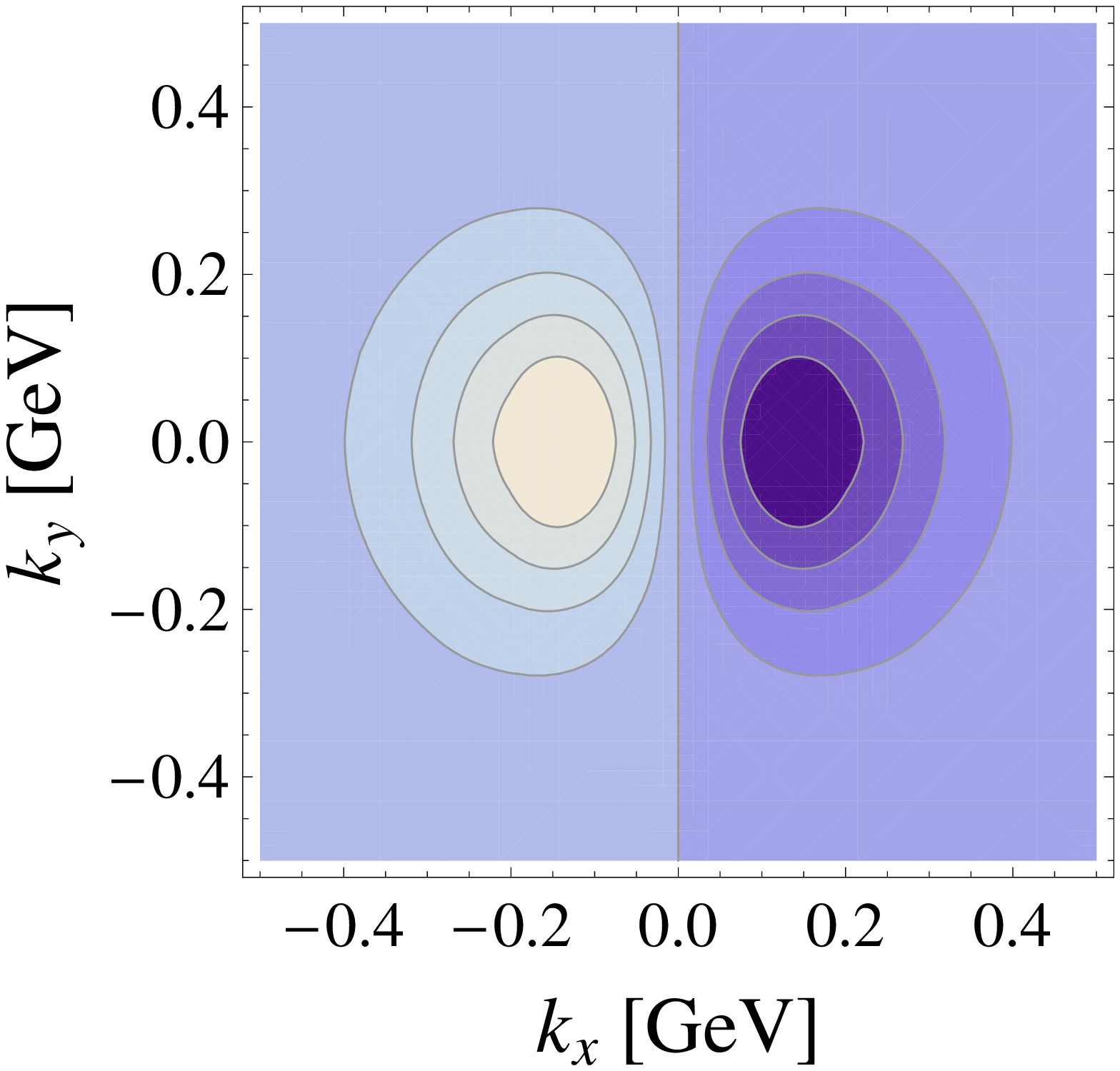, width=5.5 cm}}
\caption{Quark densities in the $\uk_\perp$ plane
for net longitudinal polarization of quarks in a transversely polarized 
proton for up (left panel ) and down (right panel) quark.
\protect\label{fig4}}
\end{figure}
\begin{figure}[hb]
\centerline{\hspace{-1 cm}\epsfig{file=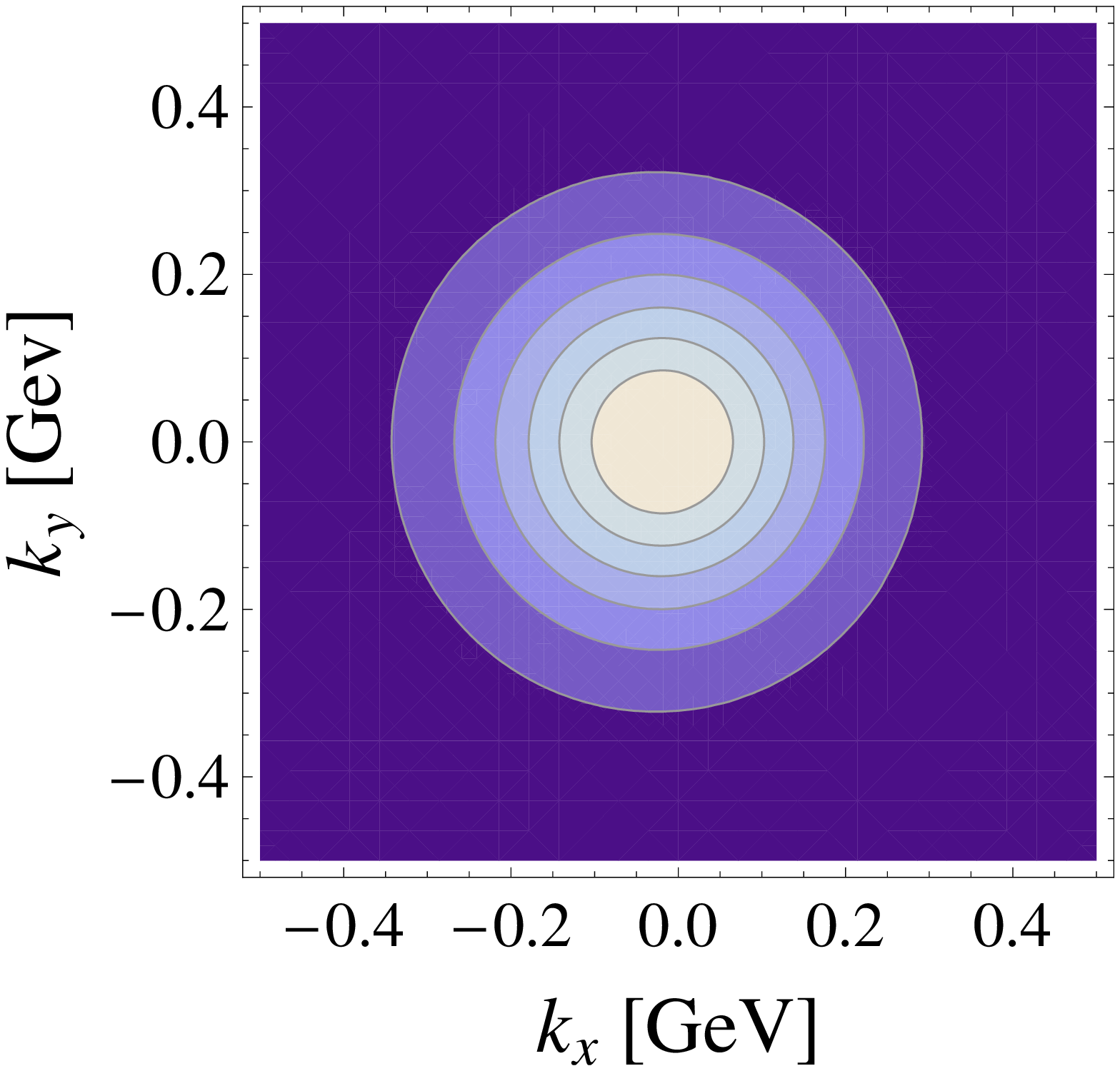, width=5.5 cm}
\epsfig{file=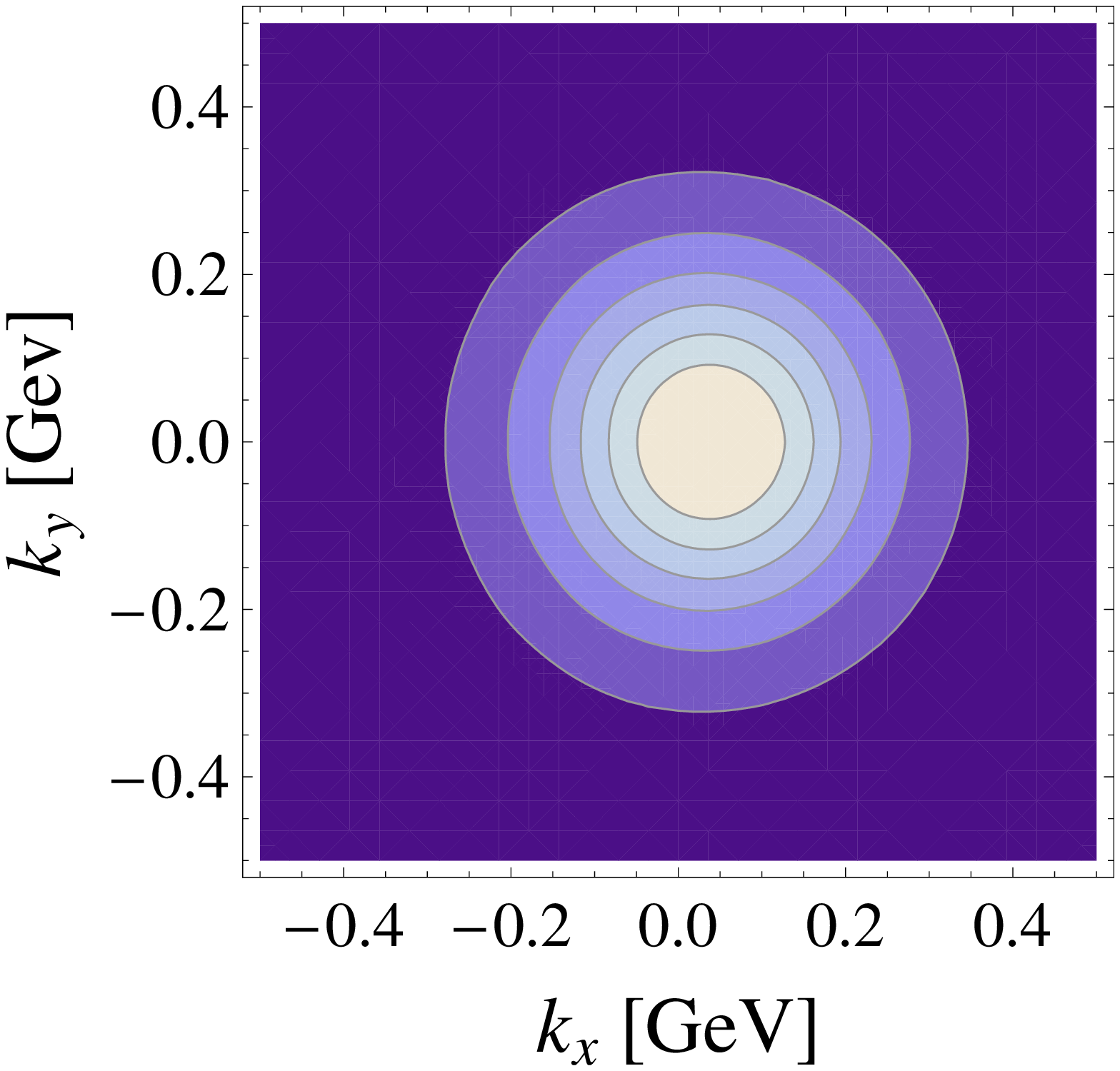, width=5.5 cm}}
\caption{Quark densities in the $\uk_\perp$ plane
for longitudinally polarized quarks in a transversely polarized 
proton for up (left panel ) and down (right panel) quark.
\protect\label{fig5}}
\end{figure}

Among the distributions in Eq.~ (\ref{piet-distr2}), the dipole 
correlations related to $g_{1T}$ and $h_{1L}^{\perp}$ have characteristic features
of intrinsic transverse momentum, since they are the only ones which have  no
analog in the spin densities
related to the GPDs in the impact parameter space~\cite{Diehl:2005jf,Pasquini:2007xz,Pasquini:2009bv}.
In particular,  $g_{1T}$  and $h_{1L}^\perp$ correspond to  quark densities
with specular configurations for the quark and nucleon spin:
$g_{1T}$ describes 
longitudinally polarized quarks in a transversely polarized nucleon, while
 $h_{1L}^\perp$ gives the distribution of transversely polarized quarks in longitudinally polarized 
nucleon.
Therefore, $g_{1T}$ requires helicity flip of the nucleon
which is not compensated by a change of the quark helicity, and vice-versa $h_{1L}^\perp$ involves helicity flip of the quarks 
but is diagonal in the nucleon helicity. 
As a result, in both cases, the LCWFs of the initial and final states
differ by one unit of orbital angular momentum and the associated 
distributions have a dipole structure.
The results in the $SU(6)$-symmetric version of the LCCQM
for the densities 
with longitudinally polarized quarks in a transversely polarized 
proton are shown in Fig.~\ref{fig4}.
The results for definite quark and nucleon polarizations, are obtained by adding the monopole contribution of $f_1$ (the contribution from the Sivers function is ignored because we do not include gauge-field degrees of freedom).
As shown in Fig.~\ref{fig5}, the sideways shift in the positive (negative) $\hat x$ direction for up (down) quark due to the dipole term $\propto \lambda S^ik^i\frac{1}{M}g_{1T}$
is sizable and corresponds to an average deformation
$\langle \vec{k}_x^u\rangle=55.8 $ MeV, and  
$\langle \vec{k}_x^d\rangle=-27.9 $ MeV. 
The dipole distortion  $\propto 
\Lambda\, s^i k^i \frac{1}{m}\, h_{1L}^\perp$ in the case of transversely polarized quarks in a longitudinally polarized proton is equal but with opposite sign, 
since in our model $h_{1L}^\perp=-g_{1T}$.
These model results are supported from a recent lattice 
calculation~\cite{Hagler:2009mb}-\cite{Musch:2009ku}
which gives, 
for the density related to $g_{1T}$,
$\langle \vec{k}_x^u\rangle=67(5) $ MeV, and  
$\langle \vec{k}_x^d\rangle=-30(5) $ MeV. For the density
related to  $h_{1L}^\perp$,  they also find shifts of similar magnitude but opposite sign:
$\langle \vec{k}_x^u\rangle=-60(5) $ MeV, and  
$\langle \vec{k}_x^d\rangle=15(5) $ MeV.

In the program of Ref.~\cite{mat1} one can also calculate and simultaneously visualize
the effects of $SU(6)$ breaking for the spin densities, and for the $x$ and
$\uk_\perp^2$ dependence of the TMDs.
For example, we summarize a few results which can be easily verified following the calculation in  the program:
first of all, one finds that the $\uk_\perp^2$ dependence of the TMDs is not Gaussian.
In particular, the $\uk^2_\perp$ dependence is the same for up and down quark 
within the $SU(6)$-symmetric version of the model, while
the admixture of mixed-symmetry components breaks this flavour independence.

\subsection{Observables}
\label{sec:observables}

One of the observables which is largely sensitive to effects of $SU(6)$ breaking is
the inclusive double spin asymmetry $A_1$ in DIS
with lepton and target nucleon longitudinally polarized.
It can be  expressed in terms of the unpolarized $f_1^q(x)$ and polarized $g_1^q(x)$ parton distributions as
\begin{equation}
\label{eq:a1}
 A_1=\frac{\sum_q e^2_q \,x \,g_1^q(x)}{\sum_q e^2_q \,x \,f_1^q(x)},
\end{equation}
which becomes, in the case  of proton and neutron target,
\begin{equation}
 A_1^p=\frac{4 g_1^u(x)+g_1^d(x)}{4 f_1^u(x)+f_1^d(x)},\quad
 A_1^n=\frac{4 g_1^d(x)+g_1^u(x)}{4 f_1^d(x)+f_1^u(x)},
\end{equation}
where we used isospin symmetry in such a way that the structure functions of the neutron are related to those of the proton by interchanging $u$ and $d$ 
flavours. 

\begin{figure}[ht]
\centerline{\epsfig{file=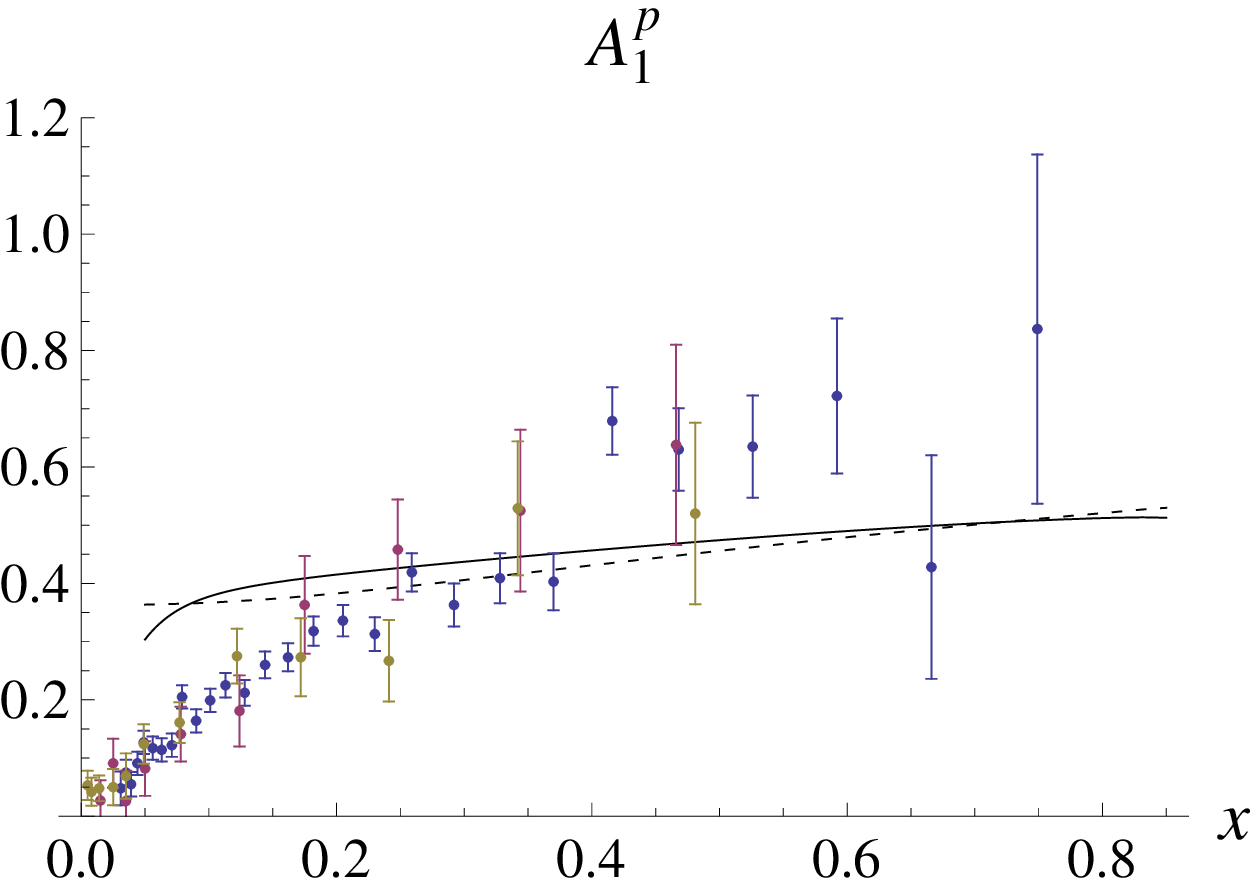, width=0.5\textwidth}
\epsfig{file=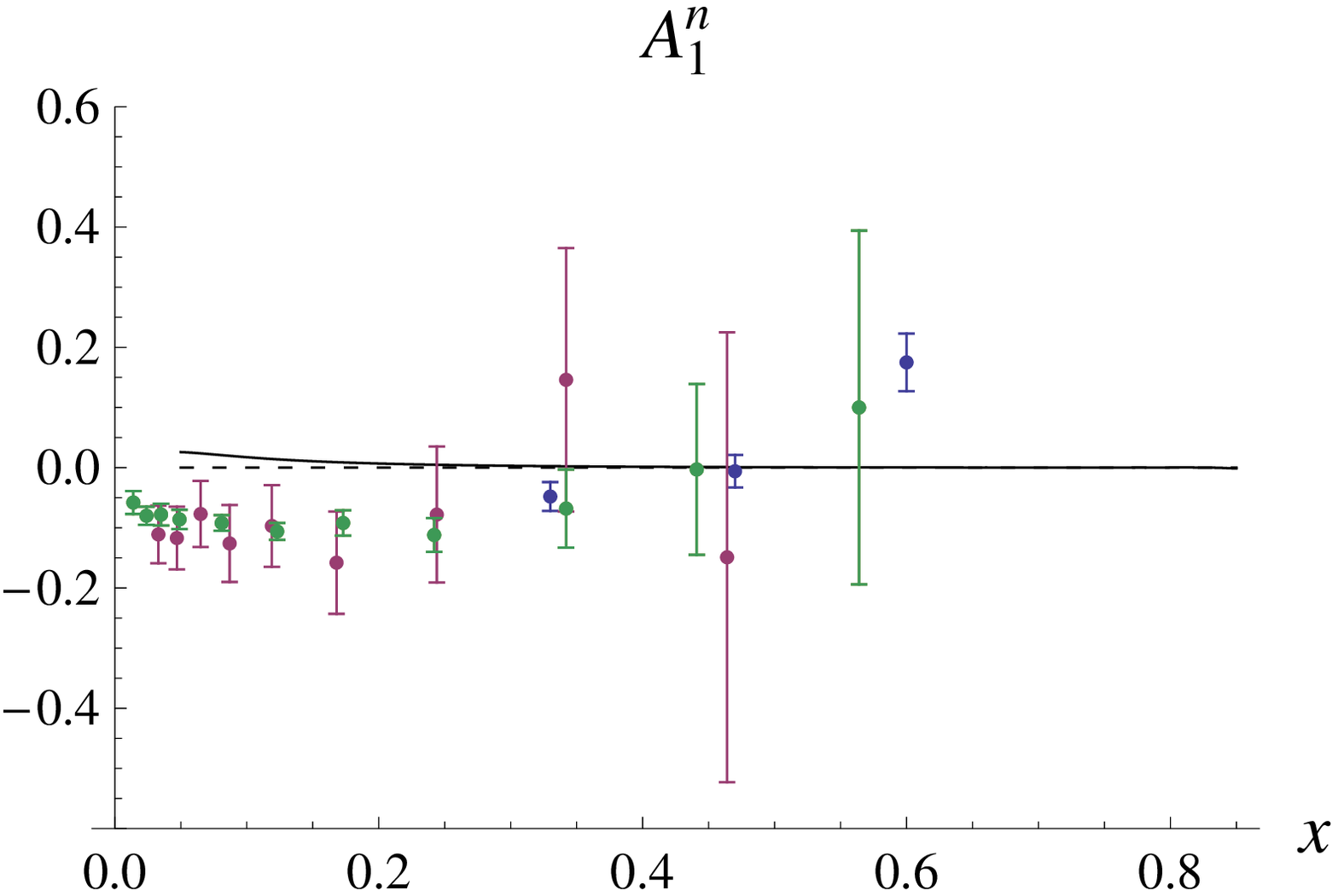, width=0.5\textwidth}}
\caption{The inclusive double spin asymmetry $A_1$ in DIS off proton 
(left panel) and neutron (right panel) as function of $x$.
The theoretical curves are obtained with $g_1^q(x)$ and $f_1^q(x)$ from the 
LCCQM with $SU(6)$ symmetry as follows: both function LO to the $\langle Q^2\rangle=2.5$ GeV$^2$ of the experiments (solid curves) and both at the low scale of the model (dashed curves). The experimental data for $A_1^p$ are from Refs.~\cite{Abe:1998wq}-\cite{hep-ex/9807015} 
and for $A_1^n$ from Refs.~\cite{nucl-ex/0308011}-\cite{hep-ph/9705344}
\protect\label{fig6}}
\end{figure}

We notice that in a $SU(6)$-symmetric model, one has $g_1^u(x)=-4g_1^d(x)$, which implies $A_1^n=0$ at the scale of the model. At higher scale, $A_1^n\ne 0$ due to evolution, but the effects remain small.
This is illustrated in Fig.~\ref{fig6} where we show the results
for the proton and neutron at the initial scale of the model (dashed curves) and after leading order (LO) evolution to $Q^2=2.5$ GeV$^2$.
We also see that experimentally $A_1^n$ (extracted by subtracting deuteron and proton data, or from $^3$He data, modulo nuclear corrections) is found clearly non zero, giving  a signature of $SU(6)$-breaking effects.
Following the calculation in the Mathematica program of Ref.~\cite{mat2}, we can use this sensitivity  to $SU(6)$ breaking to fit the parameter $\delta$ of the model in Eqs.~\eqref{eq:a-br}-\eqref{eq:b-br} to the experimental data for $A_1$.
In~\cite{mat2} we give different examples for the strategies which can be used in the fitting procedure. Here we discuss  only one of them which seems the most 
suitable for the LCCQM.
In order to decide the criteria of the fit, we should answer the following questions.
First, in which $x$-range and with what accuracy is the model 
applicable?
Second, how stable are the results under evolution? 
A related key question emerging not only here but in any nonperturbative 
calculation concerns the scale at which the model results for the parton 
distributions hold. From the point of view of
QCD where both quark and gluon degrees of freedom contribute, the role of the 
low-energy quark models is to provide initial conditions for the QCD evolution 
equations. Therefore, we assume 
the existence of a low scale $Q_0^2$ where glue and sea quark contributions 
are suppressed, and the dynamics inside the nucleon is described in terms 
of three valence (constituent) quarks confined by an effective long-range 
interaction.
In fact, glue and sea quark degrees of freedom might be thought of at this
low scale to be contained in the structure of the constituent quarks,
which are massive objects.
The actual value of $Q_0^2$ is fixed evolving back unpolarized data, 
until the  valence distribution matches the condition that the second moment
$\langle x(Q^2_0)\rangle_{{\rm val}}$,
\emph{i.e.}  the momentum fraction carried by the valence quarks, is equal to 
one~\cite{Traini:1997jz}.
Following ~\cite{arXiv:1103.5977}, we start with the LO-value of $\langle x(Q^2)\rangle_{{\rm val}}=0.35$ at 10 GeV$^2$~\cite{Martin:2009iq}   
and using LO evolution equations we find the matching condition
$\langle x(Q^2_0)\rangle_{{\rm val}}=1$,
at $Q_0=420$ MeV.
Although there is no rigorous relation between the QCD quarks and the 
constituent quarks, which requires a more fundamental description of the transition from 
soft to hard regimes, this strategy reflects the present 
state of the art for quark model 
calculations~\cite{Scopetta:1997wk}-\cite{Pasquini:2004gc},
and has been validated with a fair comparison to 
experiments~\cite{Lorce:2011dv,Boffi:2009sh,Scopetta:1997wk,Traini:1997jz}.

As can be seen in Fig.~\ref{fig6}, the description of the experimental data
is reasonable. For $x\gtrsim 0.15$ the model describes the
$A_1$ data within an accuracy of about $30\,\%$.
The description improves in the valence-$x$ region of 
$x\gtrsim 0.25$.
Since the model contains no antiquark- and gluon-degrees of freedom,
it is not surprising to observe that it does not work at small $x$.
In conclusion, it can be said that the results are weakly 
scale dependent, and the model well catches the main features of the 
observable $A_1$ in its range of applicability, namely in the valence-$x$ 
region.
Therefore, we choose to fit the parameter describing the admixture of $SU(6)$-breaking terms in the nucleon wave function using the LO results for $A_1$ of proton and neutron evolved at the average $\langle Q^2\rangle=2.5$ GeV$^2$ of the available experimental data, and restricting ourselves to  the valence region $x\ge 0.3$.\\
The results of the fit are: $\delta=7.25$ degrees with $\chi^2=52.41$ for a total of $N=28$ experimental points ($\chi^2_{{\rm red}}=\chi^2/(N-1)=1.94$).
The fitted value of $\delta$ corresponds to a percentage of $98.4\%$ ($1.6\%$) for the $SU(6)$-symmetric (-breaking) component of the nucleon wave function.
\begin{figure}[t]
\centerline{\epsfig{file=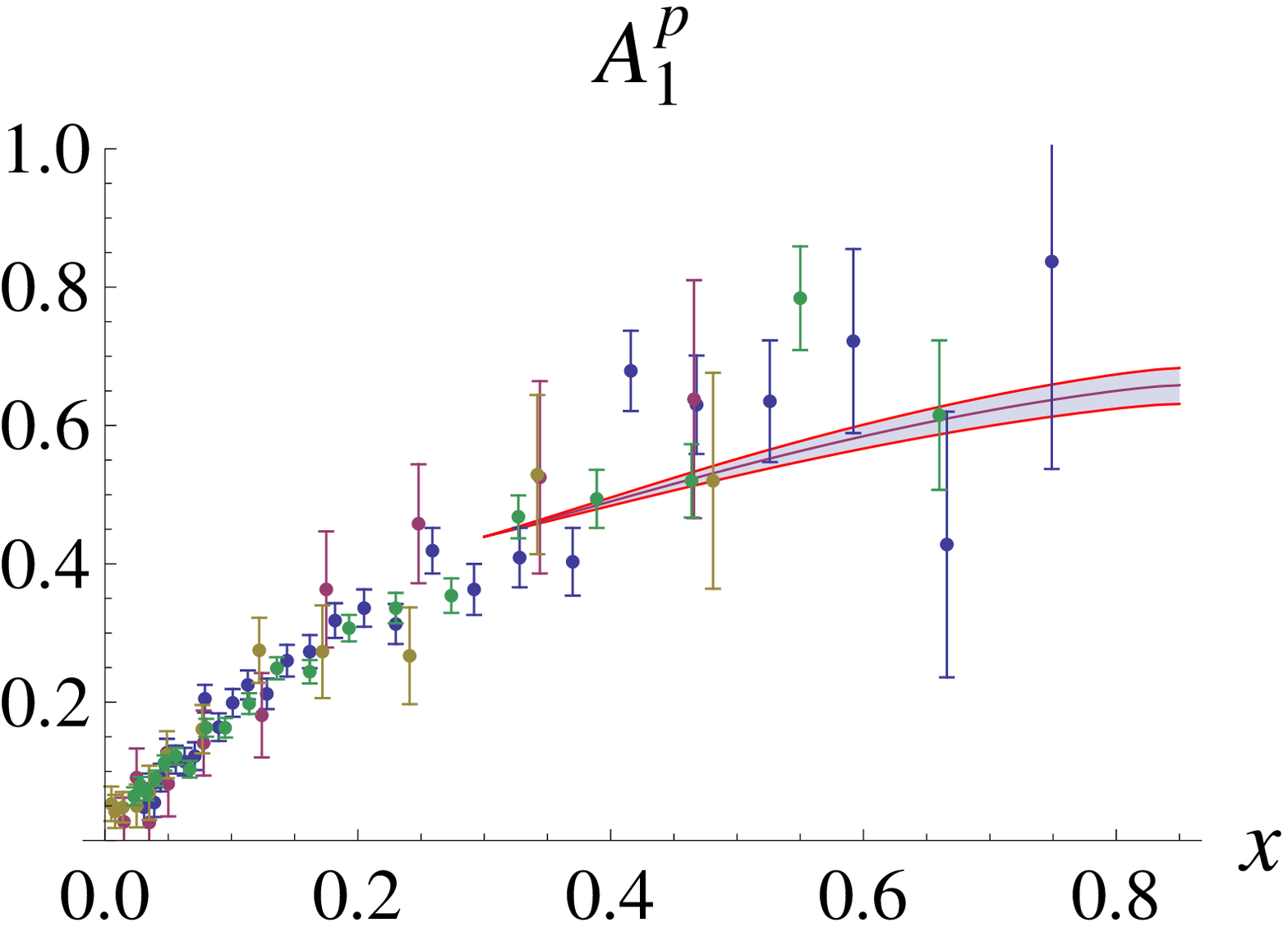, width=0.5\textwidth}
\epsfig{file=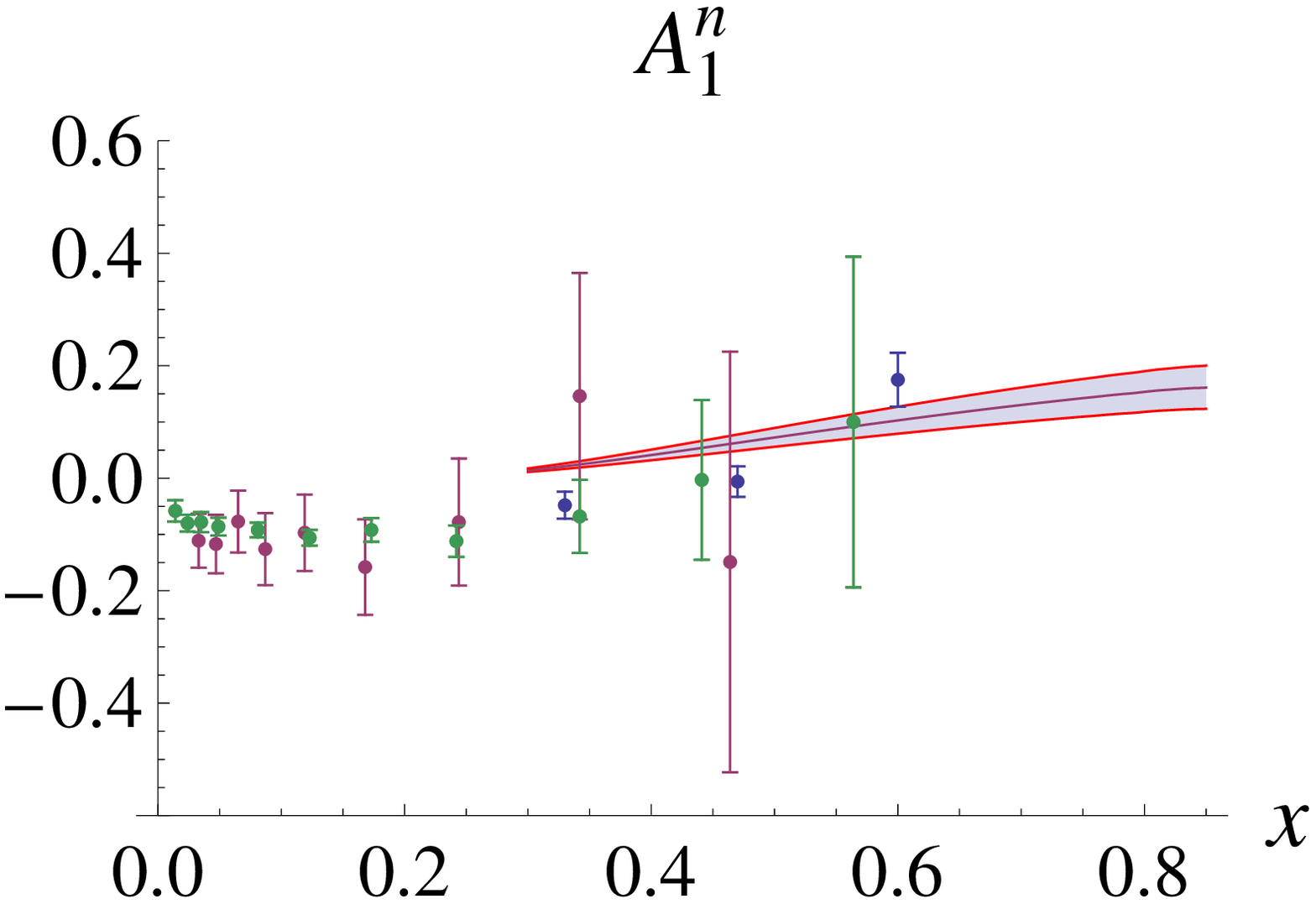, width=0.5\textwidth}}
\caption{The inclusive double spin asymmetry $A_1$ in DIS off proton 
(left panel) and neutron (right panel) as function of $x$.
The theoretical curves are obtained from LO evolution to $Q^2=2.5$ GeV$^2$ 
of the PDFs calculated in the
LCCQM with admixture of $SU(6)$ components in the nucleon wave function.
The parameter describing the percentage of the $SU(6)$-symmetric and $SU(6)$-breaking components is obtained from the fit to the experimental data, with the result $\delta=7.25$.  The experimental data for $A_1^p$ are from Refs.~\cite{Abe:1998wq}-\cite{hep-ex/9807015} 
and for $A_1^n$ from Refs.~\cite{nucl-ex/0308011}-\cite{hep-ph/9705344}
\protect\label{fig7}}
\end{figure}
The results of the fit are shown in Fig.~\ref{fig7}, with the error band 
corresponding to the $1\sigma$ region. 
We see that with a small percentage of $SU(6)$ breaking we are able to give an overall good description of the asymmetry for both proton and neutron target in the valence region.

The results of the fit can be also tested with other observables.
In particular we consider the double spin asymmetrsy $A_{LL}^p$ which can be accessed in SIDIS of hadrons off proton target. For a more extensive discussion about the calculation within the LCCQM of the azimuthal asymmetries in SIDIS related to T-even TMDs we refer to~\cite{Boffi:2009sh}, while for comprehensive reviews of recent and planned experiments to access the TMDs we refer to~\cite{Boer:2011fh,Barone:2010zz,Anselmino:2011ay}.
Assuming the Gaussian Ansatz for the $\uk^2_\perp$ dependence of the $f_1$ and $g_1$ TMDs, this asymmetry can be written as 
\begin{equation}\label{eq:ALL}
       A_{LL} = \frac{F_{LL}}{F_{UU}} 
       = \frac{\sum_a e_a^2 \,x\,g_1^a(x)\,D_1^a(z)}
              {\sum_a e_a^2 \,x\,f_1^a(x)\,D_1^a(z)}\;,
\end{equation}
which reduces to $A_1$ in Eq.~\eqref{eq:a1} if no hadrons are observed in the final state.
In Eq.~\eqref{eq:ALL}, $D_1(z)$ is the unpolarized fragmentation function describing the hadronization process 
of the struck quark decaying into the detected hadrons, with $z$ the energy fraction taken out by the detected hadron.
In the calculation we used for $D_1(z)$ the LO parametrization
 \cite{Kretzer:2000yf} at $Q^2=2.5$ GeV$^2$.
\begin{figure}[hb]
\centerline{\epsfig{file=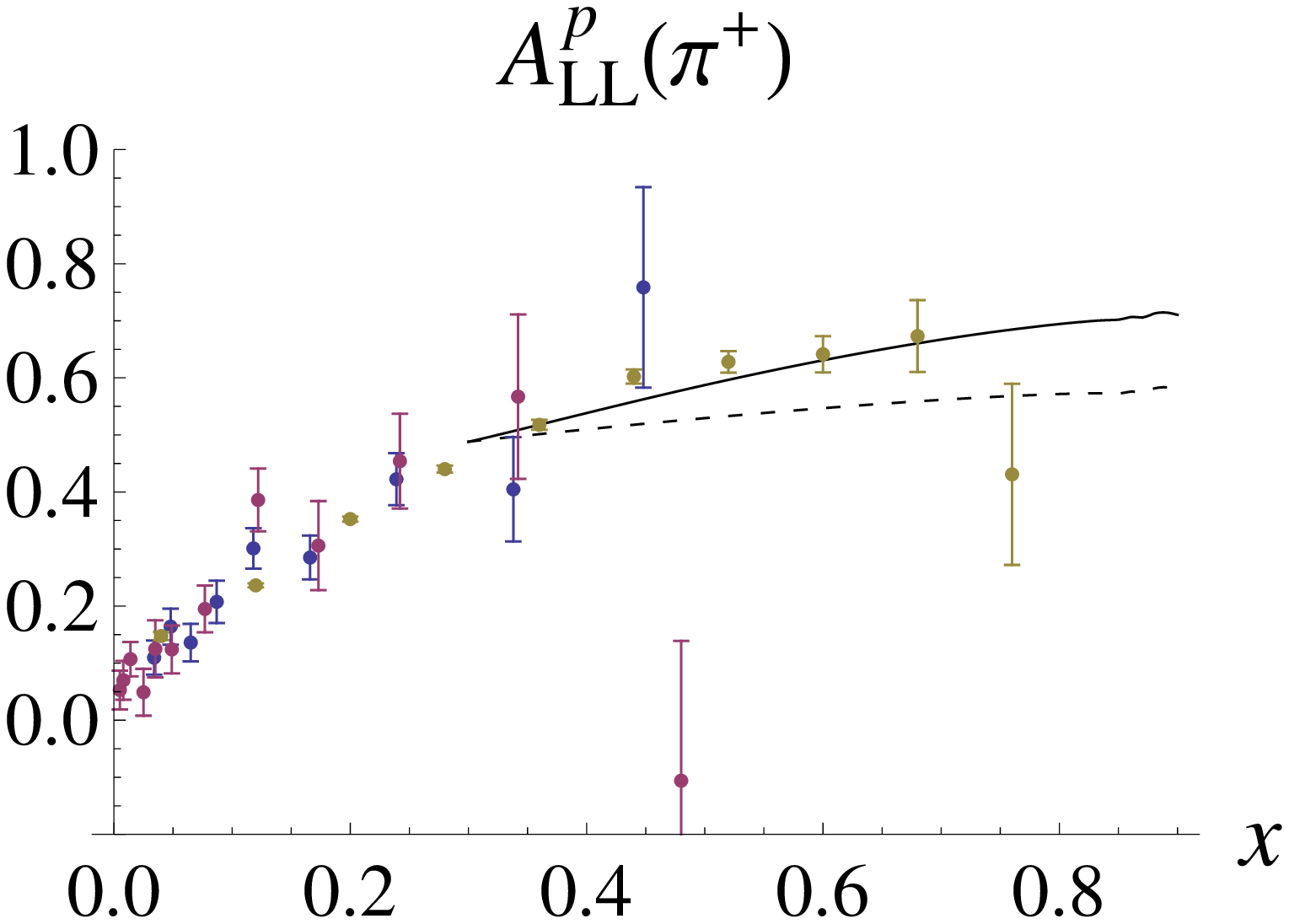, width=0.34\textwidth}
\epsfig{file=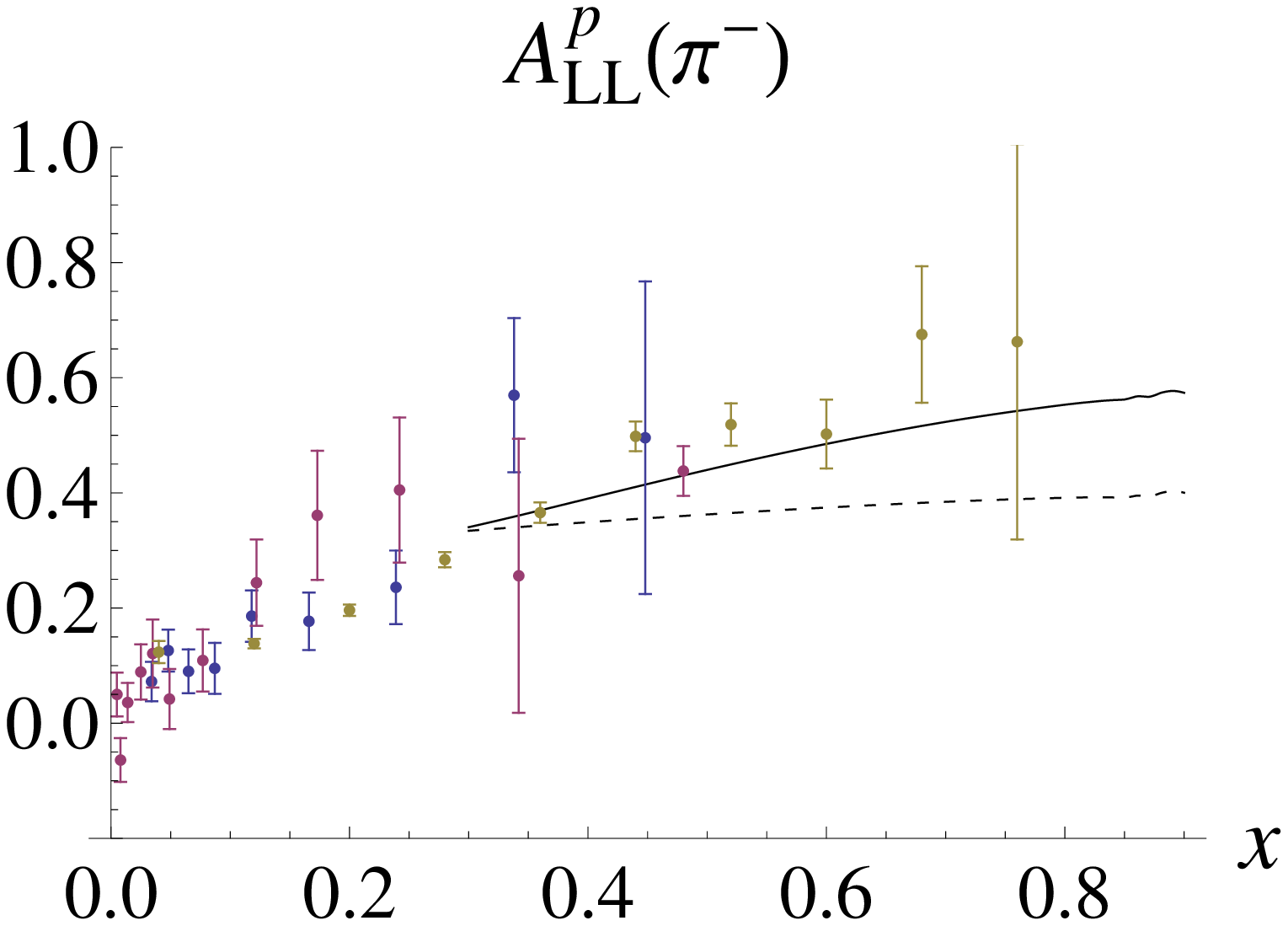, width=0.34\textwidth}
\epsfig{file=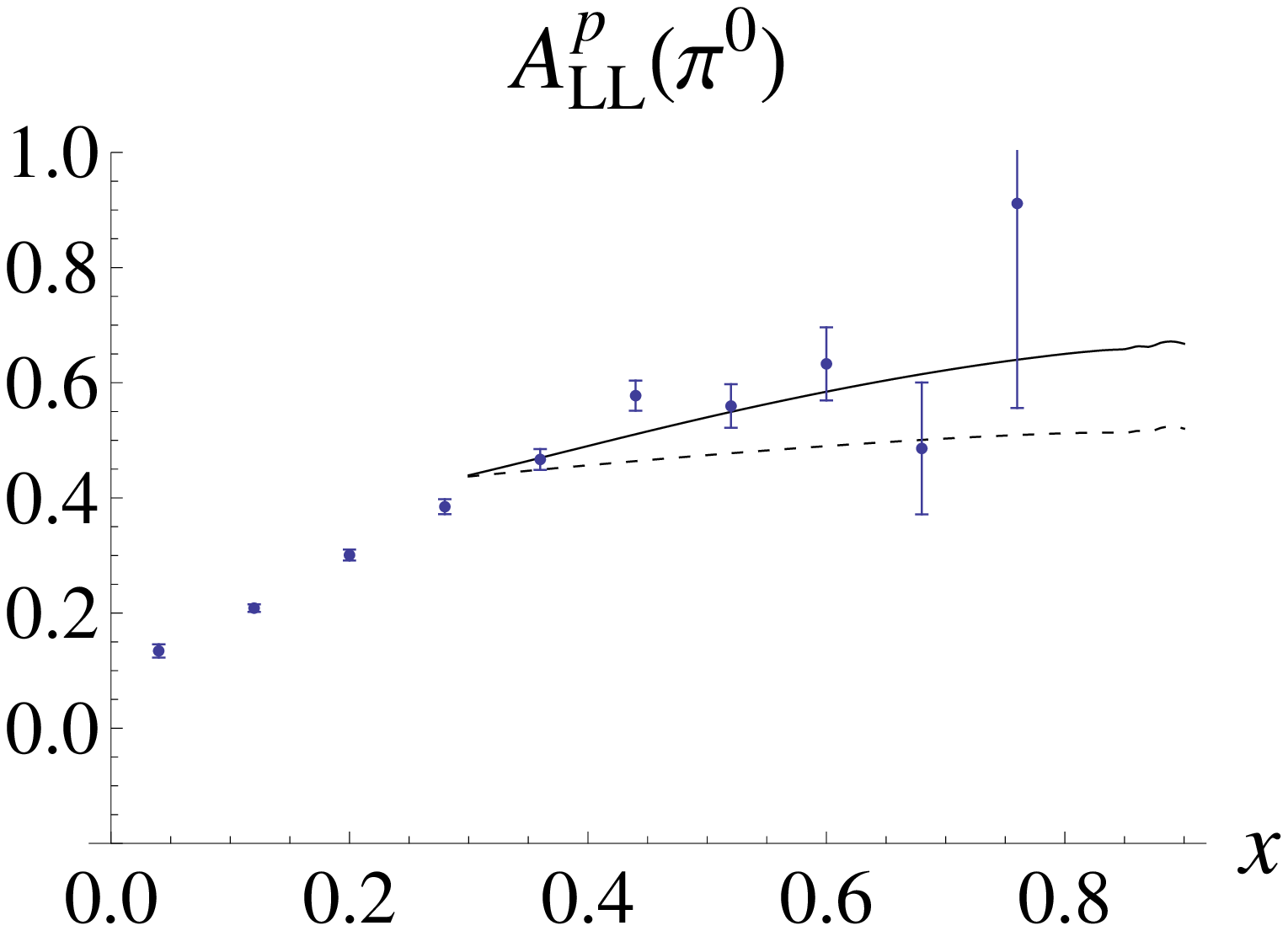, width=0.34\textwidth}
}
\caption{The semi-inclusive double spin asymmetry $A_{LL}$ in DIS of pions 
off a proton target as function of $x$.
The dashed curves correspond to the LCCQM results in the $SU(6)$-symmetric
version, while the solid curves include the contribution from $SU(6)$-breaking terms, with a percentage determined from the fit of the model to the $A_1$ (see text). For the fragmentation function $D_1$ we used the parametrization at $Q^2=2.5$ GeV$^2$ from Ref.~\cite{Kretzer:2000yf}.
The data are from Refs.~\cite{Adeva:1997qz,Airapetian:2004zf}.
\protect\label{fig8}}
\end{figure}
In Fig.~\ref{fig8}, we show the results for $A_{LL}$  in DIS production of pions off proton target.
In particular,  the theoretical curves are obtained with the model
results of $g_1(x)$ and $f_1(x)$ evolved at LO to $Q^2=2.5$ GeV$^2$.
The dashed curves show the results from the $SU(6)$-symmetric version of the LCCQM, and the solid curves correspond to the LCCQM predictions with a percentage 
of $SU(6)$-breaking terms as obtained from the fit of $A_1$.
In both cases,
we show the predictions only for $x\ge 0.3$, corresponding to the range where 
we performed the fit for $A_1$.
We clearly see that the inclusion of $SU(6)$-breaking terms 
 leads to a quite good agreement with the data in the valence region.

\begin{figure}[t]
\centerline{\epsfig{file=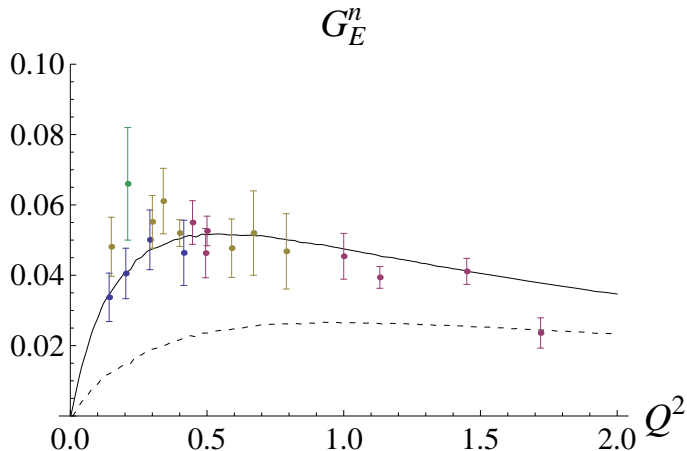,width=9 cm}
}
\caption{The electric form factor of the nucleon.
The dashed curve shows the prediction within the $SU(6)$-symmetric version of the LCCQM. The solid curve is the LCCQM results with the inclusion of
$1.6\%$ of $SU(6)$-breaking component in the nucleon wave function. The references to the data can be found in \cite{Perdrisat:2006hj}.
\protect\label{fig9}}
\end{figure}
An other observable which  is particularly sensitive to $SU(6)$ breaking is the electric neutron form factor $G_E^n$.
It is well known that in the non-relativistic $SU(6)$ constituent quark model
$G_E^n$ is  identically equal to zero.
The inclusion of relativistic effects, like in the LCCQM, produces non-zero results. However, the predictions remain small  and do not reproduce the behaviour
of the experimental data, as shown by the dashed curve in Fig.~\ref{fig9}.
Corrections can come from the meson-cloud of the nucleon~\cite{Miller:2002ig,Rinehimer:2009sz,Pasquini:2007iz}.
The meson-cloud contribution was recently calculated within the  LCCQM in Ref.~\cite{Pasquini:2007iz} and it was found quite smooth, significant only at $Q^2\le 0.5$ GeV$^2$, but not able to reproduce the trend of the data.
Only the inclusion of $SU(6)$-breaking terms can give a substantial improvement.
This is shown by the solid curve in  Fig.~\ref{fig9}, obtained within the LCCQM with the percentage of $SU(6)$-breaking terms which was obtained from the fit  to the asymmetry $A_1$. 
Remarkably, the good agreement with the experimental data provides an important 
consistency check for our estimate of the $SU(6)$-breaking contribution to the nucleon wave function.

\section{Conclusions}

In this work we presented a study of the transverse-momentum dependent parton distributions in the framework of quark models. 
In particular 
we discussed several quark models, introducing the different formalisms for the practical calculation of the TMDs and identifying the common building blocks besides the specific assumptions for modeling the quark dynamics.
We sorted these models in different classes: the light-cone models, the covariant parton model, the mean-field models and the spectator models.
Most of these models predict relations among the leading-twist T-even TMDs.
In particular, there are in total four independent relations among the leading-twist T-even TMDs: three of them are flavour independent and connect polarized TMDs, while a fourth flavour-dependent relation involves both polarized and unpolarized TMDs.
Since in QCD the eight TMDs are all independent, it is clear that such relations should be traced back to some common simplifying assumptions in the models. First of all, it was noticed that they break down in models with gauge-field degrees of freedom. Furthermore, most quark models are valid at some very low scale and these relations are expected to break under QCD evolution to higher scales. Despite these limitations, such relations are intriguing because they can provide guidelines for building parametrizations of TMDs to be tested with experimental data and can also give useful insights for the understanding of the origin of the different spin-orbit correlations of quarks in the nucleon.
We have shown that these model relations have essentially a geometrical origin, and can be traced back to properties of rotational invariance of the system. In particular, we identified the conditions which are sufficient for the existence of the flavour-independent relations. They are:
\begin{enumerate}
\item the probed quark behaves as if it does not interact directly with the other partons (\emph{i.e.} one works within the standard impulse approximation) and there are no explicit gluons;
\item the quark light-cone and canonical polarizations are related by a rotation with axis orthogonal to both the light-cone and quark transverse-momentum directions;
\item the target has spherical symmetry in the canonical-spin basis.
\end{enumerate}
For the flavour-dependent relation, one needs a further condition for the spin-flavour dependent part of the nucleon wave function. Specifically, it is required 
\begin{enumerate}[resume]
\item $SU(6)$ spin-flavour symmetry of the wave function.
\end{enumerate}
On the basis of the above assumptions, we were able to derive the model relations among TMDs within two different approaches.

The first approach is based on the representation of the quark correlator entering the definition of TMDs in terms of the polarization amplitudes of the quarks and nucleon. Such amplitudes are usually expressed in the basis of light-cone helicity. However, in order to discuss in a simple way the rotational properties of the system, we introduced the representation in the basis of canonical spin. In this framework, we showed that the conditions 1-3 are sufficient for the existence of all three flavour-independent relations. We also showed that a subset of these three relations can be derived relaxing the assumption of spherical symmetry and using the less restrictive condition of axial symmetry under a rotation around a specific direction.

The second approach is based on the representation of TMDs in terms of quark wave functions. In particular, we expressed the TMDs as overlap of light-cone wave functions, and we derived the relation with the corresponding representation in terms of overlap of wave functions in the canonical-spin basis. After discussing the consequence of spherical symmetry on the spin structure of the wave function, we were able to obtain an alternative derivation of the relations among polarized T-even TMDs. Finally, for the remaining relation among polarized and unpolarized T-even TMDs, we used the $SU(6)$ symmetry for the spin-isospin dependence of the nucleon wave function.

On the basis of this study, we  have shown how and to which extent the conditions 1-4 are realized in the different classes of quark models. In particular we verified that all the models satisfying the TMD relations also satisfy the above conditions, while models where the TMD relations do not hold fail with at least one of the above conditions.

As specific example we discussed in details 
a light-cone constituent quark model, giving the necessary ingredients to perform the calculation of the T-even TMDs.
The calculation can be reproduced numerically using two Mathematica programs which are available on-line.
In particular, the first program allows one to reproduce the model calculation for {\em i)} the T-even TMDs;
{\em ii)} the spin densities in the transverse-momentum space as function of the quark and nucleon polarizations; {\em iii)} the  $\uk^2_\perp$ dependence of the $x$ moments of the TMDs, with a comparison
of the model results with the Gaussian Ansatz.
The calculation can be reproduced using  the $SU(6)$-symmetric version of the light-cone constituent quark model as well as introducing the effects of $SU(6)$-symmetry breaking in the light-cone wave function which lead to the violation of the relation between unpolarized and polarized TMDs.
The second Mathematica program 
allows one to obtain predictions for different observables, which are particularly sensitive to $SU(6)$ breaking
and can be used to fix the free parameter of the model 
corresponding to the percentage of $SU(6)$-breaking terms in the LCWF of the nucleon.
In particular, in this program we present different strategies
for fitting the available experimental data
 of three  observables: {\em i)} the inclusive polarized asymmetry $A_1$; 
{\em ii)} the ratio $F_2^n/F_2^p$ of the neutron to proton unpolarized structure function of DIS;
{\em iii)} the nucleon electroweak form factors.
In these notes we discussed, as an example, the fit  of the double spin asymmetry $A_1$ in DIS. 
Using the results from this fit, we also discussed predictions for the 
semi-inclusive double spin asymmetry $A_{LL}$ in DIS of pions off proton target
and for the electric neutron form factor. 
The good agreement of these observables with the experimental data provides an important 
consistency check for our estimate of the $SU(6)$-breaking contribution to the nucleon wave function.

This exercise provides a good example of 
the practical value of models in phenomenological studies, showing how
model parameters related to particular assumptions on the quark dynamics 
can be tuned to describe available experimental data, and then 
 can be used to learn new information on the partonic structure of the nucleon.

\acknowledgments
We are thankful to P. Schweitzer and H. Avakian for invaluable suggestions 
in the preparation of these lectures and for stimulating discussions.
B.P. would like to express her gratitude to the organizers and participants 
of the Course ``Three-dimensional  Partonic Structure of the Nucleon'' for creating a very lively and friendly atmosphere.
This work was supported in part by the European Community Joint Research Activity ``Study of Strongly Interacting Matter'' (acronym HadronPhysics3, Grant Agreement n. 283286) under the Seventh Framework Programme of the European Community, and by the Italian MIUR through the PRIN 2008EKLACK ``Structure of the nucleon: transverse momentum, transverse spin and orbital angular momentum'' and by the P2I (``Physique des deux Infinis'') project.

\appendix

\section{Spinors and polarization four-vectors}
\label{app:1}
We collect in this Appendix the different types of free spinors and polarization vectors. The free canonical Dirac spinor $u(k,\sigma)$ and polarization four-vector $\varepsilon^\mu(k,\sigma)$ are given by
\begin{align}
u(k,\sigma)&=\begin{pmatrix}\sqrt{E+m}\,\mathds{1}\\\sqrt{E-m}\,\frac{\uk\cdot\usigma}{|\uk|}\end{pmatrix}\chi_\sigma,\\
\varepsilon^\mu(k,\sigma)&=\left(\frac{\boldsymbol\epsilon_\sigma\cdot\uk}{m},\boldsymbol\epsilon_\sigma+\frac{\uk\left(\boldsymbol\epsilon_\sigma\cdot\uk\right)}{m(E+m)}\right),
\end{align}
where $\chi_\uparrow=\left(\begin{smallmatrix}1\\0\end{smallmatrix}\right)$, $\chi_\downarrow=\left(\begin{smallmatrix}0\\1\end{smallmatrix}\right)$, and the polarization three-vectors are $\boldsymbol\epsilon_{\Uparrow,\Downarrow}=\frac{1}{\sqrt{2}}\left(\mp 1,-i,0\right)$ for $s_z=\pm 1$, and $\boldsymbol\epsilon_\odot=\left(0,0,1\right)$ for $s_z=0$. The free light-cone Dirac spinor $u_{LC}(k,\lambda)$ and polarization four-vector $\varepsilon^\mu_{LC}(k,\lambda)$ are given by
\begin{align}
u_{LC}(k,+)&=\frac{1}{\sqrt{2^{3/2}k^+}}\begin{pmatrix}\sqrt{2}\,k^++m\\k_R\\\sqrt{2}\,k^+-m\\k_R\end{pmatrix}, &u_{LC}(k,-)&=\frac{1}{\sqrt{2^{3/2}k^+}}\begin{pmatrix}-k_L\\\sqrt{2}\,k^++m\\k_L\\-\sqrt{2}\,k^++m\end{pmatrix},\\
\varepsilon^\mu_{LC}(k,\pm)&=\left[0,\frac{\boldsymbol\epsilon_{\perp\pm}\cdot\uk_\perp}{k^+},\boldsymbol\epsilon_{\perp\pm}\right], &\varepsilon^\mu_{LC}(k,0)&=\frac{1}{m}\left[k^+,\frac{\uk_\perp^2-m^2}{2k^+},\uk_\perp\right],
\end{align}
with $\boldsymbol\epsilon_{\perp\pm}=\tfrac{1}{\sqrt{2}}\left(\mp 1,-i\right)$. Both types of spinors and polarization four-vectors coincide in the rest frame $k_\text{rest}=(m,\uzero)$
\begin{align}
u(k_\text{rest},\sigma)&=u_{LC}(k_\text{rest},\sigma)=\sqrt{2m}\begin{pmatrix}\chi_\sigma\\0\end{pmatrix},\\
\varepsilon^\mu(k_\text{rest},\sigma)&=\varepsilon^\mu_{LC}(k_\text{rest},\sigma)=\left(0,\boldsymbol\epsilon_\sigma\right).
\end{align}

The ``good'' light-cone spinors are the simultaneous eigenstates of the operator $\gamma_5$ and the projector $P_+=\tfrac{1}{2}\,\gamma^-\gamma^+$ 
\begin{equation}
P_+u_G(\lambda)=u_G(\lambda),\qquad \gamma_5u_G(\lambda)=\lambda\,u_G(\lambda),\qquad u_G(\lambda)\equiv\frac{1}{\sqrt{2}}\begin{pmatrix}\mathds{1}\\\sigma_3\end{pmatrix}\chi_\lambda,
\end{equation}
and one can write
\begin{equation}
P_+=\sum_\lambda u_G(\lambda)u^\dag_G(\lambda).
\end{equation}

\section{Components of the 3Q LCWF in the light-cone and canonical polarization bases}\label{table}

Based on Eq.~\eqref{connection}, Table~\ref{3QLCWF} shows explicitly how the components of the 3Q LCWF in light-cone polarization basis are decomposed in the canonical-spin basis. 
\begin{table}
\begin{center}
\caption{\footnotesize{Decomposition in the canonical-spin basis $\psi^\uparrow_{\sigma_1\sigma_2\sigma_3}$ of the components of the 3Q LCWF in the light-cone helicity basis $\psi^+_{\lambda_1\lambda_2\lambda_3}$. The components are grouped according to the values of total orbital angular momentum $\ell_z$.\vspace{.4cm}}}\label{3QLCWF}
\begin{tabular}{lc||c|ccc}
&&$\,\,\ell_z=-1\,\,$&\multicolumn{3}{c}{$\ell_z=0$}\\
&&$\psi^\uparrow_{\uparrow\uparrow\uparrow}$&$\psi^\uparrow_{\uparrow\uparrow\downarrow}$&$\psi^\uparrow_{\uparrow\downarrow\uparrow}$&$\psi^\uparrow_{\downarrow\uparrow\uparrow}$\\\hline
$\ell_z=-1\quad$&$\psi^+_{+++}$
&$z_1z_2z_3$&$z_1z_2l_3$&$z_1l_2z_3$&$l_1z_2z_3$\\\hline
&$\psi^+_{++-}$&$-z_1z_2r_3$
&$z_1z_2z_3$
&$-z_1l_2r_3$
&$-l_1z_2r_3$
\\
$\ell_z=0$&$\psi^+_{+-+}$&$-z_1r_2z_3$
&$-z_1r_2l_3$
&$z_1z_2z_3$&$-l_1r_2z_3$\\
&$\psi^+_{-++}$&$-r_1z_2z_3$&$-r_1z_2l_3$&$-r_1l_2z_3$&$z_1z_2z_3$\\\hline
&$\psi^+_{--+}$&$r_1r_2z_3$&$r_1r_2l_3$&$-r_1z_2z_3$&$-z_1r_2z_3$
\\
$\ell_z=+1$&$\psi^+_{-+-}$&$r_1z_2r_3$&$-r_1z_2z_3$&$r_1l_2r_3$&$-z_1z_2r_3$\\
&$\psi^+_{+--}$&$z_1r_2r_3$&$-z_1r_2z_3$&$-z_1z_2r_3$&$l_1r_2r_3$\\\hline
$\ell_z=+2$&$\psi^+_{---}$&$-r_1r_2r_3$&$r_1r_2z_3$&$r_1z_2r_3$&$z_1r_2r_3$\\
\end{tabular}
\end{center}
%
\vspace{0.5 truecm}
\begin{center}
\begin{tabular}{lc||ccc|c}
&&\multicolumn{3}{c}{$\ell_z=+1$}&$\,\,\ell_z=+2\,\,$\\
&&$\psi^\uparrow_{\downarrow\downarrow\uparrow}$
&$\psi^\uparrow_{\downarrow\uparrow\downarrow}$&$\psi^\uparrow_{\uparrow\downarrow\downarrow}$&$\psi^\uparrow_{\downarrow\downarrow\downarrow}$\\
\hline\hline
$\ell_z=-1\quad$&$\psi^+_{+++}$
&$l_1l_2z_3$&$l_1z_2l_3$&$z_1l_2l_3$&$l_1l_2l_3$\\\hline
&$\psi^+_{++-}$
&$-l_1l_2r_3$&$l_1z_2z_3$&$z_1l_2z_3$&$l_1l_2z_3$\\
$\ell_z=0$&$\psi^+_{+-+}$
&$l_1z_2z_3$&$-l_1r_2l_3$&$z_1z_2l_3$&$l_1z_2l_3$\\
&$\psi^+_{-++}$
&$z_1l_2z_3$&$z_1z_2l_3$&$-r_1l_2l_3$&$z_1l_2l_3$\\\hline
&$\psi^+_{--+}$
&$z_1z_2z_3$&$-z_1r_2l_3$&$-r_1z_2l_3$&$z_1z_2l_3$\\
$\ell_z=+1$&$\psi^+_{-+-}$
&$-z_1l_2r_3$&$z_1z_2z_3$&$-r_1l_2z_3$&$z_1l_2z_3$\\
&$\psi^+_{+--}$
&$-l_1z_2r_3$&$-l_1r_2z_3$&$z_1z_2z_3$&$l_1z_2z_3$\\\hline
$\ell_z=+2$&$\psi^+_{---}$
&$-z_1z_2r_3$&$-z_1r_2z_3$&$-r_1z_2z_3$&$z_1z_2z_3$\\
\end{tabular}
\end{center}
\end{table}
We used for convenience the notations $z_i=\cos\tfrac{\theta_i}{2}$, $l_i=\hat k_{iL}\,\sin\tfrac{\theta_i}{2}$ and $r_i=\hat k_{iR}\,\sin\tfrac{\theta_i}{2}$ for the components of the rotation matrix $D^{(1/2)*}_{\sigma_i\lambda_i}$ of Eq.~\eqref{genMelosh}. For example, from the first row of Table~\ref{3QLCWF}, we have
\begin{multline}
\qquad\psi^+_{+++}=z_1z_2z_3\,\psi^\uparrow_{\uparrow\uparrow\uparrow}+z_1z_2l_3\,\psi^\uparrow_{\uparrow\uparrow\downarrow}+z_1l_2z_3\,\psi^\uparrow_{\uparrow\downarrow\uparrow}+l_1z_2z_3\,\psi^\uparrow_{\downarrow\uparrow\uparrow}\\
+l_1l_2z_3\,\psi^\uparrow_{\downarrow\downarrow\uparrow}+l_1z_2l_3\,\psi^\uparrow_{\downarrow\uparrow\downarrow}+z_1l_2l_3\,\psi^\uparrow_{\uparrow\downarrow\downarrow}+l_1l_2l_3\,\psi^\uparrow_{\downarrow\downarrow\downarrow}.\qquad
\end{multline}
It is interesting to note that any single component of the 3Q LCWF in the canonical-spin basis contributes to all components in the light-cone helicity basis, and \emph{vice-versa}. So even if one considers that the wave function has only components with $\ell_z=0$ in the canonical-spin basis, the components of the wave function in the light-cone helicity basis present all the values $\ell_z=-1,0,+1,+2$, the orbital angular momentum being generated by the rotation matrices $D^{(1/2)*}_{\sigma_i\lambda_i}$.

\section{Connection to a quark-diquark picture}

We show in this Appendix how the 3Q picture can be connected to a quark-diquark picture. In the latter, one considers the whole spectator system as an object with the quantum numbers of two quarks, namely a diquark. One may also assume that this diquark does not contain any internal orbital angular momentum. From a 3Q picture, this amounts to set $\tilde k_2=\tilde k_3=\tilde k_D/2$ and $m_D=2m$ with $\tilde k_D$ and $m_D$ the light-cone momentum and mass of the diquark, and $m$ the mass of a valence quark.

\subsection{Scalar diquark}
\label{app-c1}
The scalar diquark is obtained by coupling the two spectator quarks so to form a system with total angular momentum $j=0$. The LCWF of the scalar quark-diquark system can be written in terms of the 3Q LCWF as follows
\begin{equation}
\psi^\Lambda_\lambda(\tilde k,\tilde k_D)=\tfrac{1}{\sqrt{2}}\left[\psi^\Lambda_{\lambda+-}(\tilde k,\tfrac{\tilde k_D}{2},\tfrac{\tilde k_D}{2})-\psi^\Lambda_{\lambda-+}(\tilde k,\tfrac{\tilde k_D}{2},\tfrac{\tilde k_D}{2})\right].
\end{equation}
The total orbital angular momentum of a given component $\psi^\Lambda_\lambda$ is given by the expression $\ell_z=\Lambda-\lambda$ with $\Lambda,\lambda=\pm\tfrac{1}{2}$. 

The corresponding LCWF in the canonical-spin basis is defined through
\begin{equation}\label{connectionS}
\psi^\Lambda_\lambda=\sum_\sigma\psi^\Lambda_\sigma\,D^{(1/2)*}_{\sigma\lambda},
\end{equation}
and can consistently be written as
\begin{equation}\label{dualS}
\psi^\Lambda_\sigma(\tilde k,\tilde k_D)=\tfrac{1}{\sqrt{2}}\left[\psi^\Lambda_{\sigma\uparrow\downarrow}(\tilde k,\tfrac{\tilde k_D}{2},\tfrac{\tilde k_D}{2})-\psi^\Lambda_{\sigma\downarrow\uparrow}(\tilde k,\tfrac{\tilde k_D}{2},\tfrac{\tilde k_D}{2})\right].
\end{equation}
The explicit decomposition of Eq.~\eqref{connectionS} is displayed in Table~\ref{SLCWF}. 
\begin{table}[h!]
\begin{center}
\caption{\footnotesize{Decomposition in the canonical-spin basis $\psi^\uparrow_\sigma$ of the components of the scalar quark-diquark LCWF in the light-cone helicity basis $\psi^+_\lambda$. The components are grouped according to the values of total orbital angular momentum $\ell_z$.\vspace{.4cm}}}\label{SLCWF}
\begin{tabular}{lc||c|c}
&&$\,\,\ell_z=0\,\,$&$\,\,\ell_z=+1\,\,$\\
&&$\psi^\uparrow_\uparrow$&$\psi^\uparrow_\downarrow$\\
\hline
$\ell_z=0$&$\psi^+_+$&$z$&$l$\\\hline
$\ell_z=+1\quad$&$\psi^+_-$&$-r$&$z$
\end{tabular}
\end{center}
\end{table}

Spherical symmetry in the canonical-spin basis reads
\begin{equation}
\sum_{\Lambda'\sigma'}\left[u(\theta,\phi)\right]_{\sigma\sigma'}\left[u(\theta,\phi)\right]^*_{\Lambda\Lambda'}\psi^{\Lambda'}_{\sigma'}=\psi^\Lambda_\sigma,
\end{equation}
and in particular implies
\begin{subequations}\label{sphericalS}
\begin{align}
\psi^{-\Lambda}_{-\sigma}&=(-1)^{\Lambda-\sigma}\,\psi^\Lambda_\sigma,\\
\psi^\uparrow_\downarrow&=0,
\end{align}
\end{subequations}
in agreement with Eqs.~\eqref{rotinv1}, \eqref{rotinv2} and \eqref{dualS}.

\subsection{Axial-vector diquark}

The axial-vector diquark is obtained by coupling the two spectator quarks so to form a system with total angular momentum $j=1$. The LCWF of the axial-vector quark-diquark system can be written in terms of the 3Q LCWF as follows
\begin{equation}
\begin{split}
\psi^\Lambda_{\lambda+}(\tilde k,\tilde k_D)&=\psi^\Lambda_{\lambda++}(\tilde k,\tfrac{\tilde k_D}{2},\tfrac{\tilde k_D}{2}),\\
\psi^\Lambda_{\lambda 0}(\tilde k,\tilde k_D)&=\tfrac{1}{\sqrt{2}}\left[\psi^\Lambda_{\lambda+-}(\tilde k,\tfrac{\tilde k_D}{2},\tfrac{\tilde k_D}{2})+\psi^\Lambda_{\lambda-+}(\tilde k,\tfrac{\tilde k_D}{2},\tfrac{\tilde k_D}{2})\right],\\
\psi^\Lambda_{\lambda-}(\tilde k,\tilde k_D)&=\psi^\Lambda_{\lambda--}(\tilde k,\tfrac{\tilde k_D}{2},\tfrac{\tilde k_D}{2}).
\end{split}
\end{equation}
The total orbital angular momentum of a given component $\psi^\Lambda_{\lambda\lambda_D}$ is given by the expression $\ell_z=\Lambda-\lambda-\lambda_D$ with $\Lambda,\lambda=\pm\tfrac{1}{2}$ and $\lambda_D=+1,0,-1$. 

The corresponding LCWF in the canonical-spin basis is defined through
\begin{equation}\label{connectionAV}
\psi^\Lambda_{\lambda\lambda_D}=\sum_{\sigma\sigma_D}\psi^\Lambda_{\sigma\sigma_D}\,D^{(1/2)*}_{\sigma\lambda}\,D^{(1)*}_{\sigma_D\lambda_D},
\end{equation}
with the rotation for the axial-vector diquark given by
\begin{equation}
D^{(1)*}_{\sigma_D\lambda_D}(\tilde k_D)=\begin{pmatrix}\frac{1+\cos\theta_D}{2}&-\frac{\hat k_R}{\sqrt{2}}\,\sin\theta_D&\hat k_R^2\,\frac{1-\cos\theta_D}{2}\\
\frac{\hat k_L}{\sqrt{2}}\,\sin\theta_D&\cos\theta_D&-\frac{\hat k_R}{\sqrt{2}}\,\sin\theta_D\\
\hat k_L^2\,\frac{1-\cos\theta_D}{2}&\frac{\hat k_L}{\sqrt{2}}\,\sin\theta_D&\frac{1+\cos\theta_D}{2}
\end{pmatrix},
\end{equation}
or equivalently
\begin{equation}\label{genMelosh2}
D^{(1)*}_{\sigma_D\lambda_D}(\tilde k_D)=\begin{pmatrix}\cos^2\tfrac{\theta_D}{2}&-\sqrt{2}\,\hat k_R\,\sin\tfrac{\theta_D}{2}\cos\tfrac{\theta_D}{2}&\hat k_R^2\,\sin^2\tfrac{\theta_D}{2}\\
\sqrt{2}\,\hat k_L\,\sin\tfrac{\theta_D}{2}\cos\tfrac{\theta_D}{2}&\cos^2\tfrac{\theta_D}{2}-\sin^2\tfrac{\theta_D}{2}&-\sqrt{2}\,\hat k_R\,\sin\tfrac{\theta_D}{2}\cos\tfrac{\theta_D}{2}\\
\hat k_L^2\,\sin^2\tfrac{\theta_D}{2}&\sqrt{2}\,\hat k_L\,\sin\tfrac{\theta_D}{2}\cos\tfrac{\theta_D}{2}&\cos^2\tfrac{\theta_D}{2}
\end{pmatrix}.
\end{equation}
Provided that $\theta_D(\tilde k_D)=\theta(\tilde k_D/2)$, we can consistently write the axial-vector quark-diquark LCWF in the canonical-spin basis as
\begin{equation}\label{dualAV}
\begin{split}
\psi^\Lambda_{\sigma\Uparrow}(\tilde k,\tilde k_D)&=\psi^\Lambda_{\sigma\uparrow\uparrow}(\tilde k,\tfrac{\tilde k_D}{2},\tfrac{\tilde k_D}{2}),\\
\psi^\Lambda_{\sigma\odot}(\tilde k,\tilde k_D)&=\tfrac{1}{\sqrt{2}}\left[\psi^\Lambda_{\sigma\uparrow\downarrow}(\tilde k,\tfrac{\tilde k_D}{2},\tfrac{\tilde k_D}{2})+\psi^\Lambda_{\sigma\downarrow\uparrow}(\tilde k,\tfrac{\tilde k_D}{2},\tfrac{\tilde k_D}{2})\right],\\
\psi^\Lambda_{\sigma\Downarrow}(\tilde k,\tilde k_D)&=\psi^\Lambda_{\sigma\downarrow\downarrow}(\tilde k,\tfrac{\tilde k_D}{2},\tfrac{\tilde k_D}{2}).
\end{split}
\end{equation}
The explicit decomposition of Eq.~\eqref{connectionAV} is displayed in Table~\ref{AVLCWF}.
\begin{table}[h!]
\begin{center}
\caption{\footnotesize{Decomposition in the canonical-spin basis $\psi^\uparrow_{\sigma\sigma_D}$ of the components of the axial-vector quark-diquark LCWF in the light-cone helicity basis $\psi^+_{\lambda\lambda_D}$. The components are grouped according to the values of total orbital angular momentum $\ell_z$.\vspace{.4cm}}}\label{AVLCWF}
\begin{tabular}{lc||c|cc}
&&$\,\,\ell_z=-1\,\,$&\multicolumn{2}{c}{$\ell_z=0$}\\
&&$\psi^\uparrow_{\uparrow\Uparrow}$&$\psi^\uparrow_{\uparrow\odot}$&$\psi^\uparrow_{\downarrow\Uparrow}$
\\
\hline
$\ell_z=-1\quad$&$\psi^+_{++}$&$zz^2_D$&$\sqrt{2}\,zz_Dl_D$&$lz^2_D$
\\\hline
\multirow{2}{*}{$\ell_z=0$}&$\psi^+_{+0}$&$-\sqrt{2}\,zz_Dr_D$&$z\left(z^2_D-r_Dl_D\right)$&$-\sqrt{2}\,lz_Dr_D$
\\
&$\psi^+_{-+}$&$-rz^2_D$&$-\sqrt{2}\,rz_Dl_D$&$zz^2_D$
\\\hline
\multirow{2}{*}{$\ell_z=+1$}&$\psi^+_{-0}$&$\sqrt{2}\,rz_Dr_D$&$-r\left(z^2_D-r_Dl_D\right)$&$-\sqrt{2}\,zz_Dr_D$
\\
&$\psi^+_{+-}$&$zr^2_D$&$-\sqrt{2}\,zz_Dr_D$&$lr^2_D$
\\\hline
$\ell_z=+2$&$\psi^+_{--}$&$-rr^2_D$&$\sqrt{2}\,rz_Dr_D$&$zr^2_D$
\end{tabular}
\end{center}
\vspace{0.5 truecm}
\begin{center}
\begin{tabular}{lc||cc|c}
&&\multicolumn{2}{c|}{$\,\,\ell_z=+1\,\,$}&$\,\,\ell_z=+2\,\,$\\
&&$\psi^\uparrow_{\downarrow\odot}$
&$\psi^\uparrow_{\uparrow\Downarrow}$&$\psi^\uparrow_{\downarrow\Downarrow}$\\
\hline
$\ell_z=-1\quad$&$\psi^+_{++}$
&$\sqrt{2}\,lz_Dl_D$&$zl^2_D$&$ll^2_D$\\\hline
\multirow{2}{*}{$\ell_z=0$}&$\psi^+_{+0}$
&$\,l\left(z^2_D-r_Dl_D\right)$&$\sqrt{2}\,zz_Dl_D$&$\sqrt{2}\,lz_Dl_D$\\
&$\psi^+_{-+}$
&$\sqrt{2}\,zz_Dl_D$&$-rl^2_D$&$zl^2_D$\\\hline
\multirow{2}{*}{$\ell_z=+1$}&$\psi^+_{-0}$
&$\,z\left(z^2_D-r_Dl_D\right)$&$-\sqrt{2}\,rz_Dl_D$&$\sqrt{2}\,zz_Dl_D$\\
&$\psi^+_{+-}$
&$-\sqrt{2}\,lz_Dr_D$&$zz^2_D$&$lz^2_D$\\\hline
$\ell_z=+2$&$\psi^+_{--}$
&$-\sqrt{2}\,zz_Dr_D$&$-rz^2_D$&$zz^2_D$
\end{tabular}
\end{center}
\end{table}

Spherical symmetry in the canonical-spin basis reads
\begin{equation}
\sum_{\Lambda'\sigma'\sigma'_D}\left[u(\theta,\phi)\right]_{\sigma\sigma'}\left[U(\theta,\phi)\right]_{\sigma_D\sigma'_D}\left[u(\theta,\phi)\right]^*_{\Lambda\Lambda'}\psi^{\Lambda'}_{\sigma'\sigma'_D}=\psi^\Lambda_{\sigma\sigma_D},
\end{equation}
where
\begin{equation}
U(\theta,\phi)=\begin{pmatrix}\frac{1+\cos\theta}{2}\,e^{-i\phi}&-\frac{1}{\sqrt{2}}\,\sin\theta\,e^{-i\phi}&\frac{1-\cos\theta}{2}\,e^{-i\phi}\\
\frac{1}{\sqrt{2}}\,\sin\theta&\cos\theta&-\frac{1}{\sqrt{2}}\,\sin\theta\\
\frac{1-\cos\theta}{2}\,e^{i\phi}&\frac{1}{\sqrt{2}}\,\sin\theta\,e^{i\phi}&\frac{1+\cos\theta}{2}\,e^{i\phi}
\end{pmatrix},
\end{equation}
and in particular implies 
\begin{subequations}\label{sphericalAV}
\begin{align}
\psi^{-\Lambda}_{-\sigma-\sigma_D}&=(-1)^{\Lambda+\sigma+\sigma_D}\,\psi^\Lambda_{\sigma\sigma_D},\\
\psi^\uparrow_{\uparrow\Uparrow}&=\psi^\uparrow_{\downarrow\odot}=\psi^\uparrow_{\uparrow\Downarrow}=\psi^\uparrow_{\downarrow\Downarrow}=0,\\
\psi^\uparrow_{\downarrow\Uparrow}&=-\sqrt{2}\,\psi^\uparrow_{\uparrow\odot},
\end{align}
\end{subequations}
in agreement with Eqs.~\eqref{rotinv1}-\eqref{rotinv3} and \eqref{dualAV}.

\end{document}